\documentclass[fleqn,usenatbib]{mnras}

\usepackage{newtxtext,newtxmath}

\usepackage[T1]{fontenc}

\DeclareRobustCommand{\VAN}[3]{#2}
\let\VANthebibliography\thebibliography
\def\thebibliography{\DeclareRobustCommand{\VAN}[3]{##3}\VANthebibliography}

\usepackage{graphicx}	
\usepackage{amsmath}	
\newcommand{\colibre}{\textsc{colibre}}

\title[Progenitors of $z\gtrsim10$ JWST galaxies in COLIBRE]{The progenitors of $z\gtrsim10$ JWST galaxies in the COLIBRE simulations}

\author[E. Chaikin et al.]{Evgenii Chaikin,$^{1,2}$\thanks{E-mail: evgenii.chaikin@durham.ac.uk} Andrew Pontzen,$^{1}$ Carlos S. Frenk,$^{1}$ Joop Schaye,$^{2}$ Shengdong Lu,$^{1}$ \newauthor Robert A. Crain,$^{3}$ and Anna Durrant$^{3}$
\\
$^{1}$Institute for Computational Cosmology, Department of Physics, University of Durham, South Road, Durham, DH1 3LE, UK\\
$^{2}$Leiden Observatory, Leiden University, PO Box 9513, 2300 RA Leiden, the Netherlands \\
$^{3}$Astrophysics Research Institute, Liverpool John Moores University, 146 Brownlow Hill, Liverpool L3 5RF, UK
}

\date{Accepted XXX. Received YYY; in original form ZZZ}
 
\pubyear{\the\year{}}

\begin{document}
\label{firstpage}
\pagerange{\pageref{firstpage}--\pageref{lastpage}}
\maketitle

\begin{abstract}
\textit{JWST} has revealed a large population of luminous galaxies ($M_{\rm UV}\lesssim -20$) at redshifts $z \gtrsim 10$, widely interpreted as posing a challenge to models of galaxy formation within the $\Lambda$CDM cosmology. Here, we search for counterparts of the \textit{JWST} galaxies in the \colibre{} simulations of galaxy formation. Although these simulations have not been tuned to reproduce any $z > 0$ observations, we find a population of \colibre{} galaxies with properties similar to those of the \textit{JWST} galaxies, and trace them to their earliest evolutionary phases, $z\simeq25$, to investigate the onset of galaxy formation. We study the evolution of galaxy stellar masses, sizes, star formation rates, UV magnitudes, metallicities, central black hole masses, and molecular gas and dust content, finding good agreement with observationally inferred properties at $z > 10$, except for UV magnitudes and dust masses, which \colibre{} underpredicts and overpredicts, respectively. Our results indicate that the standard galaxy formation physics and $\Lambda$CDM cosmology adopted in \colibre{} are sufficient to reproduce a broad range of properties of the most extreme $z > 10$ \textit{JWST} galaxies — including their compact sizes, stellar masses, gas content, and metallicities. We show that the discrepancies with the UV magnitudes and dust masses can both be attributed to the uncertain rate of grain growth at high redshift, possibly alongside a top-heavy stellar initial mass function. These findings provide strong evidence that the standard cosmological model can naturally explain even the most extreme galaxies in the early Universe.
\end{abstract}

\begin{keywords}
methods: numerical – galaxies: formation – galaxies: evolution – galaxies: high-redshift
\end{keywords}

\section{Introduction}

One of the major goals of modern astronomy is to understand the astrophysical processes in the early Universe that give rise to galaxies. It is well established that galaxies form by the infall and condensation of baryons into dark matter (DM) haloes \citep{1978MNRAS.183..341W}, which originate from primordial density fluctuations generated during cosmic inflation and grow hierarchically through gravitational instability driven by cold dark matter  \citep[CDM;][]{1991ApJ...379...52W,1993MNRAS.264..201K,2000MNRAS.319..168C}. Starting from the $\Lambda$CDM cosmological model, modern cosmological simulations of galaxy formation in representative volumes, such as \textsc{Eagle} \citep{2015MNRAS.446..521S,2015MNRAS.450.1937C}, \textsc{IllustrisTNG} \citep{2018MNRAS.475..648P,2018MNRAS.475..624N}, and \textsc{Simba} \citep{2019MNRAS.486.2827D}, as well as semi-analytical models such as the upgraded version of \textsc{Galform} \citep{2016MNRAS.462.3854L} and \textsc{Shark} \citep{2018MNRAS.481.3573L}, broadly reproduce a wide range of observed galaxy properties in the local Universe \citep[see][for a recent review]{2023ARA&A..61..473C}. However, matching observational data at higher redshift has proven more difficult, especially in light of the recent constraints from \textit{JWST}.

Soon after its launch at the end of 2021, \textit{JWST} discovered a large population of luminous ($M_{\rm UV}\lesssim -20$) galaxies at redshifts $z\gtrsim 10$ \citep[e.g.][]{2022ApJ...938L..15C,2022ApJ...940L..55F,2022ApJ...940L..14N,2023NatAs...7..611R}, with measured UV luminosity functions (UVLFs) extending to $12 \lesssim z \lesssim 16$ \citep{2023ApJ...951L...1P,2023MNRAS.518.6011D,2023MNRAS.523.1036B,2023ApJS..265....5H}. The majority of high-$z$ \textit{JWST} galaxies were first detected in \textit{JWST}/NIRCam imaging data through the Lyman-dropout technique, with the most promising objects subsequently followed up with deep \textit{JWST}/NIRSpec and/or \textit{JWST}/MIRI observations \citep[e.g.][]{2023NatAs...7..622C,2023A&A...677A..88B, 2023ApJ...957L..34W, 2024ApJ...976..160H,2024ApJ...960...56H, 2024Natur.633..318C}, as well as with the Atacama Large Millimeter/submillimeter Array \citep[ALMA; e.g.][]{2025A&A...696A..87C, 2026ApJ..1000..159M}. Although some of the early \textit{JWST}/NIRCam-selected galaxy candidates later turned out to be low-redshift interlopers \citep[e.g.,][]{2023ApJ...943L...9Z,2023Natur.622..707A}, the consensus that there is a high abundance of bright galaxies at high redshift has only strengthened \citep{2025ApJ...980..138H,2026ApJ..1002..136W}.

Currently, more than 30 galaxies at $z>10$ have spectroscopically confirmed redshifts \citep{2026ApJ..1001...38T,2026MNRAS.548ag701R}. Among the most notable examples are JADES-GS-z14-0 \citep{2025A&A...696A..87C} and MoM-z14 \citep{2026OJAp....956033N}. These galaxies have spectroscopic redshifts of $z=14.18$ and $z=14.44$, respectively, and absolute UV magnitudes of $M_{\rm UV}=-20.81$ and $M_{\rm UV}=-20.23$, making them the two highest-redshift luminous galaxies known to date. Their stellar masses inferred from stellar population synthesis analysis, $10^8 \lesssim M_*/\mathrm{M_\odot} \lesssim 10^9$, imply extremely rapid stellar mass assembly at early times, since $z\approx14$ corresponds to only $\approx 300$ Myr after the Big Bang. 

The other properties of MoM-z14, JADES-GS-z14-0 and most other luminous ($M_{\rm UV}\lesssim -20$) and relatively massive ($10^{8}\lesssim M_*/\mathrm{M_\odot}\lesssim 10^{9}$) galaxies at $z>10$ \citep[see][]{2023Natur.622..707A,2023NatAs...7..611R,2023ApJ...952...74T,2025NatAs...9..155Z,2026arXiv260121833H,2026ApJ..1002..134D} include (i) a wide range of sizes (half-light radii $50 \lesssim r_{\rm e}/\mathrm{pc} \lesssim 500$), (ii) low $V$-band dust attenuation ($A_{\rm V}\lesssim 0.3$) and very steep UV slopes ($\beta_{\rm UV} \lesssim -2.2$), both indicative of little to no dust content, and (iii) sub-solar metallicities ($Z\lesssim 0.5\,\rm Z_{\odot}$). Different ansatz have been advanced to reproduce these observed properties, including: (i) elevated star-formation efficiencies at high redshift \citep{2023MNRAS.523.3201D,2024A&A...690A.108L,2025MNRAS.544.3774S,2025MNRAS.538.3210B}, (ii) a top-heavy stellar initial mass function (IMF; \citealt{2022ApJ...938L..10I,2024MNRAS.534..523C,2025MNRAS.536.1018L,2026arXiv260526209F}), (iii) bursty star formation \citep{2023MNRAS.521..497M,2024ApJ...975..192G,2025ApJ...989..219S}, (iv) the efficient removal of interstellar dust to large scales by radiation-driven outflows \citep{2023MNRAS.522.3986F,2023ApJ...943L..27F,2024A&A...684A.207F}, and even (v) modifications to the $\Lambda$CDM cosmology \citep{2024MNRAS.533.3923S,2025arXiv250919427S}. 

While a number of theoretical models predicted fewer bright galaxies at high redshift than were observed by \textit{JWST} (see e.g. fig.~17 of \citealt{2023ApJS..265....5H}), the pre-\textit{JWST} predictions of \citet{2018MNRAS.474.2352C}, based on the \textsc{Galform} semi-analytical model with a top-heavy IMF in starbursts  \citep{2016MNRAS.462.3854L}, provided a very good match to the observations up to $z=14$, as subsequently shown by \citet{2025MNRAS.536.1018L}. In spite of the agreement between the predictions of \citet{2018MNRAS.474.2352C} and the \textit{JWST} data, the discrepancies with the data found in other models, together with the very high inferred stellar masses of high-redshift \textit{JWST} galaxies, have sparked intense discussion of whether their existence requires exotic astrophysics or modifications to $\Lambda$CDM \citep{2023NatAs...7..731B,2024ApJ...969L...2F,2023MNRAS.518.2511L,2024ApJ...965...98C,2024MNRAS.527.5929Y}. 

Theoretical predictions from cosmological hydrodynamics galaxy simulations are particularly important for comparisons with \textit{JWST} due to their ability to predict the non-linear, self-consistent evolution of gas, stars, and DM. However, the very low number density of the observed luminous $z>10$ galaxies ($\sim 10^{-6}-10^{-5}~\mathrm{cMpc}^{-3}$; e.g. \citealt{2024ApJ...965...98C}) requires simulated volumes of at least $\sim 100^3~\mathrm{cMpc}^3$ to obtain a sufficient number of simulated counterparts. While there are ongoing efforts to compare predictions from very high-resolution simulations with the high-redshift \textit{JWST} data  \citep[e.g.][]{2024MNRAS.531.3406B,2025OJAp....8E.153K,2025MNRAS.536..988F,2025arXiv251005201K}, such simulations typically have (effective) volumes too small to contain the massive, luminous galaxies observed by \textit{JWST}, and are often not performed down to redshift $z = 0$ for validation. In contrast, previous simulations of sufficiently large (effective) volumes that have been validated at $z = 0$ \citep[e.g.][]{2023MNRAS.524.2594K,2023MNRAS.519.3118W} do not self-consistently model the cold phase of the interstellar medium (ISM) or the evolution of interstellar dust, both of which are important for understanding early galaxy formation.

In this work, we use the \colibre{} simulations of galaxy formation \citep{2026MNRAS.548ag375S,2026MNRAS.548ag300C} to predict the properties of massive, luminous galaxies at $z>10$ and compare them with \textit{JWST} data. The \colibre{} simulations are available at three resolutions, with gas and DM particle masses of $m_{\rm gas} \approx m_{\rm dm} \sim 10^{7}$, $10^{6}$, and $10^{5}~\mathrm{M_\odot}$, in cosmological volumes of up to $400^3$, $200^3$, and $100^3~\mathrm{cMpc}^{3}$, respectively. The \colibre{} model includes a multiphase ISM with non-equilibrium chemistry for hydrogen and helium \citep{2025MNRAS.543..891P}, and employs a detailed self-consistent prescription for the formation and evolution of interstellar dust \citep{2026MNRAS.545f2040T}. The strength of SN and AGN feedback in the simulations was calibrated using machine learning to reproduce the observationally inferred galaxy stellar mass function (GSMF) and size--stellar mass relation in the local Universe \citep{2026MNRAS.548ag300C};  no high-redshift data were used to constrain the model. 

\colibre{} has been shown to reproduce the observed $z=0$ luminosity functions from the UV to near-IR and from the far-IR to the submillimetre \citep{2026arXiv260502022L}, the evolution of the GSMF and star formation rates (SFRs) \citep{2026MNRAS.548ag740C}, the galaxy size--stellar mass relation \citep{2026arXiv260326200L}, the mass--metallicity relation \citep{2026arXiv260625995S}, the integrated and spatially resolved Kennicutt-Schmidt relations \citep{2026MNRAS.549ag947L}, the abundance of massive quenched galaxies at high redshift \citep{2025arXiv251216208C,2026MNRAS.548ag740C},  black hole (BH) masses, as well as molecular and atomic gas fractions \citep{2026MNRAS.548ag375S}. 

Here, we investigate the growth and properties of massive galaxies between $z=25$ and 10, predicting the evolution of their stellar masses, halo masses, sizes, SFRs, UV magnitudes, metallicities, central BH masses, and molecular gas and dust content. We compare the \colibre{} predictions with existing \textit{JWST} observations at $z>10$, and make predictions where observational data are not yet available (up to $z = 25$). This paper is structured as follows. In Section~\ref{section: methods}, we describe our methodology; in Section~\ref{section: results}, we present our results; in Section~\ref{section: discussion}, we discuss their implications; and in Section~\ref{section: conclusions}, we summarise our conclusions.

\section{Methods}
\label{section: methods}

\subsection{The COLIBRE simulations}

The \colibre{} simulations of galaxy formation are detailed in \citet{2026MNRAS.548ag375S}. Here we provide a summary.

All simulations were performed using the astrophysical code \textsc{Swift} \citep{2024MNRAS.530.2378S}. The equations of hydrodynamics are solved using the density-energy smoothed particle hydrodynamics (SPH) scheme \textsc{Sphenix} \citep{2022MNRAS.511.2367B}. Gravity is solved using a Fourier-space particle-mesh method for long-range forces and the Fast Multipole Method for short-range forces, with a fixed gravitational softening length for baryons and DM.

The initial conditions (ICs) of the simulations were generated at $z=63$ using the \textsc{monofonIC} code \citep{2020ascl.soft08024H,2021MNRAS.500..663M}, based on second-order Lagrangian perturbation theory. All simulations assume a $\Lambda$CDM cosmology with parameter values from \citet{2022PhRvD.105b3520A} (named `3x2pt + all external constraints'): $\Omega_{\rm m,0} = 0.306$, $\Omega_{\rm b,0} = 0.0486$, $\sigma_8 = 0.807$, $h = 0.681$, $n_{s} = 0.967$, alongside a single massive neutrino species with a mass of 0.06~eV.

The \colibre{} simulations were performed at three resolution levels: m7 (gas and DM particle masses of, respectively, $m_{\rm gas} = 1.47 \times 10^7~\mathrm{M_\odot}$ and $m_{\rm dm} = 1.94 \times 10^7~\mathrm{M_\odot}$), m6 ($m_{\rm gas} = 1.8 \times 10^6~\mathrm{M_\odot}$, $m_{\rm dm} = 2.4 \times 10^6~\mathrm{M_\odot}$), and m5 ($m_{\rm gas} = 2.3 \times 10^5~\mathrm{M_\odot}$, $m_{\rm dm} = 3.0 \times 10^5~\mathrm{M_\odot}$). \colibre{} uses four DM particles for each baryonic particle in the ICs to reduce spurious energy transfer from dynamically hot DM to dynamically cool stars, which can have undesirable effects on galaxy stellar components \citep{2021MNRAS.508.5114L,2023MNRAS.525.5614L}. At m7 resolution, the Plummer-equivalent gravitational softening length of baryons and DM is set to the minimum of $1.4$~proper kpc and $3.6$~ckpc, and at m6 and m5 resolutions it is reduced by a factor of 2 and 4, respectively, relative to that of m7.

\subsubsection{Cooling, chemistry, and dust}

The gas cooling and heating rates are calculated using the non-equilibrium thermochemistry solver \textsc{chimes} \citep{2014MNRAS.440.3349R,2014MNRAS.442.2780R} for hydrogen and helium and their free electrons ($10$ species in total, including the $\mathrm H_2$ molecule). The cooling rates for nine metals (C, N, O, Ne, Mg, Si, S, Ca, and Fe) are instead calculated using \textsc{hybrid-chimes} \citep{2025MNRAS.543..891P}, which are based on species fractions calculated by \textsc{chimes} assuming ionization equilibrium and steady-state chemistry, but rescaled to account for the difference between the non-equilibrium and equilibrium free electron number densities. Unlike previous galaxy simulations of representative volumes that typically did not allow interstellar gas to cool below $T\sim 10^{4}~\mathrm{K}$, in \colibre{} the gas can naturally reach temperatures as low as $\sim 10~\mathrm{K}$.

The radiative cooling and heating rates account for the presence of a modified version of the \citet{2020MNRAS.493.1614F} redshift-dependent homogeneous radiation background from distant galaxies and quasars, following \citet{2020MNRAS.497.4857P}, with hydrogen and helium reionizations completed at $z \approx 7$ and $z \approx 3$, respectively. Inside the ISM (roughly corresponding to gas densities $n_{\rm H} \gtrsim 10^{-2}~\mathrm{cm^{-3}}$ and temperatures $T \lesssim 10^4~\mathrm{K}$), the cooling and heating rates further account for an interstellar radiation field (ISRF), an ionizing cosmic ray rate, and dust. Since \colibre{} does not model radiative transfer, the strength of the ISRF and the cosmic ray ionization rate are approximated to scale with a column density raised to the power\footnote{The power of 1.4 comes from \citet{1998ApJ...498..541K}, who showed that the observed SFR surface density scales with the gas surface density approximately as a power law with an exponent of 1.4 in nearby galaxies.} of 1.4. The column density is calculated from the gas density and temperature using the local Jeans length approximation, accounts for turbulence (assuming a 1D turbulent velocity dispersion of $6$~km~s$^{-1}$), and is capped at high column densities. The ISRF, the cosmic ray rate, and the background radiation from distant galaxies and quasars are shielded based on the column density of gas and dust. Further details can be found in \citet{2025MNRAS.543..891P}.

\begin{figure*}
    \centering
    \includegraphics[width=0.99\linewidth]{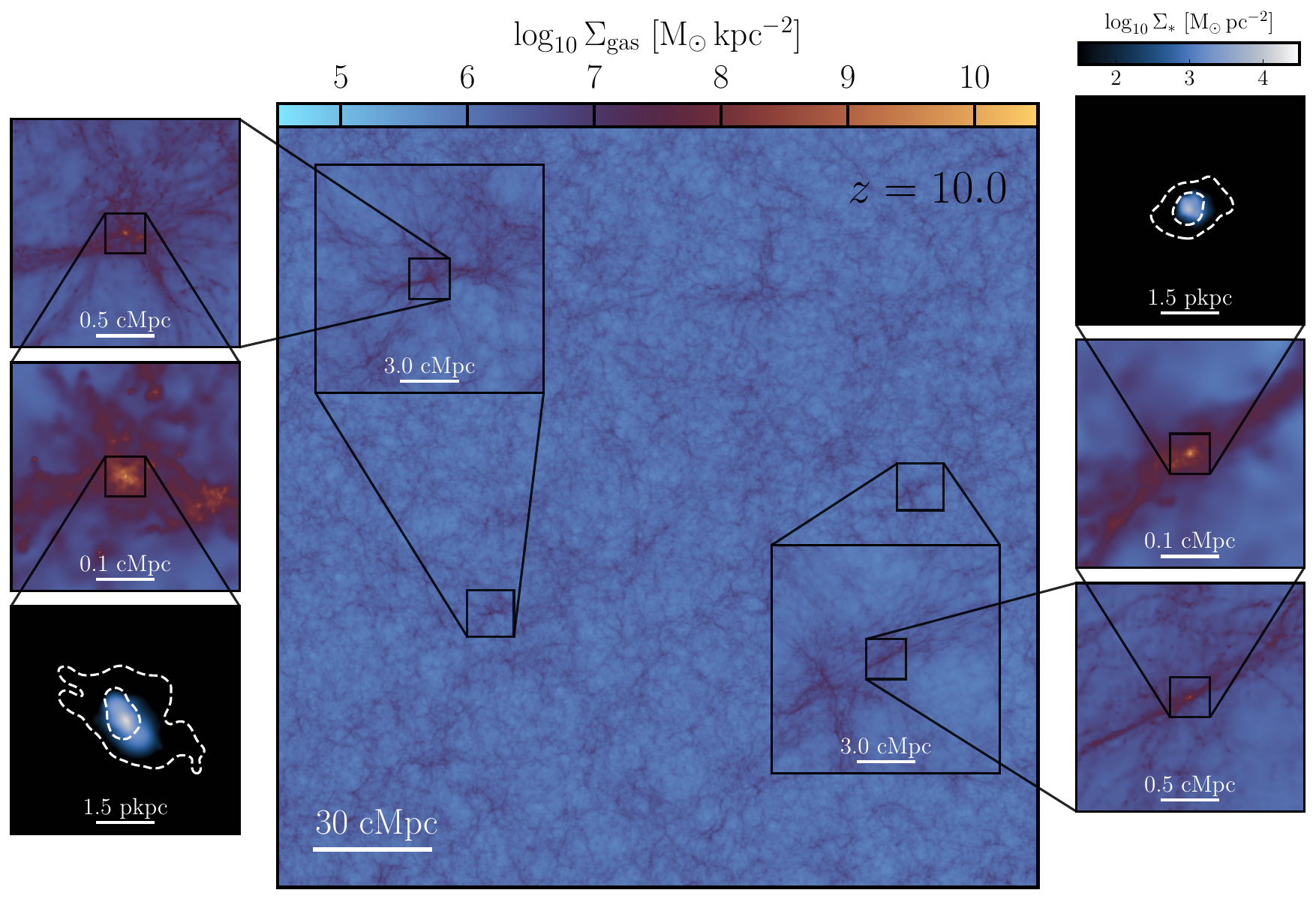}
    \caption{Visual impression of the massive galaxies selected from the \colibre{} L200m6 simulation at redshift $z = 10$. The large central panel shows the projected gas surface density at $z = 10$, computed in a $(200~\mathrm{cMpc})^2$ slab with a depth of $2$~cMpc. The connected smaller panels on the left- and right-hand sides show four successive zoom-ins on two massive galaxies identified within this slab, with stellar masses at $z=10$ of $M_*\approx10^{9.5}~\mathrm{M_\odot}$ and $M_*\approx10^{9.0}~\mathrm{M_\odot}$, respectively. The colours in all but the final zoom-in panel indicate the gas surface density, whereas the final zoom-in panel (shown using a different colour map) displays the stellar surface density. In these panels, we also overplot the dust distribution, indicated by two white dashed contours corresponding to dust surface density thresholds of $0.1$ and $1~\mathrm{M_\odot\,pc^{-2}}$.}
    \label{fig:visualisation}
\end{figure*}

The gas chemical composition is tracked by modelling the abundances of 12 individual elements (H, He, C, N, O, Ne, Mg, Si, Fe, Sr, Ba, and Eu), which are diffused among SPH gas neighbours using a velocity shear-based subgrid model for turbulent mixing \citep{2026MNRAS.548ag645C}. 

The \colibre{} model includes a subgrid prescription for the formation and evolution of dust grains \citep{2026MNRAS.545f2040T}, which is coupled to the \textsc{chimes} solver. The dust model tracks two grain species, graphite and silicates, with the latter further subdivided into Fe- and Mg-rich flavours, as well as two grain sizes, $0.01$ and $0.1~\mu$m (assuming spherical grains), resulting in six grain species in total. Dust grains are produced in winds from asymptotic giant branch (AGB) stars and by core-collapse supernovae (CCSNe), using the dust yields from \citet{2017MNRAS.467.4431D} and \citet{2008A&A...479..453Z}, respectively, with 90~per cent of the dust mass initially in large grains.

Dust grains grow by accreting material from the gas phase and can be eroded as a result of thermal sputtering in hot gas, destroyed by shocks from supernova (SN) and AGN feedback, or removed through astration (i.e. when the gas particle carrying the dust becomes a star). When dust is destroyed, the constituent metal elements are released back into the gas phase. The dust grains also undergo changes in size through shattering and coagulation. Because the resolution of the \colibre{} simulations is insufficient to resolve the dense regions inside giant molecular clouds where grain growth is most efficient, grain growth is boosted by assuming a clumping factor that monotonically increases with gas density up to a value of 100 \citep[see equation (2) in][]{2026MNRAS.548ag375S}.

\subsubsection{Star formation and stellar feedback}

Star-forming gas particles are identified using the gravitational instability criterion following \citet{2024MNRAS.532.3299N} and assigned an SFR based on the \citet{1959ApJ...129..243S} law, assuming a fixed star formation efficiency per free-fall time of $\varepsilon=0.01$. Star-forming gas particles are converted into stellar particles stochastically. Due to the limited resolution of the \colibre{} simulations, star formation in mini-haloes is unresolved.

Newly formed stellar particles represent simple stellar populations characterized by a \citet{2003PASP..115..763C} IMF. Following \citet{2026MNRAS.548ag645C}, stellar particles enrich their surrounding gas with metals through six chemical enrichment channels: AGB stars, type-Ia SNe, CCSNe, neutron star mergers, common envelope jet SNe, and collapsars.

Energy feedback from CCSNe is modelled using the thermal-kinetic stochastic model \citep{2012MNRAS.426..140D,2023MNRAS.523.3709C}, with modifications as described in \citet{2026MNRAS.548ag375S}. Furthermore, \colibre{} includes stochastic energy feedback from type-Ia SNe, as well as three early stellar feedback processes: H~\textsc{ii} regions, stellar winds, and radiation pressure, all of which are also implemented stochastically \citep{2026MNRAS.546ag268B}.

\subsubsection{Black holes and AGN feedback}

Supermassive BHs are modelled using BH particles, which are seeded at high redshift within Friends-of-Friends (FoF) haloes by converting the densest gas particle in a halo into a collisionless BH particle. Seeding occurs when the FoF halo mass, calculated on the fly, exceeds a threshold of $5 \times 10^{10}~\mathrm{M_\odot}$ at m7 resolution or $10^{10}~\mathrm{M_\odot}$ at m6 and m5 resolutions, provided that the FoF halo does not already host a BH.

Once seeded, BHs can grow in mass through accretion of surrounding gas and mergers with other BHs. The BH accretion rate is computed using the Bondi-Hoyle-Lyttleton formula, including turbulence and vorticity corrections from \citet{Krumholz_et_al_2006}. Dynamical friction and BH--BH mergers are modelled following the prescriptions of \citet{2022MNRAS.516..167B}.

At each resolution, the \colibre{} model includes simulations with the fiducial thermal AGN feedback model \citep{2009MNRAS.398...53B}, as well as simulations with a hybrid AGN feedback model that combines kinetic jets with thermal energy injections \citep{2026MNRAS.547ag324H}. The largest \colibre{} volumes are available only for the thermal AGN feedback model \citep[see table~2 in][]{2026MNRAS.548ag375S}. In this work, we therefore exclusively use the simulations with thermal AGN feedback to maximize the statistics of massive galaxies at $z>10$.

Thermal AGN feedback is implemented deterministically following \citet{2009MNRAS.398...53B}, but with BH particles heating neighbouring gas by a temperature increment proportional to the BH (subgrid) mass, $\Delta T_{\rm AGN} \propto m_{\rm BH}$, rather than by a fixed temperature increment. This BH-mass-dependent heating ensures that the frequency of AGN feedback events remains independent of BH mass at fixed Eddington fraction. The BH accretion rate is allowed to exceed the Eddington limit by up to a factor of 100, accelerating BH growth at high redshift and resulting in earlier quenching of massive galaxies \citep{2026MNRAS.550g1148C}.

\subsubsection{Subhalo identification and properties}

Substructures are identified by running a FoF halo finder on the DM particles, using a linking length of 0.2 times the mean separation between DM particles. Baryonic particles, including stars, gas, and BHs, are then attached to the nearest DM particle within the linking length, if any. 

We identify subhaloes using the HBT-HERONS subhalo finder\footnote{\url{https://hbt-herons.strw.leidenuniv.nl/}} \citep{2025MNRAS.543.1339F}, an updated version of the Hierarchical Bound Tracing algorithm (HBT+; \citealt{2018MNRAS.474..604H}). HBT-HERONS identifies subhaloes using an iterative unbinding algorithm applied to particles within spatial FoF groups, starting from the highest-redshift simulation output ($z=30$ in \colibre{}). Once self-bound haloes are found (requiring at least 20 bound particles), their 10 most bound particles (DM or stars) are tracked across subsequent snapshots to follow the evolution of these subhaloes forward in time. At each subsequent snapshot, every candidate subhalo undergoes checks for self-boundness and phase-space consistency to determine whether it remains resolved, has merged with another subhalo, or has been disrupted.

\begin{table*}
\caption{Observational data used in this work. Column (1): name of the observed galaxy; column (2): observed redshift; column (3): absolute UV magnitude; column (4): stellar mass; column (5): SFR over the shortest available timescale; column (6): stellar age; column (7): half-light radius; column (8): gas-phase metallicity; column (9): dust mass; column (10): gas mass; column (11): BH mass. Each row corresponds to an observed galaxy (top to bottom): GN-z11, CEERS2-588, GHZ2, PAN-z14-1, GS-z14-0, and MoM-z14. The properties of GHZ2 and GS-z14-0 have been corrected for gravitational lensing amplification using $\mu=1.3$ \citep{2025NatAs...9..155Z} and $\mu=1.17$ \citep{2024Natur.633..318C}, respectively. The stellar masses and SFRs of PAN-z14-1 \citep{2026ApJ..1002..134D} and MoM-z14 \citep{2026OJAp....956033N}, whose original values are based on the \citet{2001MNRAS.322..231K} and \citet{1955ApJ...121..161S} IMFs, respectively, have been converted to values corresponding to the \citet{2003PASP..115..763C} IMF. The IR-based upper limits on the dust masses are taken from \citet{2024MNRAS.530..340F} for GN-z11 and from \citet{2026ApJ..1000..159M} for GHZ2 and GS-z14-0; we also estimate dust masses from the $V$-band attenuation following \citet{2025A&A...694A.286F} (see $\S$\ref{subsubsection: dust mass} for details).}
\centering
\begin{tabular}{lrrrrrrrrrr}
\hline
Name & Redshift & $M_{\rm UV}$  & $\log_{10}M_*$ & SFR & $t_{\rm age}$  & $r_{\rm e}$  & $12+$ & $ \log_{10}M_{\rm dust}$ &  $\log_{10} M_{\rm gas}$  & $\log_{10} M_{\rm BH}$ \\
&  &  & [$\rm M_\odot$] & [$\rm M_\odot \, yr^{-1}$] & [Myr]  & [pc] & $\log_{10}$(O/H) & [$\rm M_\odot$] & [$\rm M_\odot$] &  [$\rm M_\odot$] \\
\hline

GN-z11 & $^{\rm a}10.60$ & $^{\rm a}$$-21.58^{+0.03}_{-0.04}$ & $^{\rm a}9.1^{+0.3}_{-0.4}$& $^{\rm a}12^{+10}_{-3}$ & $^{\rm a}24^{+20}_{-10}$ & $^{\rm a}64^{+20}_{-20}$ & $^{\rm b}7.91^{+0.07}_{-0.07}$ & $^{\rm c}<6.9$ & & $^{\rm d}6.2^{+0.3}_{-0.3}$ \\

CEERS2-588 & $^{\rm e}11.04$ & $^{\rm e}$$-20.4$ & $^{\rm f}9.05^{+0.07}_{-0.09}$ & $^{\rm f}<4.4$ & $^{\rm f}82^{+33}_{-30}$& $^{\rm g}453^{+120}_{-82}$  & $^{\rm f}8.55^{+0.10}_{-0.13}$ &  & & \\

GHZ2 & $^{\rm h}12.33$ & $^{\rm h}$$-20.53^{+0.01}_{-0.01}$ & $^{\rm i}8.27^{+0.23}_{-0.18}$ & $^{\rm i}9^{+3}_{-3}$ & $^{\rm i}30^{+40}_{-20}$ & $^{\rm j}105^{+9}_{-9}$ & $^{\rm i}7.40^{+0.52}_{-0.37}$ & $^{\rm k}<5.0$ & $^{\rm k}8.76^{+0.21}_{-0.44}$ & \\

PAN-z14-1 & $^{\rm l}13.53$ & $^{\rm l}$$-20.6^{+0.2}_{-0.2}$ & $^{\rm l}8.20^{+1.14}_{-0.21}$ & $^{\rm l}4.5^{+12.8}_{-4.5}$ & $^{\rm l}5^{+61}_{-4}$ & $^{\rm l}233^{+10}_{-10}$ & $^{\rm l}7.29^{+1.21}_{-0.60}$ &  &  & \\
    
GS-z14-0 & $^{\rm m}14.18$ & $^{\rm n}$$-20.81^{+0.16}_{-0.16}$ & $^{\rm m}8.29^{+0.14}_{-0.09}$ & $^{\rm m}14.5^{+1.8}_{-2.2}$ & $^{\rm m}40^{+5}_{-5}$& $^{\rm n}260^{+20}_{-20}$ & $^{\rm m}7.91^{+0.03}_{-0.03}$ & $^{\rm k}<5.3$ & $^{\rm k}8.41^{+0.48}_{-3}$  &\\

MoM-z14 & $^{\rm o}14.44$ & $^{\rm o}$$-20.23^{+0.06}_{-0.06}$ & $^{\rm o}7.89^{+0.3}_{-0.2}$ & $^{\rm o}8.19^{+2.3}_{-2.2}$ & $^{\rm o}4^{+10}_{-1.4}$ & $^{\rm o}74^{+15}_{-12}$ & $^{\rm o}7.33^{+0.65}_{-0.56}$ & & & \\
\hline
\end{tabular}

References: 
$^{\rm a}$\citet{2023ApJ...952...74T}, 
$^{\rm b}$\citet{2025A&A...695A.250A}, 
$^{\rm c}$\citet{2024MNRAS.530..340F}, 
$^{\rm d}$\citet{2024Natur.627...59M}, 
$^{\rm e}$\citet{2024ApJ...960...56H}, 
$^{\rm f}$\citet{2026arXiv260121833H}, 
$^{\rm g}$\citet{2023ApJ...951...72O}, 
$^{\rm h}$\citet{2024ApJ...972..143C}, 
$^{\rm i}$\citet{2025NatAs...9..155Z}, 
$^{\rm j}$\citet{2022ApJ...938L..17Y}, 
$^{\rm k}$\citet{2026ApJ..1000..159M}, 
$^{\rm l}$\citet{2026ApJ..1002..134D}, 
$^{\rm m}$\citet{2025A&A...696A..87C}, 
$^{\rm n}$\citet{2024Natur.633..318C}, 
$^{\rm o}$\citet{2026OJAp....956033N}.
\label{table:observations}
\end{table*}

We process the output of HBT-HERONS using the Spherical Overdensity and Aperture Processor (SOAP; \citealt{McGibbon_2025}) to obtain a list of subhalo and galaxy properties used in this work. Unless otherwise specified, we compute the galaxy stellar mass as the total mass of all gravitationally bound stellar particles within a 50 proper kpc 3D aperture\footnote{We use this aperture as it is the default choice in \colibre{} \citep[see, e.g.,][]{2026MNRAS.548ag375S}, though we note that, for the high-redshift galaxies studied here, it effectively corresponds to selecting \textit{all} bound stellar mass because these galaxies have stellar half-mass radii $\lesssim 1$~kpc (as we will show in Fig.~\ref{fig:matching_size_z})} centred on the most-bound particle of the galaxy's host halo, while the halo mass includes all gravitationally bound particles of any type and at any distance.

\subsection{Radiative transfer post-processing}
\label{subsection: radiative transfer}

In this work, in addition to halo and galaxy properties directly predicted by the \colibre{} simulations, we analyse absolute UV magnitudes and half-light radii. To compute these two properties, we perform 3D radiative transfer calculations in post-processing on the \colibre{} output data using the calibration-free \colibre-\textsc{skirt} pipeline \citep{2026arXiv260714901G}. As shown by the authors, this pipeline reproduces the observed cosmic spectral energy distribution at $z \approx 0$.

Briefly, \textsc{skirt} is a 3D radiative transfer code that self-consistently propagates stellar and nebular radiation through a dusty ISM, including dust absorption, scattering, heating, and thermal re-emission \citep{2011ApJS..196...22B,2015A&C.....9...20C}. We follow the setup presented by \citet{2026arXiv260714901G}, which was also adopted in \citet{2026arXiv260502022L,2026arXiv260506782L}. For a given galaxy, the sources of radiation consist of evolved stars and star-forming regions. Star particles older than $10$~Myr are treated as evolved stars, with their luminosities computed using \textsc{bpass} (v2.2.1; \citealt{2017PASA...34...58E,2018MNRAS.479...75S}) assuming a \citet{2003PASP..115..763C} IMF. The emission from star-forming regions is modelled using (i) existing gas particles with non-zero SFRs averaged over a 10-Myr interval of lookback time and (ii) the parent gas particles of star particles with ages less than 10~Myr.  As discussed in \citet{2026arXiv260714901G}, the inclusion of star-forming gas improves the sampling of young stellar populations, which would otherwise be overly coarse due to the limited numerical resolution of the simulations. The emission from the star-forming regions is calculated using the \textsc{toddlers} library \citep{2023MNRAS.526.3871K,2024A&A...692A..79K}, which is itself based on \textsc{bpass} (v2.2.1), consistent with the treatment of evolved stellar populations. The adopted version of \textsc{toddlers} includes nebular line emission but omits nebular continuum emission \citep[see][for details]{2026arXiv260714901G}. Furthermore, no additional dust below the resolution of the \colibre{} simulations is assumed, implying that the attenuation by dust in \textsc{skirt} radiative transfer calculations is based entirely on the dust abundances and grain size distribution predicted by \colibre{}. 

The radiation spectra produced by evolved stars and star-forming regions span $0.09$ to $1.0~\mu\mathrm{m}$ in wavelength and are sampled using 40 logarithmically spaced bins. This wavelength range is sufficient to compute rest-frame UV magnitudes and half-light sizes in the \textit{JWST} F444W filter using the redshifted spectrum, which are the only \textsc{skirt}-derived properties studied in this work. A total of $10^{7.5}$ photon packets are launched for each galaxy. The galaxies are viewed from a distance of 10~Mpc along the native $z$-axis of the simulation volume, which effectively corresponds to a random orientation of the galaxy. No point-spread function convolution is applied, and no noise is added to the flux maps generated by \textsc{skirt}. 

Galaxy luminosities are measured within a fixed projected circular aperture with a proper radius of 10~kpc, considering only particles gravitationally bound to the galaxy. The 10-kpc aperture is more than 1~dex larger than the projected half-mass radii of massive \colibre{} galaxies at $z>10$ (see Fig.~\ref{fig:matching_size_z}), ensuring that the full galaxy flux is captured. UV magnitudes are calculated at $1500$~\AA{} by convolving the galaxy SED with a top-hat filter covering the wavelength range $1450$--$1550$~\AA. Throughout this work, we report UV magnitudes as rest-frame absolute AB magnitudes.

To estimate galaxy half-light radii, we use the luminosities measured within projected circular apertures logarithmically spaced from 10~proper~pc to 10~proper~kpc with a step of 0.1~dex. For each galaxy, we interpolate the enclosed luminosity as a function of aperture radius and determine the radius containing half of the total luminosity, where the total luminosity is defined as that measured within the largest aperture. To facilitate a consistent comparison with observations, the luminosities are obtained by redshifting the galaxy SED into the observer's frame and convolving it with the transmission curve of the \textit{JWST} F444W filter. As in the case of the UV magnitudes, all galaxies are viewed along the $z$-axis of the simulation domain.

We perform radiative transfer calculations both with and without dust attenuation, allowing us to assess the impact of dust on the UV magnitudes and half-light radii of massive galaxies at high redshift.

\subsection{Observational data}
\label{subsection: observational data}

\begin{figure}
    \centering
    \includegraphics[width=0.99\linewidth]{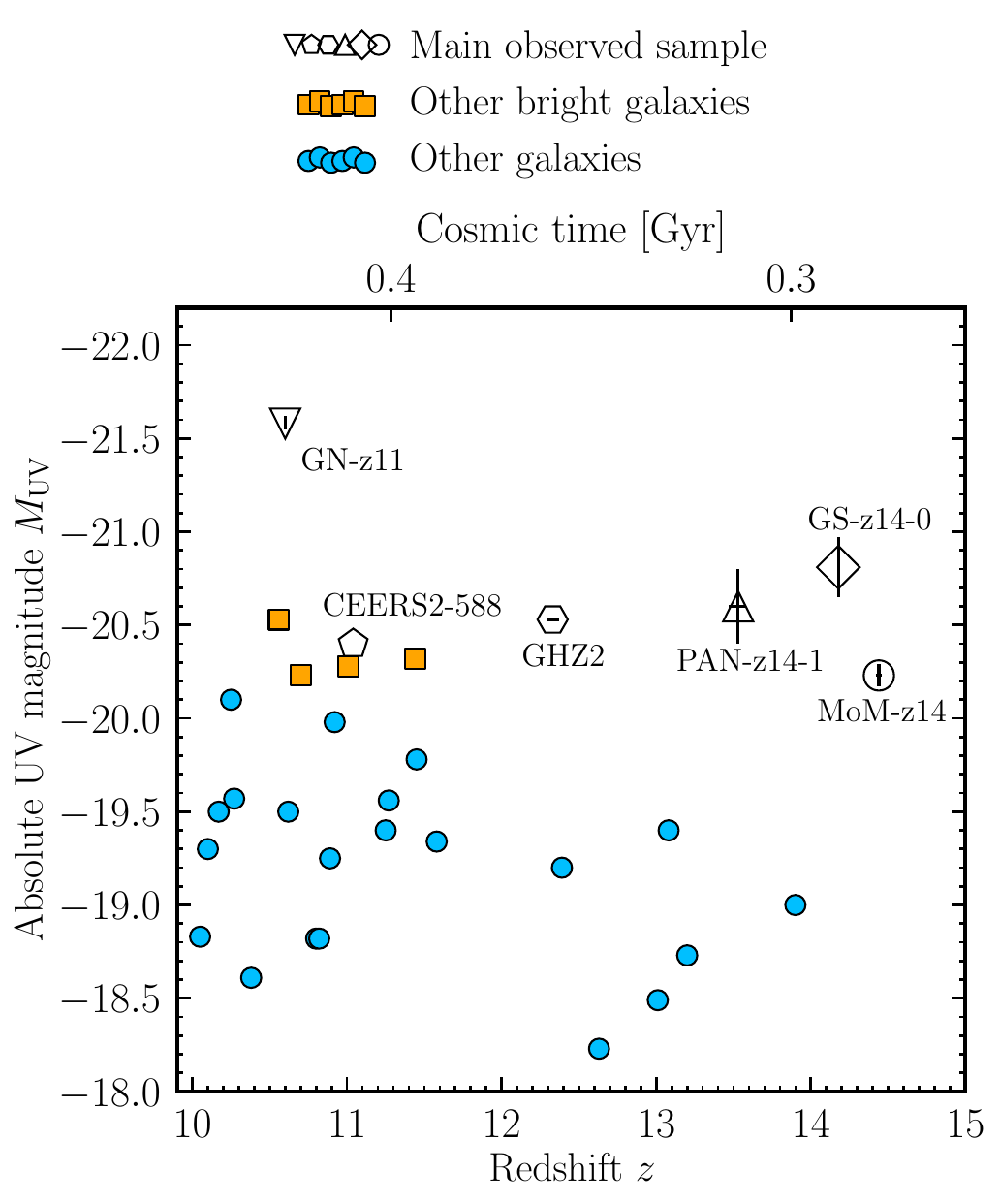}
    \caption{Spectroscopically confirmed \textit{JWST} galaxies at $z > 10$ (the sample is taken from \citealt{2026arXiv260522914F}). From these, we select the six most luminous ($M_{\rm UV}<-20.2$) galaxies at their respective redshifts (black symbols; see Table~\ref{table:observations} for details) as our main observed sample for detailed comparison with predictions from the \colibre{} simulations throughout this work. The other four galaxies that also satisfy $M_{\rm UV}<-20.2$ (orange squares) are studied in Appendix~\ref{appendix:comparison_to_other_bright_galaxies}. The remaining galaxies are shown as blue circles.}
    \label{fig:selection}
\end{figure}

In Section $\S$\ref{subsection:comparison with JWST}, we compare \colibre{} predictions to the properties of the most luminous (absolute UV magnitude\footnote{We chose the UV magnitude cut of $M_{\rm UV}< -20.2$ to ensure that the highest-redshift galaxy identified to date (MoM-z14, $z=14.44$, $M_{\rm UV}=-20.23$) satisfies our selection criterion.} $M_{\rm UV} < -20.2$) and well-studied high-redshift ($z>10$) galaxies with available spectroscopic measurements. Based on the adopted redshift and UV magnitude cuts, we restrict the main comparison to six observed galaxies: GN-z11, CEERS2-588, GHZ2, PAN-z14, GS-z14-0, and MoM-z14, which are the most luminous spectroscopically confirmed galaxies at their respective redshifts (see Fig.~\ref{fig:selection}).  Below we summarise the main properties of these six objects:

\begin{itemize}
    \item  \textit{GN-z11} is a very bright galaxy ($M_{\rm UV} = -21.6$) discovered in the GOODS-North field \citep{2010ApJ...725.1587B,2014ApJ...786..108O}. It was initially reported at a spectroscopically confirmed redshift of $z\approx 11$ based on the detection of the Lyman break \citep{2016ApJ...819..129O}. Its redshift was later refined to $z = 10.6$ \citep{2023A&A...677A..88B} based on multiple high signal-to-noise emission lines detected in the \textit{JWST}/NIRSpec spectrum. \citet{2023ApJ...952...74T} performed SED fitting of \textit{JWST}/NIRCam observations of this galaxy with the \textsc{Prospector} code \citep{2021ApJS..254...22J}, assuming a bursty continuity prior on the star formation history (SFH). They derived a stellar mass of $M_*\approx 10^{9.1}~\mathrm{M_\odot}$, an SFR of $\approx 12~\mathrm{M_\odot}\,\mathrm{yr}^{-1}$ averaged over the past 10~Myr, and found an extremely compact morphology, with a half-light radius of only $\approx 64$~pc.

    \item \textit{CEERS2-588} is a galaxy in the CEERS field \citep{2023ApJ...946L..13F} with a spectroscopic redshift of $z=11.04$, confirmed by the detection of the [O\textsc{ii}]$\lambda3727$ emission line in the \textit{JWST}/NIRSpec PRISM spectrum, together with the observed Lyman break \citep{2026arXiv260121833H}. It has a UV magnitude of $M_{\rm UV}=-20.4$ and an effective radius of $\approx450$~pc. The H$\alpha$ line was not detected in the \textit{JWST}/MIRI data, yielding an H$\alpha$-based SFR upper limit of $\approx4.4~\mathrm{M_\odot\,yr^{-1}}$, while the UV-based SFR is $\approx 8.2~\mathrm{M_\odot\,yr^{-1}}$, suggestive of a recent quenching event. SED fitting was performed using \textsc{Bagpipes} \citep{2018MNRAS.480.4379C} with a continuity prior on the SFH, yielding a stellar mass of $M_*\approx10^{9.1}~\mathrm{M_\odot}$.

    \item \textit{GHZ2} (also known as GLASS-z12) was independently discovered in the GLASS-\textit{JWST} Early Release Science NIRCam field by \citet{2022ApJ...938L..15C} and \citet{2022ApJ...940L..14N}. Its effective radius was measured to be $\approx 100$~pc by \citet{2022ApJ...938L..17Y} using \textit{JWST}/NIRCam imaging. Its spectroscopic redshift was confirmed to be $z=12.33$ by \citet{2024ApJ...972..143C} through the detection of [Ne \textsc{iii}]$\lambda3868$ in the NIRSpec spectrum, and they also reported a UV magnitude of $M_{\rm UV}=-20.53$ for this source. Later, \citet{2025NatAs...9..155Z} detected H$\alpha$ and doubly ionized oxygen ([O\textsc{iii}]$\lambda\lambda4959,5007$) in the \textit{JWST}/MIRI spectra. Using \textsc{Bagpipes} to perform SED modelling and assuming a bursty continuity SFH model, they inferred a stellar mass of $M_*\approx 10^{8.27}~\mathrm{M_\odot}$ and, based on the H$\alpha$ flux, an SFR of $\approx 9~\mathrm{M_\odot~yr^{-1}}$.
    
    \item \textit{PAN-z14} is a luminous galaxy from the PANORAMIC survey \citep{2025ApJ...979..140W}, with a spectroscopically confirmed redshift $z=13.53$ based on fitting the Lyman break in the \textit{JWST}/NIRSpec PRISM spectrum \citep{2026ApJ..1002..134D}. The authors measured a UV magnitude of $M_{\rm UV}=-20.6$ and used \textsc{Bagpipes} to perform SED modelling, yielding a stellar mass of $\approx 10^{8.23}~\mathrm{M_\odot}$, and a 10 Myr-averaged SFR of $\approx 5~\mathrm{M_\odot \, yr^{-1}}$  for a continuity bursty SFH model. The authors also inferred a circularised half-light radius of 233~pc.

    \item \textit{GS-z14-0} was discovered in the \textit{JWST} Advanced Deep Extragalactic Survey (JADES; \citealt{2023ApJS..269...16R,2026ApJS..283....6E}). It has an absolute UV magnitude of $M_{\rm UV}=-20.81$, a spectroscopically confirmed redshift of $z=14.32$ based on the Lyman break in the \textit{JWST}/NIRSpec spectrum, and a half-light radius of $\approx260$~pc \citep{2024Natur.633..318C}. By combining \textit{JWST}/NIRSpec and MIRI data with ALMA observations, \citet{2025A&A...696A..87C} later found a more precise redshift of $z=14.1796$ based on the candidate C{\sc iii}] line and the detected [O\,{\sc iii}] 88~$\mu$m line. By performing SED fitting to all these data with \textsc{Prospector}, assuming a rising prior on the SFH from \citet{2025MNRAS.537.1826T}, the authors inferred a stellar mass of $M_*\approx 10^{8.29}~\mathrm{M_\odot}$ and a 10-Myr averaged SFR of $\approx 14.5~\mathrm{M_\odot}~\mathrm{yr}^{-1}$.

    \item \textit{MoM-z14} is a bright ($M_{\rm UV}=-20.23$) galaxy in the COSMOS Legacy field, spectroscopically confirmed at redshift $z = 14.44$ through the detection of multiple rest-frame UV emission lines in the \textit{JWST}/NIRSpec PRISM spectrum \citep{2026OJAp....956033N}. Using the SED fitting code \textsc{Prospector} with a bursty continuity SFH prior, the authors inferred a stellar mass of $M_*\approx10^{8.1}~\mathrm{M_\odot}$ and SFRs of $\approx 13$ and 2.2~$\mathrm{M_\odot\,yr^{-1}}$ over 5 and 50~Myr timescales, respectively. They also measured a circularised half-light radius of $\approx 74$~pc based on S\'{e}rsic profile fits to imaging of the source.
\end{itemize}

The properties of these six galaxies are summarised in Table~\ref{table:observations}. The quoted uncertainties refer only to the internal uncertainties of the SED-fitting model (i.e. they do not account for systematic errors). More specifically, while UV magnitudes and half-light sizes can be measured directly, inferring physical properties such as stellar masses, SFRs, and stellar ages through SED fitting to the available photometric and spectroscopic data relies on multiple assumptions — including the adopted stellar libraries, stellar evolutionary tracks, nebular continuum, dust attenuation curve, SFH, and IMF — and can therefore significantly bias the inferred values. In particular, uncertainties on stellar mass estimates (including random and systematic uncertainties at fixed IMF) span a range from $0.2$–$0.5$~dex \citep{2023Natur.622..707A,2025ApJ...978L..42C} to $0.6$–$1$~dex \citep{2025NatAs...9..729H,2025NatAs...9..155Z,2026A&A...707A..29C,2026ApJ..1002..134D}.

\begin{table*}
\caption{The outcome of matching simulated galaxies from the \colibre{} L200m6 simulation to luminous \textit{JWST} galaxies at $z>10$ from Table~\ref{table:observations}. The matching was performed in stellar mass at the observed redshift. For each observed galaxy, six matched galaxies were identified in the simulation (see $\S$\ref{subsection: matching} for details). Column (1): name of the observed galaxy; column (2): observed redshift; column (3): observationally inferred stellar mass (best-fit value); columns (4) and (5): median stellar and halo mass, respectively, at the observed redshift in the matched samples of simulated galaxies, together with the differences between the median and the minimum and maximum values in the matched samples; columns (6) and (7): as columns (4) and (5), but at $z=10$.}
	\centering
	\begin{tabular}{lrrrrrr}
   \hline
   Name & $z_{\rm obs}$ & $\log_{10}M_{*,\rm obs}$ & $\log_{10}M_{*,\rm sim}(z_{\rm obs})$ &  $\log_{10}M_{\rm halo,\rm sim}(z_{\rm obs})$  &  $\log_{10}M_{*,\rm sim}(z=10)$ & $\log_{10}M_{\rm halo,\rm sim}(z=10)$  \\
       & & [$\rm M_\odot$] &  [$\rm M_\odot$] &  [$\rm M_\odot$] &  [$\rm M_\odot$] &  [$\rm M_\odot$] \\
   \hline
GN-z11 & 10.60 & 9.10 & $9.11^{+0.19}_{-0.07}$ & $10.94^{+0.07}_{-0.08}$ &$9.41^{+0.14}_{-0.26}$ &$11.02^{+0.24}_{-0.07}$ \\
CEERS2-588 & 11.04 & 9.05 & $9.04^{+0.34}_{-0.11}$ & $10.88^{+0.27}_{-0.21}$ &$9.49^{+0.27}_{-0.26}$ &$11.12^{+0.14}_{-0.10}$ \\
GHZ2 & 12.33 & 8.27 & $8.32^{+0.22}_{-0.15}$ & $10.44^{+0.16}_{-0.07}$ &$8.73^{+0.71}_{-0.22}$ &$10.87^{+0.16}_{-0.25}$ \\
PAN-z14-1 & 13.53 & 8.20 & $8.03^{+0.68}_{-0.08}$ & $10.27^{+0.45}_{-0.18}$ &$8.64^{+1.13}_{-0.35}$ &$10.90^{+0.28}_{-0.27}$ \\
GS-z14-0 & 14.18 & 8.29 & $8.08^{+0.70}_{-0.37}$ & $10.29^{+0.32}_{-0.31}$ &$9.00^{+0.76}_{-0.77}$ &$10.99^{+0.26}_{-0.51}$ \\
MoM-z14 & 14.44 & 7.89 & $7.70^{+0.37}_{-0.05}$ & $9.95^{+0.59}_{-0.06}$ &$8.35^{+1.41}_{-0.30}$ &$10.67^{+0.51}_{-0.19}$ \\
 \hline
\end{tabular}
\label{table:matching}
\end{table*}

For observations that report multiple estimates of inferred physical properties based on different SFH priors, we adopt the values derived using the \textit{bursty continuity} prior from \citet{2022ApJ...927..170T}, if available. Among the advantages of this prior are (i) its non-parametric form (as opposed to, e.g., a delayed exponential prior), which allows for greater flexibility in modelling the shape of the SFH, and (ii) its suitability by design for capturing bursty SFHs, which are expected to occur regularly at high redshift \citep[e.g.][]{2023MNRAS.524.2312E,2024MNRAS.533.1111E,2025A&A...697A..88L,2025ApJ...980..138H,2026ApJ..1000..285R}. This prior has also been shown to perform well in recovering the true stellar mass in mock tests \citep{2025ApJ...978L..42C}. We discuss the impact of using other SFH priors on our results in Appendix~\ref{appendix: SED assumptions}.

Although our sample of the six observed galaxies is relatively small, these galaxies span a broad range of redshifts ($10.6<z<14.44$), stellar masses ($10^{7.9}\lesssim M_*/\mathrm{M_\odot}\lesssim 10^{9.1}$), SFRs ($0<\mathrm{SFR}/\mathrm{M_\odot \, yr^{-1}}<14.5$), sizes ($64<r_{\rm e}/\mathrm{pc}<453$), and metallicities ($7.33<12+\log_{10}(\mathrm{O/H})<8.55$), thereby sampling a representative range of the observed properties of high-redshift galaxies. Among the other well-studied galaxies at $z>10$ with spectroscopic data and inferred physical properties (e.g. table~1 in \citealt{2026arXiv260522914F}), there are four that formally satisfy our adopted UV magnitude threshold ($M_{\rm UV}<-20.2$): EGS-22637 \citep{2026MNRAS.548ag701R}, CAPERS\_UDS\_z10 \citep{2025ApJ...988L..10K}, CAPERS\_UDS\_z11 \citep{2025ApJ...988L..10K}, and MAISIE \citep{2022ApJ...940L..55F,2023Natur.622..707A}, which are denoted by the orange squares in Fig.~\ref{fig:selection}. Although still high, the UV luminosities and inferred stellar masses of these galaxies are somewhat less extreme than those of the six galaxies in our main sample at comparable redshifts. For completeness, in Appendix~\ref{appendix:comparison_to_other_bright_galaxies} we compare \colibre{} predictions with the properties of these four additional galaxies, finding results similar to those obtained for our main observed sample.

\subsection{Matching simulated galaxies to observed galaxies}
\label{subsection: matching}

To predict the properties of the six observed galaxies summarised in Table~\ref{table:observations}, we construct a sample of simulated galaxies from the L200m6 simulation, matched to the observed galaxies in stellar mass at their observed redshifts. We use the L200m6 simulation because it provides a good balance between a sufficiently large volume, required to build a representative sample of matched galaxies, and sufficiently high resolution, enabling the study of star formation at very high redshift. Numerical convergence of the main galaxy properties studied in this work is investigated in Appendix~\ref{appendix: convergence}, where we show that our main results are not driven by numerical effects.

For each observed galaxy, we select six matched galaxies from the L200m6 simulation based on their stellar masses at the observed redshift. The choice of six reflects a compromise between the number of close analogues available in the L200m6 simulation and the need to minimize systematic deviations from the observationally inferred stellar masses. In particular, for CEERS2-588 and GS-z14-0, we find only two simulated galaxies in the entire L200m6 simulation with stellar masses exceeding the observationally inferred values at the observed redshift. Consequently, selecting substantially more than six matched galaxies would require including progressively more lower-mass counterparts, thereby systematically biasing the matched sample towards lower stellar masses.

Since \colibre{} snapshots do not coincide exactly with the observed redshifts, we estimate the stellar masses of simulated galaxies by linearly interpolating their $\log_{10}$ stellar masses between the two adjacent snapshots that bracket the observed redshift (the snapshot spacing is $\Delta z=0.5$ at $10 < z < 15$). To identify the best six matched galaxies, we first select all simulated galaxies whose interpolated stellar masses lie within 0.6~dex of the observed value. This tolerance is adopted to prevent the selection of galaxies with significantly different stellar masses and is motivated by the systematic uncertainties in stellar masses derived from SED fitting for high-redshift galaxies \citep{2025NatAs...9..729H,2025NatAs...9..155Z,2026ApJ..1002..134D}. Because of the shape of the GSMF, the number of lower-mass galaxies satisfying our tolerance is systematically larger than the number of higher-mass galaxies. To minimize this bias, we select three galaxies with stellar masses above and three below the observationally inferred stellar mass whenever possible, always choosing those with the smallest absolute deviations from the observed value. If only two higher-mass galaxies are available within the adopted tolerance, the remaining four galaxies are selected from the lower-mass side.\footnote{We have verified that the conclusions of this work are insensitive to the specific matching procedure adopted and to the number of matched simulated galaxies used. For example, using a slightly different stellar-mass tolerance or selecting all galaxies within the tolerance, rather than restricting the sample to six matched galaxies, yields qualitatively unchanged results.} The final matched samples are summarised in Table~\ref{table:matching}.

\begin{figure*}
    \centering
    \includegraphics[width=0.95\linewidth]{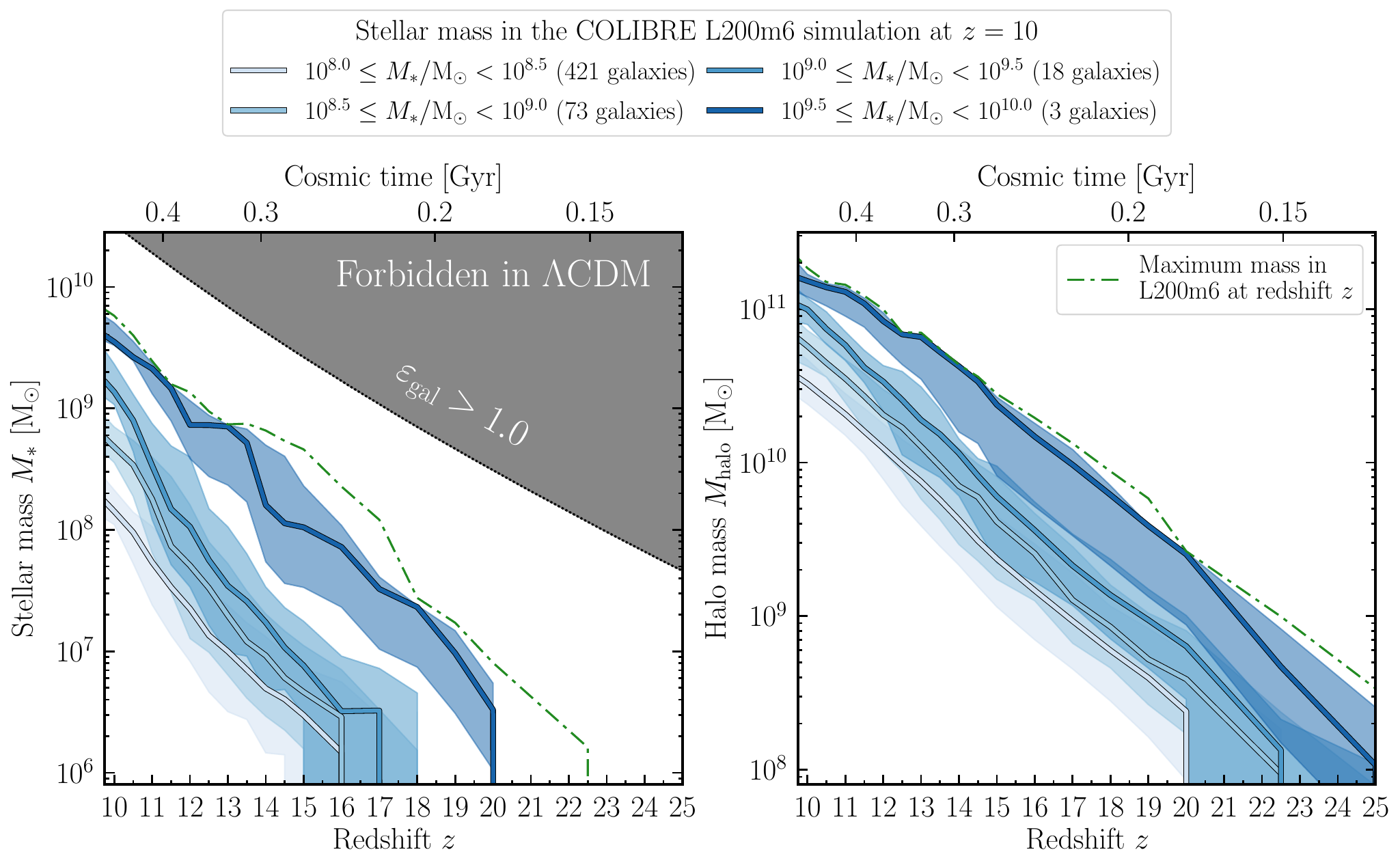}
    \caption{Stellar mass (left) and halo mass (right) versus redshift for massive galaxies from the \colibre{} L200m6 simulation. The differently coloured solid curves show the median values in four $z=10$ stellar mass bins, computed by selecting galaxies according to their stellar mass at $z=10$ and following their main progenitor branch back in time. The shaded regions indicate the 16$^{\rm th}$ to 84$^{\rm th}$ percentiles for the same four bins. The green dash-dotted lines indicate the maximum stellar and halo masses in the simulation at a given redshift (which may correspond to different objects at different times). For reference, the grey region in the left panel shows the area corresponding to galaxy formation efficiencies greater than 100~per cent (see text for details). Galaxies that are most massive at $z=10$ form their first stars as early as $z\approx20$, in haloes of mass $M_{\rm halo}\sim 10^{9}~\mathrm{M_\odot}$. At $10<z<15$, these galaxies reach stellar masses $10^8\lesssim M_*/\mathrm{M_\odot}\lesssim10^{9.5}$ and reside in haloes of mass $10^{10}\lesssim M_{\rm halo}/\mathrm{M_\odot}\lesssim10^{11}$.}
    \label{fig:evolution_mstar}
\end{figure*}

We do not perform matching based on any galaxy properties other than stellar mass, as our goal is to maximize the predictive power of the simulations. Specifically, we ask the following question: if we take the stellar masses inferred for observed high-$z$ galaxies at face value, can we identify galaxies with comparable stellar masses in \colibre{} at the same redshift, and do those simulated galaxies naturally reproduce their other observed properties? We emphasize that \citet{2026MNRAS.548ag740C} showed that \colibre{} reproduces the observed evolution of the GSMF over the redshift range $0 < z < 12$, implying a plausible abundance of massive galaxies at high redshift. We do not repeat those comparisons here and instead refer the reader to that work for further details.

Lastly, we note that UV magnitude could in principle be used as the matching variable instead of stellar mass. However, we do not select galaxies based on their UV magnitudes because the dust-attenuated UVLF predicted by \colibre{} undershoots the observed UVLF at the bright end for $z>7$ \citep{2026arXiv260506782L}. As shown by \citet{durrant2026cosmologicalsimulationshighredshiftgalaxy}, one way to alleviate this discrepancy within the \colibre{} model is by adopting a variable IMF that becomes top-heavy at high redshift, in place of the \citet{2003PASP..115..763C} IMF used in the fiducial \colibre{} simulations. Another (complementary) solution may be to reduce the (potentially overly strong) dust attenuation in \colibre{}, as we will discuss in Section~\ref{section: discussion}.

\section{Results}
\label{section: results}

We begin this section by showing the evolutionary tracks of stellar and halo masses of galaxies in the \colibre{} L200m6 simulation at $z>10$, selected at $z=10$ to have stellar masses $M_* \geq 10^8~\mathrm{M_\odot}$ (\S\ref{subsection: massive galaxies}). We then present a detailed comparison between the properties of a subsample of these galaxies and observed luminous galaxies at $z>10$ with \textit{JWST} spectroscopic data (\S\ref{subsection:comparison with JWST}). The subsample of simulated galaxies is selected following the procedure described in \S\ref{subsection: matching}, while the observational data used for the comparison are introduced in \S\ref{subsection: observational data}. Finally, in \S\ref{subsection: extra_predictions}, we present \colibre{} predictions for additional properties, such as central gas and DM fractions and galaxy formation efficiencies, which currently lack strong observational constraints.

\subsection{The evolution of stellar mass and halo mass of massive COLIBRE galaxies at high redshifts}
\label{subsection: massive galaxies}

\begin{figure*}
    \centering
    \includegraphics[width=0.95\linewidth]{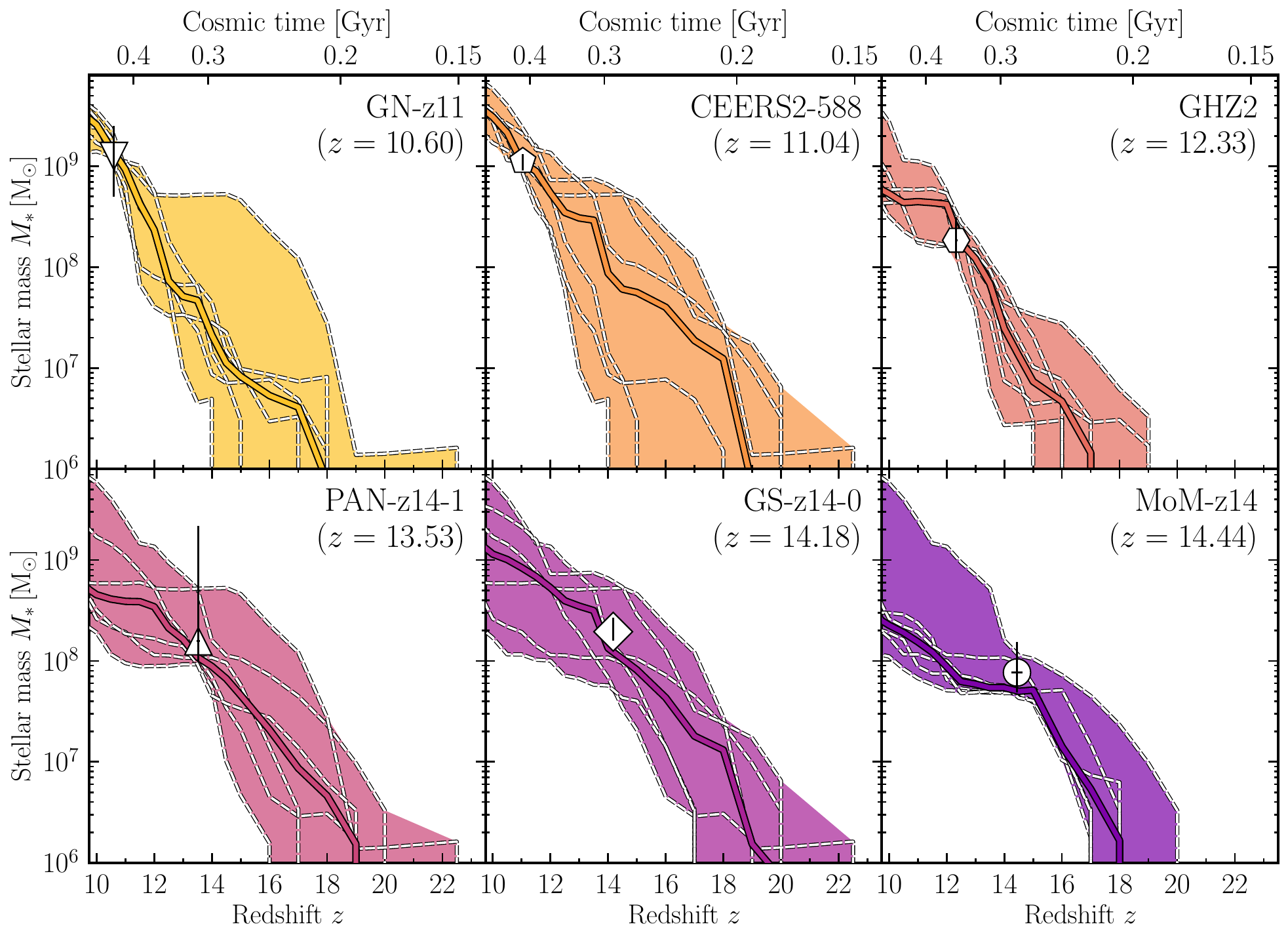}
    \caption{The predicted evolution of stellar mass for six luminous ($M_{\rm UV}<-20.2$) galaxies observed by \textit{JWST} at $z>10$ (six different panels). The inferred stellar masses of the observed galaxies are indicated by the white symbols. Each observed galaxy is matched with six \colibre{} galaxies from the L200m6 simulation, requiring the stellar mass difference at the observed redshift to be no more than 0.6~dex (see text for details). The white dashed lines show the stellar mass evolution along the branch of the main progenitor for individual \colibre{} galaxies that were matched to each observed galaxy. The thick solid line indicates the evolution of the median stellar mass of the matched \colibre{} samples, while the boundaries of the shaded regions show the minimum and maximum stellar masses in these samples. The name and redshift of each observed galaxy are indicated in the top right corner of each panel. All observed galaxies are well matched by \colibre{} in terms of their stellar mass at the observed redshift.}
    \label{fig:matching_Mstar_z}
\end{figure*}

Fig.~\ref{fig:evolution_mstar} shows the redshift evolution of the stellar and halo masses of massive \colibre{} galaxies selected at $z=10$ from the L200m6 simulation to have stellar masses of $M_* \geq 10^8~\mathrm{M_\odot}$. There are $515$ such galaxies in the simulation. We divide these galaxies into four $z=10$ stellar mass bins spanning $10^8 \leq M_*/\mathrm{M_\odot} <10^{10}$, each with a width of 0.5~dex, yielding, from the lowest- to highest-mass bin, 421, 73, 18, and 3 galaxies, respectively. Each galaxy is then tracked back in time along its main progenitor branch, allowing us to determine its stellar and halo masses at earlier cosmic times. The solid curves in the left (right) panel show the median stellar (halo) mass, computed over the galaxies in each bin, with different colours corresponding to the different $z=10$ stellar mass bins. The shaded regions denote the 16$^{\rm th}$ to 84$^{\rm th}$ percentiles for the same four bins.

For reference, in the left panel we overlay a grey region indicating a galaxy formation efficiency greater than 100~per~cent. The lower boundary of this region is computed using the halo mass function fitting function from \citet{2026arXiv260705505Z}, which is calibrated using the Voids-within-Voids-within-Voids \citep{2020Natur.585...39W} and P-Millennium \citep{2019MNRAS.483.4922B} cosmological simulations and provides an accurate description of the halo population across a wide range of halo masses ($10^{-6}<M_{\rm halo}/\mathrm{M_\odot}<10^{15.5}$) and redshifts ($0<z<30$). Specifically, for a simulation volume of $V=(200~\mathrm{cMpc})^3$ at redshift $z$, we first estimate the maximum halo mass, $M_{\rm halo,max}$, by requiring
\begin{equation}
V\int_{M_{\rm halo,max}(z)}^\infty \frac{\mathrm{d}n(M_{\rm halo},z)}{\mathrm{d} M_{\rm halo}} \, \mathrm{d} M_{\rm halo} = 1 \, ,
\label{eq:m_halo_max}
\end{equation}
where $\mathrm{d}n(M_{\rm halo},z)/\mathrm{d}M_{\rm halo}$ is the halo mass function from \citet{2026arXiv260705505Z}. We then multiply $M_{\rm halo,max}$ by $\Omega_{\rm b,0}/\Omega_{\rm m,0} \approx 0.16$ to derive the expected maximum stellar mass assuming a galaxy formation efficiency of 100~per~cent.

As expected, galaxies selected in higher $z=10$ stellar-mass bins reach systematically higher stellar masses and reside in more massive haloes at fixed redshift. The formation of the first star particles in \colibre{}, that end up in galaxies in the highest $z=10$ stellar-mass bin, begins as early as $z \approx 20$, with a minimum non-zero stellar mass of $M_*\sim10^6~\mathrm{M_\odot}$, set by the mass of a single baryonic resolution element in the simulation. The corresponding halo masses of these objects at $z \approx 20 $ are $M_{\rm halo}\sim10^{9}~\mathrm{M_\odot}$. The range of halo masses with non-zero SFRs extends to somewhat lower masses, reaching $M_{\rm halo}\sim10^{8}~\mathrm{M_\odot}$ (not shown), comparable to the critical halo mass at this redshift above which H~\textsc{i} cooling is efficient \citep[e.g.][]{2020MNRAS.498.4887B}.

As the redshift decreases, galaxies and their host haloes undergo rapid mass growth. By $z=14$ and $z=12$, the most massive galaxies reach stellar masses of $M_* \sim 10^{8.0-8.5}~\mathrm{M_\odot}$ and $M_* \sim 10^{8.5-9.0}~\mathrm{M_\odot}$, respectively. These values are comparable to the highest observationally inferred stellar masses at these redshifts, corresponding to the spectroscopically confirmed \textit{JWST} galaxies GS-z14-0 \citep{2024Natur.633..318C} and GHZ2 \citep{2025NatAs...9..155Z}, respectively. At $z=14$, the most massive \colibre{} galaxies reside in haloes with masses of $M_{\rm halo}\sim 10^{10.2-10.6}~\mathrm{M_\odot}$, increasing to $M_{\rm halo}\sim 10^{10.8-11.0}~\mathrm{M_\odot}$ by $z=12$. Finally, by the selection redshift of $z=10$, the galaxies in the highest $z=10$ stellar-mass bin reach, by construction, stellar masses within the range $10^{9.5} \leq M_* < 10^{10.0}~\mathrm{M_\odot}$, while their host haloes grow to masses of $M_{\rm halo}\sim 10^{11.0-11.2}~\mathrm{M_\odot}$.

\begin{figure*}
    \centering
    \includegraphics[width=0.95\linewidth]{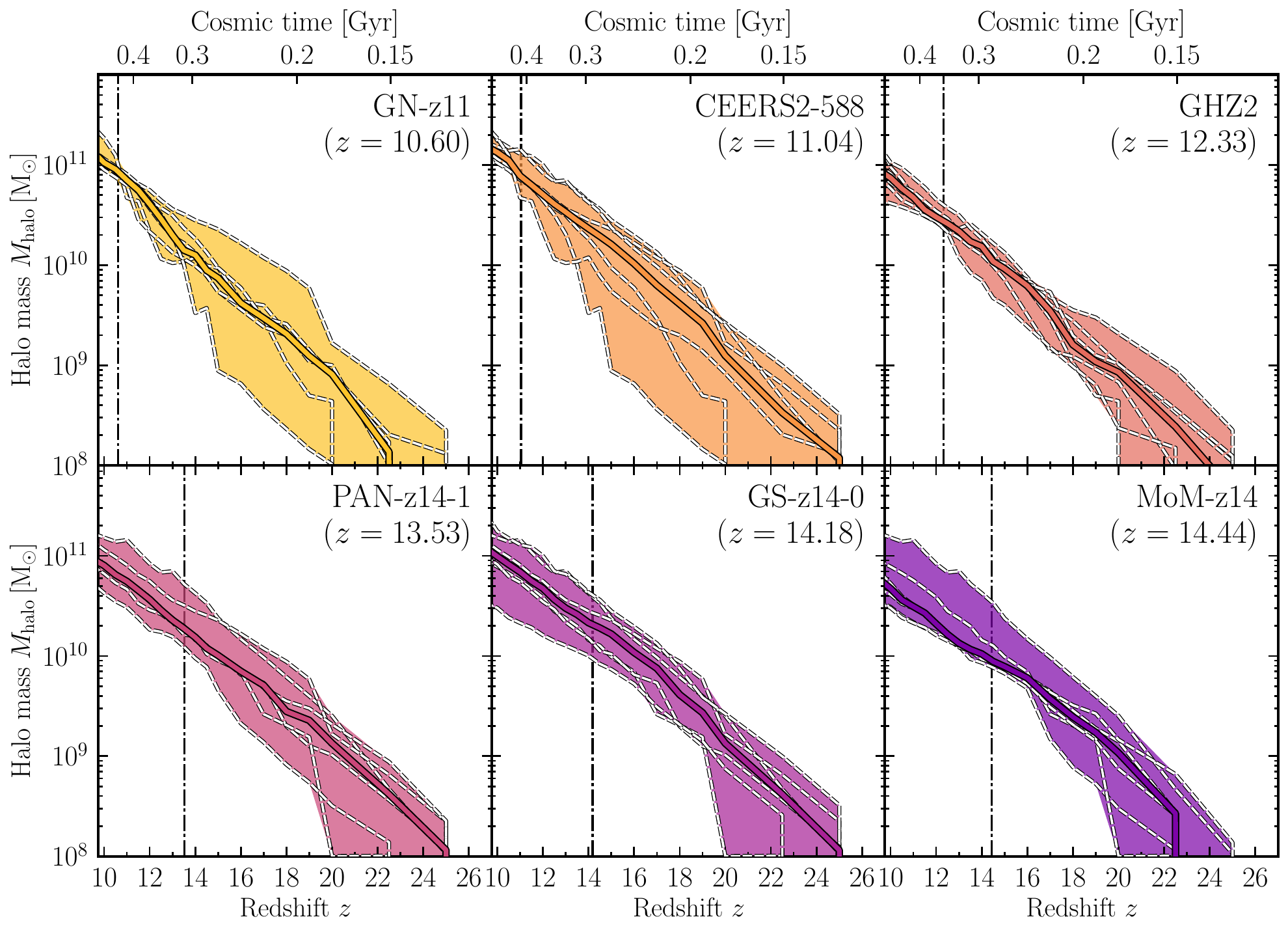}
    \caption{As Fig.~\ref{fig:matching_Mstar_z}, but showing the evolution of the halo mass of the \colibre{} galaxies that were matched to the six observed galaxies from Table~\ref{table:observations}. Black vertical lines indicate the redshifts of the observed galaxies. For all six observed galaxies, the halo mass of the matched galaxies in the \colibre{} L200m6 simulation increases from $M_{\rm halo}\sim 10^8~\mathrm{M_\odot}$ at $z=24$ to $M_{\rm halo}\sim 10^{11}~\mathrm{M_\odot}$ by $z=10$.}
    \label{fig:matching_Mhalo_z}
\end{figure*}

\subsection{Comparison with high-redshift JWST data}
\label{subsection:comparison with JWST}

Having explored the fundamental properties of massive galaxies and their host haloes at $z>10$, we now turn to a comparison between \colibre{} predictions and the properties of six spectroscopically confirmed most luminous \textit{JWST} galaxies at $z>10$ (see Table~\ref{table:observations} and $\S$\ref{subsection: observational data} for details). As detailed in $\S$\ref{subsection: matching}, for each observed galaxy, we have identified six counterparts in the \colibre{} L200m6 simulation that match it in stellar mass at the observed redshift.

\subsubsection{The evolution of stellar mass and halo mass}

Fig.~\ref{fig:matching_Mstar_z} shows stellar mass as a function of redshift. The six panels correspond to the six observed galaxies at $z>10$ listed in Table~\ref{table:observations}. The inferred stellar masses of these galaxies at their observed redshifts are shown as the white symbols with error bars. For each observed galaxy, the white dashed lines indicate the evolution of the stellar mass of the individual matched \colibre{} galaxies along the main progenitor branch. The thick solid lines indicate the evolution of the median stellar mass computed over the matched samples of \colibre{} counterparts, while the lower and upper boundaries of the shaded regions denote the minimum and maximum stellar mass in these samples, respectively.

On average, the stellar masses of the matched galaxies increase from $M_*\sim10^6~\mathrm{M_\odot}$ at $z\approx20$ to $\sim10^9~\mathrm{M_\odot}$ by $z=10$, corresponding to a time interval of only $\approx300$ Myr. All six observed galaxies are well matched by \colibre{} in terms of their stellar masses at the observed redshifts. The first stars in the matched galaxies form, on average, at $z\approx 20$, although the scatter from galaxy to galaxy is significant. The two highest-redshift observed galaxies ($z>14$; GS-z14-0 and MoM-z14) provide the strongest constraints on the SFHs of the matched galaxies, requiring all matched galaxies to form their first stars at $z\geq 16.5$. This is expected, as the cosmic time at $z=14$ is only $\approx 300$~Myr, necessitating very early onset of star formation to reach the required stellar mass of $M_*\sim 10^{8.0-8.5}~\mathrm{M_\odot}$ by the observed redshift.

For the two observed galaxies at the lowest redshifts, GN-z11 and CEERS2-588, the constraints on the SFH are weaker. Although both require the matched galaxies to grow to stellar masses of $M_* \sim 10^{9}~\mathrm{M_\odot}$, the fact that this condition is applied at $z\approx 11$ (rather than $z\approx 14$) allows for greater diversity in the growth histories of the matched galaxies. As can be seen in the top left and middle panels, even though all individual matched galaxies form $M_* \sim 10^{9}~\mathrm{M_\odot}$ by $z\approx 11$, the trajectories by which they reach this value differ dramatically, with some galaxies forming their first stars as early as $z=22$, while others do so only at $z=14$. The difference in stellar mass among the matched galaxies peaks at intermediate redshifts ($15 \lesssim z\lesssim 18$), reaching values as large as $2.5$~dex.

Fig.~\ref{fig:matching_Mhalo_z} shows the evolution of halo mass for the same \colibre{} galaxies as in Fig.~\ref{fig:matching_Mstar_z}, using the same layout and style. In all cases, the median halo mass increases from roughly $M_{\rm halo}\sim 10^8~\mathrm{M_\odot}$ at $z=24$ to $M_{\rm halo}\sim 10^{10}~\mathrm{M_\odot}$ by $z=14$, and further to $M_{\rm halo}\sim 10^{11}~\mathrm{M_\odot}$ by $z=10$. For a given observed galaxy, the spread in halo masses among the matched galaxies at fixed redshift is somewhat smaller than the corresponding spread in stellar mass shown in Fig.~\ref{fig:matching_Mstar_z}, although the overall trends are similar. For example, for GN-z11 and CEERS2-588, while the difference in halo mass is smallest at $z=10$ (within $\approx0.3$~dex), it increases to about $1.5$~dex by $z=16$, consistent with the divergence of their SFHs at high redshift in Fig.~\ref{fig:matching_Mstar_z}, where the difference in stellar mass increases from $\approx0.5$~dex at $z=10$ to $\approx2.5$~dex by $z=16$. This implies that the large differences at $z>15$ in the SFHs of the simulated galaxies matched to GN-z11 and CEERS2-588 are likely a consequence of different formation times of their host haloes.

The black vertical lines in Fig.~\ref{fig:matching_Mhalo_z} denote the redshifts of the six observed galaxies. By comparing them to the predicted halo mass evolution of the matched \colibre{} galaxies, we infer halo masses of $M_{\rm halo}\approx10^{10.9\text{--}11.0}~\mathrm{M_\odot}$ for GN-z11 and $M_{\rm halo}\approx10^{10.7\text{--}11.2}~\mathrm{M_\odot}$ for CEERS2-588, consistent with previous estimates for these galaxies \citep{2023ApJ...952...74T,2026arXiv260121833H}. For GHZ2, we infer $M_{\rm halo}\approx10^{10.4\text{--}10.6}~\mathrm{M_\odot}$; for PAN-z14-1, $M_{\rm halo}\approx10^{10.1\text{--}10.7}~\mathrm{M_\odot}$; for GS-z14-0, $M_{\rm halo}\approx10^{10.0\text{--}10.6}~\mathrm{M_\odot}$; and for MoM-z14, $M_{\rm halo}\approx10^{9.9\text{--}10.5}~\mathrm{M_\odot}$.

\subsubsection{The evolution of star formation rates}
\label{subsubsection:evolution_sfr_matching}

\begin{figure}
    \centering
    \includegraphics[width=0.99\linewidth]{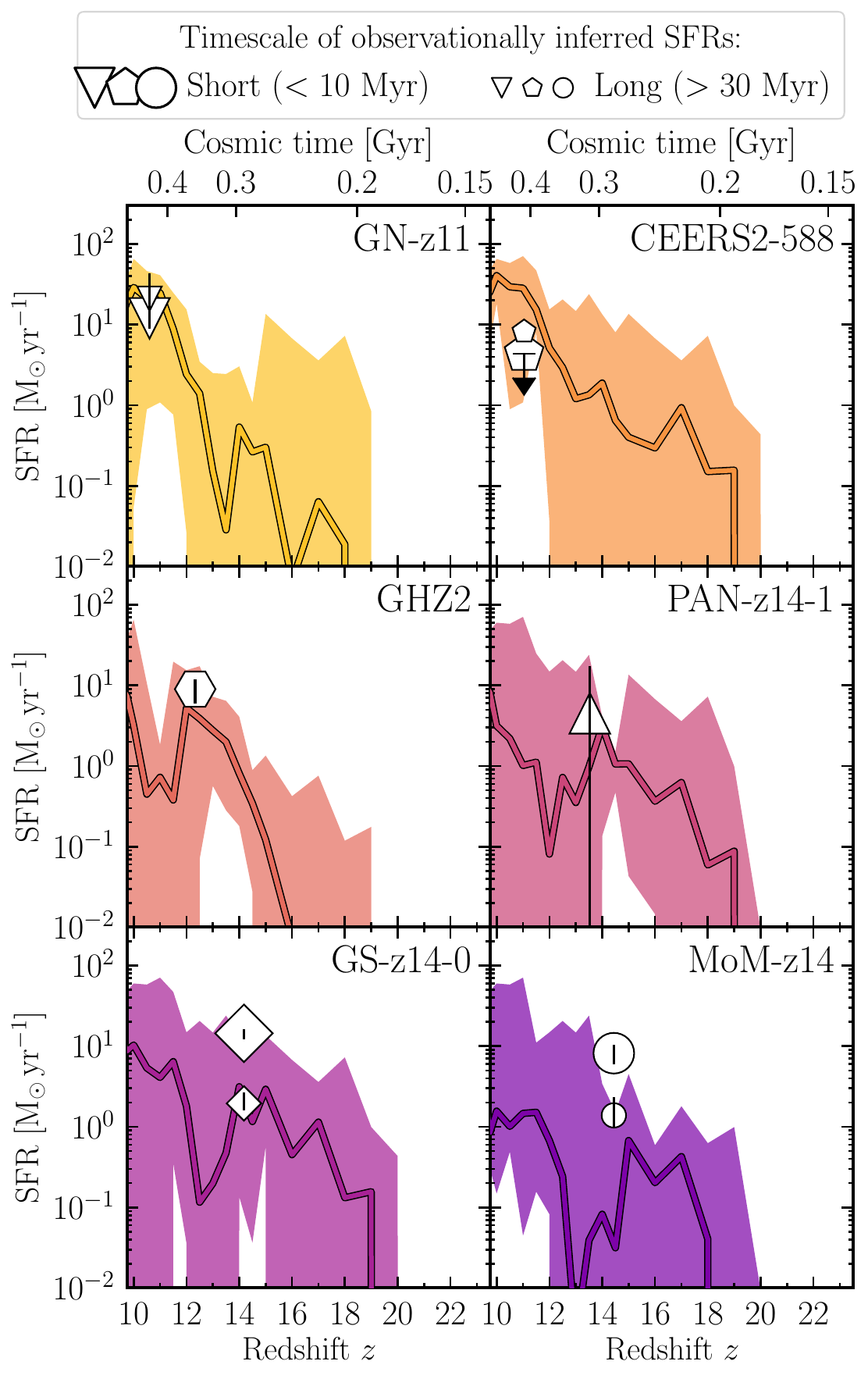}
    \caption{SFR as a function of redshift for six luminous $z>10$ galaxies observed by \textit{JWST} (different panels). Solid curves indicate the median SFR predicted by the simulations, while the boundaries of the shaded regions are defined by the minimum and maximum SFRs, all computed over the samples of six \colibre{} galaxies matched to each observation. The simulated SFRs are instantaneous SFRs measured within 3D apertures of twice the stellar half-mass radius. The large and small white symbols indicate observationally inferred SFRs averaged over short ($<10$~Myr) and long ($>30$~Myr) timescales, respectively. Although the \colibre{} galaxies were matched to the observed galaxies based only on stellar mass, they are also in overall good agreement with the inferred SFRs of their observational counterparts, except for MoM-z14, whose SFR is approximately 0.5~dex higher than predicted.}
    \label{fig:matching_SFR_z}
\end{figure}

Fig.~\ref{fig:matching_SFR_z} shows the evolution of galaxy SFRs as a function of redshift. The figure has the same style as Fig.~\ref{fig:matching_Mstar_z}, with white symbols denoting the observational data for the six observed galaxies (different panels). Unlike in Figs.~\ref{fig:matching_Mstar_z} and \ref{fig:matching_Mhalo_z}, here for clarity we do not show predictions for individual galaxies, as the SFR evolution is highly variable compared to relatively smoothly evolving quantities such as stellar and halo mass. The simulated SFRs are instantaneous, and computed within 3D apertures of size $2R_{\rm *,1/2}$, where $R_{\rm *,1/2}$ is the 3D stellar half-mass radius. The larger white symbols denote the observationally inferred SFRs over short timescales ($\lesssim 10$~Myr), while the smaller white symbols denote those over long timescales ($\gtrsim 30$~Myr). For four observed galaxies (GN-z11, CEERS2-588, GS-z14-0, and MoM-z14), we show both SFR estimates. For CEERS2-588, the SFR over the short timescale is an upper limit, based on the non-detection of the H$\alpha$ line \citep{2026arXiv260121833H}.

\begin{figure}
    \centering
    \includegraphics[width=0.99\linewidth]{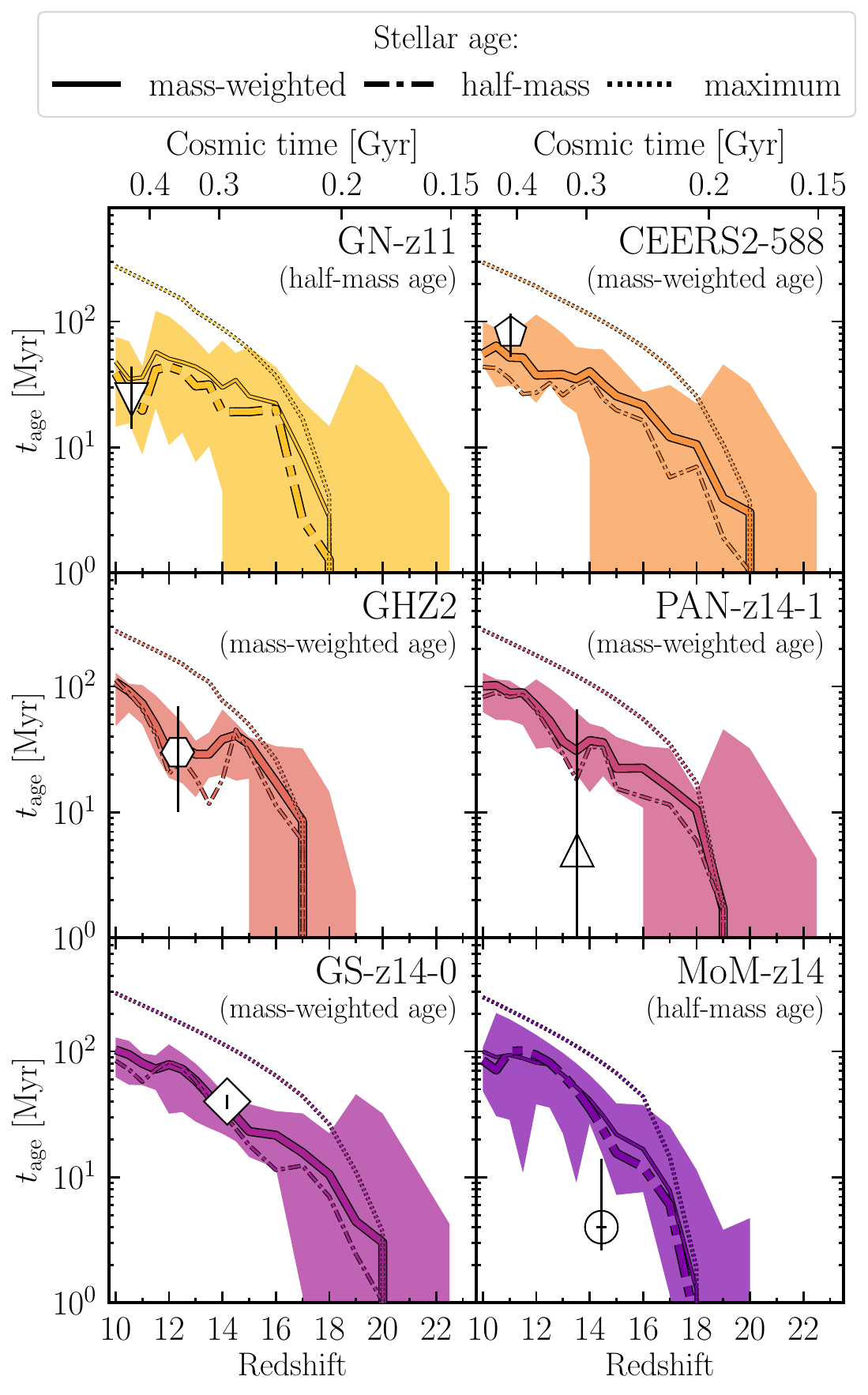}
    \caption{As Fig.~\ref{fig:matching_SFR_z}, but showing stellar age versus redshift. The stellar age in the simulations is computed in three different ways: the mass-weighted stellar age (solid lines), the age within which half of the stellar mass has formed (dash-dotted lines), and the age of the oldest star particle (dotted lines). Observations report one of the first two of these three quantities, inferred from reconstructed SFHs, as indicated in the top right corner of each panel. The line corresponding to the median stellar age for the definition most similar to that used in the observations is shown thicker, and the minimum and maximum ages in the matched samples, given by the boundaries of the shaded regions, are computed using the same definition. \colibre{} is in good agreement with the stellar ages inferred from all observations, except for MoM-z14.}
    \label{fig:matching_tage_z}
\end{figure}

\begin{figure*}
    \centering
    \includegraphics[width=0.95\linewidth]{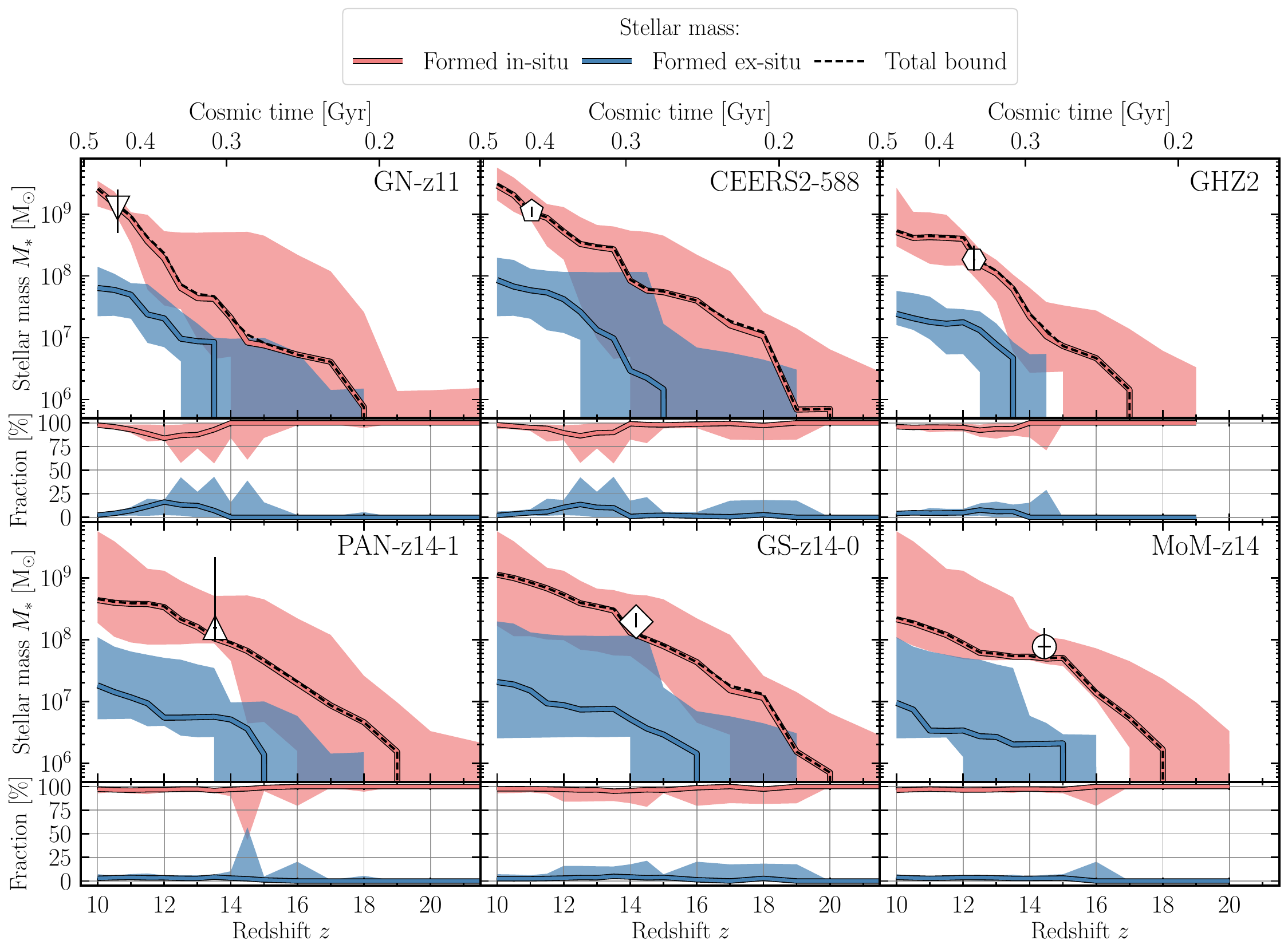}
    \caption{Decomposition of stellar mass into in-situ and ex-situ components. The six large panels correspond to the six observed galaxies from Table~\ref{table:observations}. In each large panel, we show the evolution of the total bound stellar mass (black) and the stellar mass formed in situ (red) and ex situ (blue). The six smaller panels show the fraction of the total bound stellar mass at a given redshift that was formed in situ and ex situ. The curves show the median values computed over the samples of \colibre{} galaxies matched to each observed galaxy, while the lower and upper boundaries of the shaded regions indicate the minimum and maximum values in these samples, respectively. For all objects and at all redshifts, the in-situ mass is significantly larger than the ex-situ component.}
    \label{fig:insitu-exsitu_decomposition}
\end{figure*}

The median SFHs of the simulated galaxies have a rising shape, increasing steeply with cosmic time. At fixed redshift, the scatter in SFR within a given sample of simulated galaxies can exceed 3~dex, reflecting the stochastic nature of star formation. Never the less, as we show in Appendix~\ref{appendix: smfs and sfe}, on average, massive \colibre{} galaxies evolve along the star-forming main sequence as they grow in mass. The $\approx 30$--$50$~Myr dip in the median simulated SFRs around $z\approx11$--$14$, which is particularly pronounced for the two galaxies at the highest redshift, GS-z14-0 and MoM-z14, corresponds to `mini-quenching' episodes caused by feedback from the first stars formed in the simulated galaxies \citep[see also][]{2024A&A...689A.310F}.

Even though the simulated galaxies were matched to observed galaxies using solely the stellar mass at the observed redshift, their SFRs are roughly consistent with the observationally inferred values. One exception is MoM-z14, whose 5-Myr-averaged SFR ($\approx 8~\mathrm{M_\odot}~\mathrm{yr}^{-1}$; the large circular symbol in the lower right panel) is about a factor of four higher than the maximum SFR of the matched \colibre{} sample. However, at redshifts that are only $\Delta z = 0.5$ lower or higher than the observed one, the predicted maximum SFR increases by a factor of $\approx 2.5$. Therefore, the discrepancy with the SFR of MoM-z14 at the observed redshift may be exacerbated by small-number statistics (recall that, for each observed galaxy, we have only six matched simulated counterparts). Furthermore, \citet{2026OJAp....956033N} report that MoM-z14 was observed during a starburst phase and that its SFR likely increased by an order of magnitude over the past 10 Myr. Indeed, its 50-Myr average SFR drops to $2.2~\mathrm{M_\odot}~\mathrm{yr}^{-1}$ (the small circular symbol), which is consistent with the \colibre{} predictions. 

\subsubsection{The evolution of stellar age}

Fig.~\ref{fig:matching_tage_z} shows the evolution of the ages of galaxy stellar populations. For each observed galaxy, we display \colibre{} predictions for three definitions of stellar age: the mass-weighted age (solid lines), the age at which 50~per~cent of the stellar mass has formed (dash-dotted lines), and the age of the oldest star particle (dotted lines). For each simulated galaxy and at each redshift, these three quantities are computed using all stellar particles gravitationally bound to the galaxy's host subhalo. Observational studies estimate stellar ages from SFHs inferred through SED fitting. The adopted age definition in the observations (mass-weighted or half-mass) is indicated in the top-right corner of each panel.

The typical ages of stellar populations of the simulated galaxies at $z>10$, as traced by the mass-weighted and half-mass stellar ages, remain within $\sim 100$~Myr. More specifically, for both age definitions, an age of $\sim 100$ Myr is reached by $z=10$, while at $z=12$ and $z=14$ the median ages are roughly $50$ and $30$~Myr, respectively. The mass-weighted and half-mass stellar ages closely track one another, with the former being systematically, though only marginally, larger at fixed redshift. This behaviour is expected for SFHs of the form $\mathrm{SFR}(t)\propto t^n$, where $t$ is cosmic time: the two age definitions coincide for $n=0$, while for $n>0$ the mass-weighted age exceeds the half-mass age by a factor of up to $\approx 1.4$. The age of the oldest star particle (dotted lines) is significantly larger, reaching $\approx300$~Myr at $z=10$. At the highest redshifts corresponding to the onset of star formation, all three age definitions converge, as expected.

The \colibre{} simulations reproduce the observationally inferred stellar ages of five of the six galaxies within the observational uncertainties. This agreement is anticipated, as the simulations have already been shown to reproduce both the stellar mass (Fig.~\ref{fig:matching_Mstar_z}) and SFR (Fig.~\ref{fig:matching_SFR_z}) of the observed sources, thereby placing strong constraints on the stellar age. Indeed, as in Fig.~\ref{fig:matching_SFR_z}, the only exception is again MoM-z14, whose observationally inferred stellar age is slightly lower than predicted by the simulations, consistent with its inferred SFR in Fig.~\ref{fig:matching_SFR_z}, which is higher than predicted. As discussed in $\S$\ref{subsubsection:evolution_sfr_matching}, this discrepancy is likely driven by a recent starburst event in MoM-z14 coupled to the small-number statistics in the simulation. 

Lastly, we caution that, as with stellar masses, the results of our comparison of stellar ages are subject to systematic uncertainties in the observationally inferred values, with inferred stellar ages particularly sensitive to the SFH priors adopted in SED fitting. Priors allowing burstier SFHs, such as the bursty continuity prior, yield systematically younger ages and lower stellar masses and are typically favoured over continuity priors, which sometimes result in stellar masses and ages in tension with the $\Lambda$CDM cosmological model \citep[e.g.][]{2022ApJ...927..170T,2025NatAs...9..729H}. We discuss these implications in Appendix~\ref{appendix: SED assumptions}.

\subsubsection{In-situ and ex-situ star formation}

The next question we address is what fraction of the stellar mass shown in Fig.~\ref{fig:matching_Mstar_z} formed in situ versus ex situ. These results are presented in Fig.~\ref{fig:insitu-exsitu_decomposition}. In each large panel, we show the evolution of the total bound stellar mass (black), the stellar mass formed in situ (red), and the stellar mass formed ex situ (blue). The six smaller panels show the fraction of the total bound stellar mass formed in situ and ex situ, expressed as a percentage, at each redshift. In all panels, the curves show the median values computed from the matched samples of simulated galaxies, while the shaded regions indicate the minimum and maximum values in these samples. For clarity, the shaded regions are shown only for the in-situ and ex-situ components, and not for the total bound stellar mass. For reference, in the large panels, we also overplot the observationally inferred stellar masses as white symbols.

For a given galaxy, the in-situ mass at redshift $z$ is computed by summing the masses of star particles that have been bound to the galaxy in every \colibre{} output since their birth. The ex-situ mass at redshift $z$ is computed by considering star particles that (i) are bound to the galaxy at redshift $z$ and (ii) were either bound to a different galaxy or unbound in at least one \colibre{} output at redshifts higher than $z$.

For all observed galaxies and at all redshifts, the median predicted in-situ mass is significantly larger than the median ex-situ component, contributing more than 75~per~cent of the total bound stellar mass in all cases, and approaching nearly 100~per~cent at some redshifts. The extent of the shaded regions, however, reveals a few exceptions in which the ex-situ fraction can be higher. In particular, some simulated galaxies have 25--40~per~cent of their stellar mass at $12<z<15$ formed ex-situ, with the maximum value reaching $\approx55$~per~cent at $z=14.5$ for one of the galaxies matched to PAN-z14-1. 

\subsubsection{The evolution of gas-phase metallicities}

\begin{figure}
    \centering
    \includegraphics[width=0.99\linewidth]{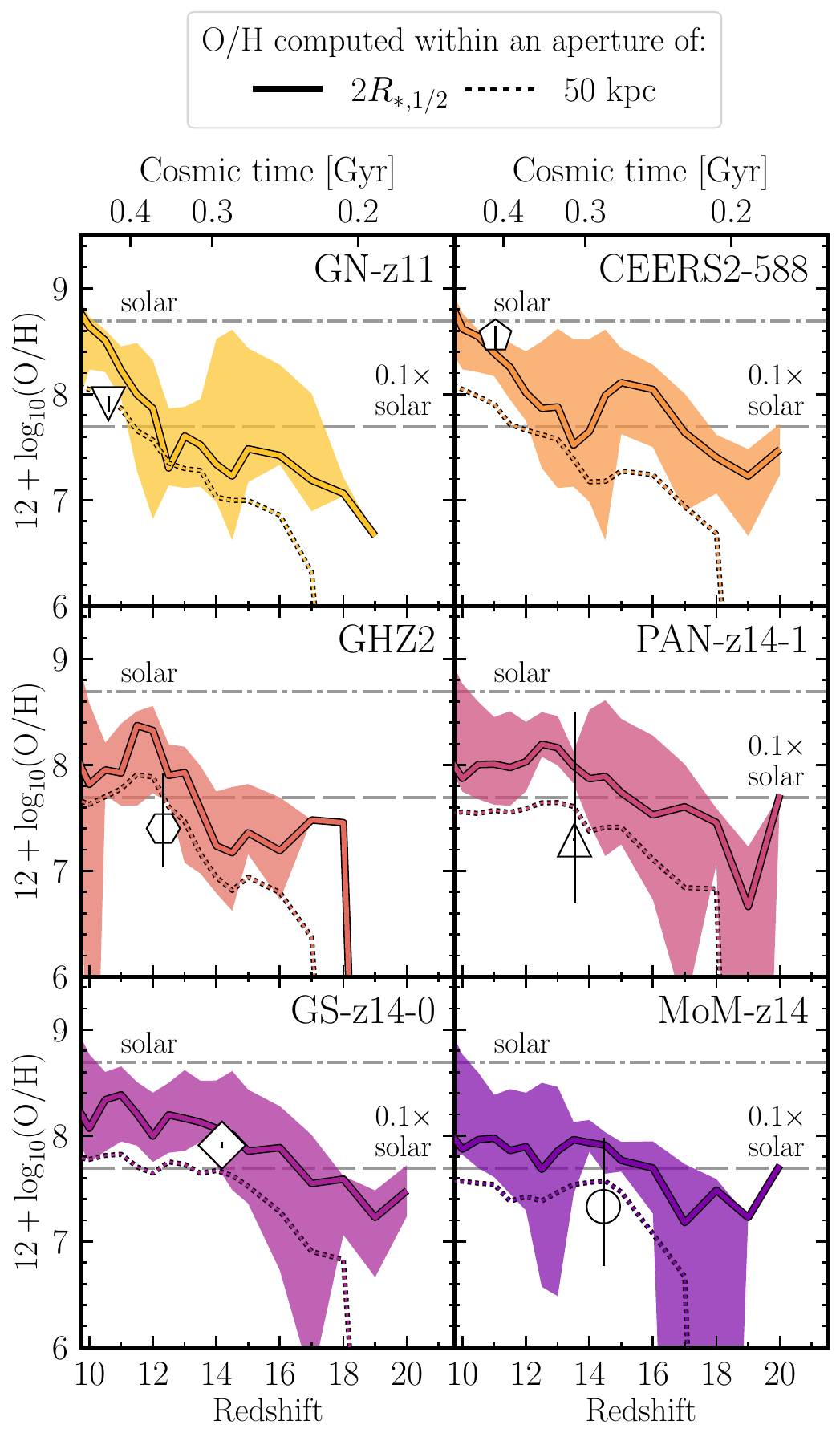}
    \caption{As Fig.~\ref{fig:matching_SFR_z}, but showing gas-phase metallicity versus redshift. The solid curves indicate the metallicity measured within twice the stellar half-mass radius ($2R_{\rm *,1/2}$), whereas the thin dotted curves show the measurements within 50-kpc 3D apertures. In both cases, only bound gas that is both cool ($T<10^{4.5}~\mathrm{K}$) and dense ($n_{\rm H}>0.1~\mathrm{cm}^{-3}$) is considered. The metallicity is expressed in units of $12 + \log_{10}(\rm O/H)$, excluding metals depleted onto dust grains. The shaded regions indicate the minimum and maximum gas metallicities of the matched galaxies measured within $2R_{\rm *,1/2}$. To guide the eye, the grey horizontal lines indicate solar metallicity and 10 per cent solar metallicity. The \colibre{} model predicts rapid metal enrichment, reaching 10 per cent of the solar value already by $z\approx 15-17$, and is in overall good agreement with the observational constraints.}
    \label{fig:matching_Z_z}
\end{figure}

\begin{figure}
    \centering
    \includegraphics[width=0.99\linewidth]{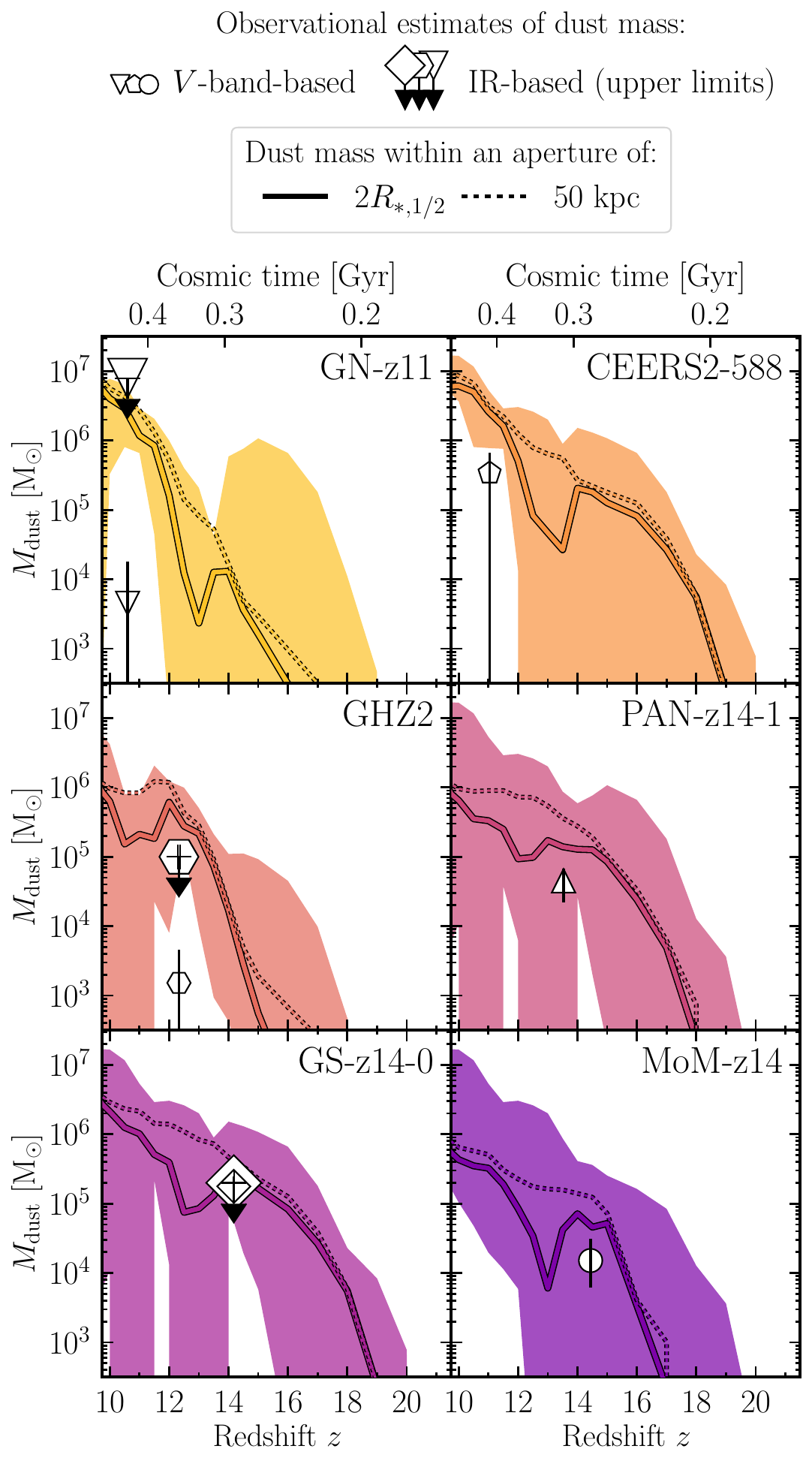}
    \caption{As Fig.~\ref{fig:matching_SFR_z}, but showing dust mass as a function of redshift. The solid and dotted curves indicate the dust mass measured within $2R_{\rm *,1/2}$ and 50-kpc 3D apertures, respectively, with the boundaries of the shaded regions indicating the minimum and maximum dust mass for the former definition. The small white symbols indicate observational constraints on the dust mass inferred from $V$-band attenuation, while the large white symbols with downward-pointing arrows indicate IR-based upper limits on the dust mass. The \colibre{} simulations predict rapid grain growth, with most of the dust concentrated within twice the stellar half-mass radius and dust masses of $\sim10^5~\mathrm{M_\odot}$ already in place by $z\approx15-16$. Overall, grain growth in \colibre{} tends to be more efficient than suggested by the observational constraints.}
    \label{fig:matching_dust_z}
\end{figure}

Fig.~\ref{fig:matching_Z_z} shows the redshift evolution of gas-phase metallicity, expressed as $12 + \log_{10}(\rm O/H)$. Following \citet{2026MNRAS.548ag375S} and \cite{2026arXiv260625995S}, the metallicity is measured in the gas with densities $n_{\rm H}>0.1~\mathrm{cm}^{-3}$ and temperatures $T < 10^{4.5}~\mathrm{K}$, considering only bound gas particles. For a given galaxy, the O/H abundance\footnote{Individual elemental abundances are recorded only in the \colibre{} snapshots available at redshifts $z=10$, $12$, $15$, and $20$. At intermediate redshifts with reduced \colibre{} outputs, we instead estimate O/H using $\log_{10}(\rm O/H) = \log_{10}(Z_{\rm gas}/\mathrm{Z_\odot}) + \log_{10}(O/H)_\odot + \varepsilon_{Z}$ where $Z_{\rm gas}$ is the recorded total gas metallicity, $\mathrm{Z_\odot}=0.0134$ and $12 + \log_{10}(\rm O/H)_\odot=8.69$ \citep{2009ARA&A..47..481A}, and $\varepsilon_{Z}=0.1$. The value of $\varepsilon_{Z}$ was calibrated using the snapshots for which both O/H and $Z_{\rm gas}$ are available.} is calculated as the mass-weighted average of the ratios of oxygen to hydrogen nuclei carried by individual gas particles bound to its host halo, excluding oxygen locked up in dust grains. The metallicity is shown for 3D apertures with radii equal to $2R_{\rm *,1/2}$ (thick solid curves) and for larger 50-kpc 3D apertures (dashed curves).

In the observations, the metallicity of GN-z11 was estimated using the direct $T_{\rm e}$ method \citep{2025A&A...695A.250A}. The metallicities of GHZ2 \citep{2025NatAs...9..155Z} and CEERS2-588 \citep{2026arXiv260121833H} were estimated by combining empirical calibrations of metallicity indicators with the measured ratios of the strong oxygen lines to the upper limit on the H$\beta$ line, assuming zero dust attenuation. For PAN-z14-1 \citep{2026ApJ..1002..134D}, no oxygen emission lines were detected; we therefore adopt the metal mass fraction inferred from SED fitting with the best-fitting \textsc{Bagpipes} model, which we convert to O/H using $\log_{10}(\rm O/H) = \log_{10}(Z/\mathrm{Z_\odot}) + \log_{10}(O/H)_\odot$, where the solar values $Z_{\rm \odot}=0.0134$ and $12+\log_{10}(\rm O/H)_\odot = 8.69$ are taken from \citet{2009ARA&A..47..481A}. The metallicity of GS-z14-0 was inferred from SED fitting with \textsc{Prospector} applied to \textit{JWST}/NIRSpec, NIRCam, MIRI, and ALMA observations, including the detected [O~\textsc{iii}]~$88~\mu\mathrm{m}$ line \citep{2025A&A...696A..87C}. Finally, the metallicity of MoM-z14 \citep{2026OJAp....956033N} was constrained using the emission-line modelling code \textsc{Cue} \citep{2025ApJ...986....9L}, applied to multiple blended emission lines detected in the PRISM spectrum of the galaxy.

The ISM of the matched \colibre{} galaxies undergoes very rapid metal enrichment, with metallicities measured within a sphere of $2R_{\rm *,1/2}$ reaching 10~per~cent of the solar value by redshift $z\approx 15$--$17$. The gas metallicity continues to rise at lower redshifts as increasing amounts of metals are ejected into the ISM by newly detonated CCSNe. For the two observed galaxies with the highest stellar masses (GN-z11 and CEERS2-588; $M_{*}\sim 10^9~\mathrm{M_\odot}$), the median metallicity of the matched counterparts reaches solar values by $z=10$. For the other four observed galaxies, the median metallicity of the matched counterparts remains sub-solar by $z=10$, as expected from the mass--metallicity relation, given that their median stellar masses are lower than those of the galaxies matched to GN-z11 and CEERS2-588. At fixed redshift, the difference in the predicted values of $\log_{10}(\mathrm{O/H})$ can exceed $0.5$--$1$~dex. For example, among the galaxies matched to PAN-z14-1, some reach nearly solar metallicity by $z=14$, while others remain below 10~per~cent of the solar value. This is consistent with the large scatter in SFHs seen in Fig.~\ref{fig:matching_Mstar_z} at similar redshifts.

When measured within a radius of 50~kpc, the metallicity is lower by $0.5$--$1$~dex than that within $2R_{\rm *,1/2}$, owing to the presence of more pristine gas at larger radii. For some simulated galaxies, the gas-phase metallicity measured within $2R_{\rm *,1/2}$ can temporarily decrease with decreasing redshift. This can happen if metal-enriched gas is pushed out of the ISM by stellar feedback, and/or if the pristine gas supply temporarily dominates over the metals produced by stellar feedback, thereby diluting the metal-enriched ISM. 

Finally, the metallicities of the matched \colibre{} galaxies are in broad agreement with the observational constraints for all six galaxies, including reproducing the nearly solar metallicity of CEERS2-588 at $z\approx11$. These results are consistent with \citet{2026arXiv260625995S}, who showed that \colibre{} reproduces the observed mass--metallicity relation across cosmic time and that this relation is already in place in the simulations by $z=10$. 

\subsubsection{The evolution of dust mass}
\label{subsubsection: dust mass}

Fig.~\ref{fig:matching_dust_z} shows the evolution of dust mass predicted by \colibre{} and compares it to observational constraints on the dust abundance at high redshift. We emphasize that \colibre{} includes a prescription for the formation and evolution of dust grains, which is self-consistently coupled to the \colibre{} chemistry solver \citep{2026MNRAS.545f2040T}, and has been shown to broadly reproduce the observed dust scaling relations across cosmic time \citep{2026arXiv260726058V}.

The large white symbols with downward-pointing arrows show IR-based observational constraints on the dust mass for GN-z11 (top-left panel), GHZ2 (middle-left panel), and GS-z14-0 (bottom-left panel). GN-z11 was observed with the Northern Extended Millimetre Array (NOEMA) by \citet{2024MNRAS.530..340F}. Based on the non-detection of dust continuum emission, the authors derived an upper limit on the dust mass of $M_{\rm dust}\approx10^{6.9}~\mathrm{M_\odot}$. For GHZ2 and GS-z14-0, the constraints correspond to dust mass upper limits of $\sim10^{5}~\mathrm{M_\odot}$ for both galaxies, derived from the non-detection of dust thermal emission in deep ALMA observations \citep{2026ApJ..1000..159M}.

For these three galaxies, as well as the other three galaxies in our observed sample, we also show dust masses estimated from the observationally inferred $V$-band dust attenuation, $A_{\rm V}$, and the galaxy half-light radius (small white symbols), following \citet{2023MNRAS.520.2445Z,2025A&A...694A.286F}. Specifically, we use equation~(1) from \citet{2025A&A...694A.286F}, assuming spherical geometry and adopting the dust mass absorption coefficient for a Milky Way extinction curve from \citet{2001ApJ...548..296W}. Using $A_{\rm V}=0.08$, $0.12$, $0.01$, $0.06$, $0.19$, and $0.2$ for GN-z11 \citep{2023ApJ...952...74T}, CEERS2-588 \citep{2026arXiv260121833H}, GHZ2 \citep{2025NatAs...9..155Z}, PAN-z14-1 \citep{2026ApJ..1002..134D}, GS-z14-0 \citep{2025A&A...696A..87C}, and MoM-z14 \citep{2026OJAp....956033N}, respectively, yields dust masses of order $10^{3}$--$10^{5}~\mathrm{M_\odot}$. These estimates are roughly consistent with the NOEMA- and ALMA-based upper limits for GN-z11 and for GHZ2 and GS-z14-0, respectively.

The thick solid and dotted lines show the median dust masses in the simulation, measured within 3D apertures of size $2R_{\rm *,1/2}$ and 50~kpc, respectively. In line with the rapid metal enrichment shown in Fig.~\ref{fig:matching_Z_z}, \colibre{} predicts very efficient dust grain production and growth, starting as early as $z\approx18$. By $z=14$ ($z=12$), the median dust mass in both apertures increases to about $10^{5}~\mathrm{M_\odot}$ ($10^{5.5}~\mathrm{M_\odot}$) and exhibits large scatter at fixed redshift within the matched samples. The predicted median values exceed the observational constraints for five of the six observed galaxies, except for GS-z14-0, for which the median predicted dust mass is both consistent with the $V$-band-based constraint and marginally consistent with the ALMA upper limit. Furthermore, for GN-z11, the simulations predict significantly more dust than inferred from the $V$-band attenuation, but remain consistent with the NOEMA upper limit, although this upper limit has relatively weak constraining power. By $z=10$, the median dust masses predicted by \colibre{} lie in the range $10^{6} \lesssim M_{\rm dust}/\mathrm{M_\odot} \lesssim 10^7$. 

Overall, \colibre{} tends to produce more dust at high redshifts than suggested by the observations, potentially indicating either overly efficient grain growth in the simulation or inefficient dust removal from star-forming regions by stellar feedback. Moreover, the observational constraints on the six galaxies we compare to are consistent with the broader picture that the majority of luminous \textit{JWST} galaxies at $z>10$ show signatures of little to no dust attenuation \citep[e.g.][]{2024ApJ...964L..24M,2026MNRAS.548ag701R}, although see \citet{2025ApJ...988L..10K,2026arXiv260315841R} for the few notable exceptions. 

\subsubsection{The evolution of absolute UV magnitudes}  

\begin{figure}
    \centering
    \includegraphics[width=0.99\linewidth]{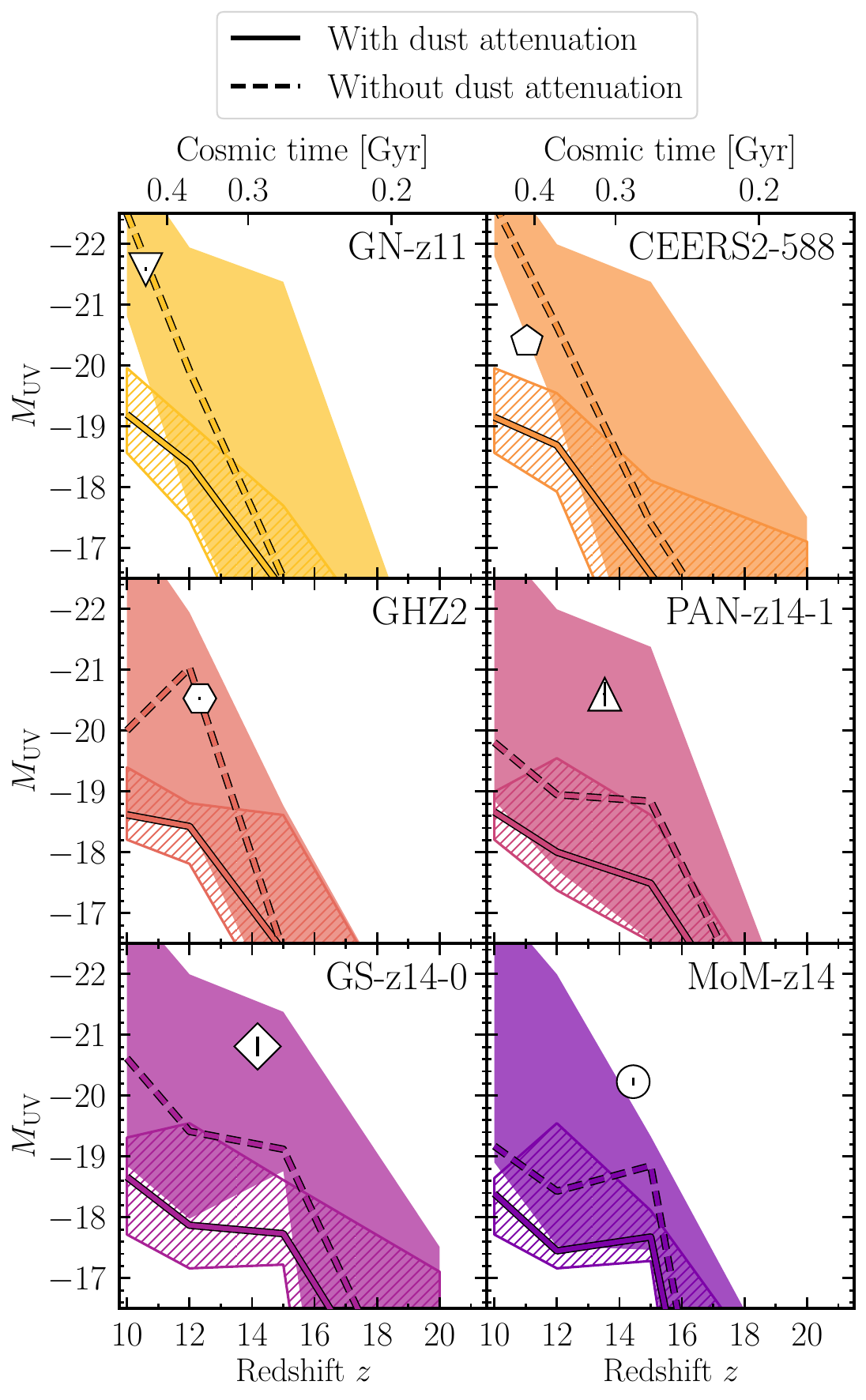}
    \caption{As Fig.~\ref{fig:matching_SFR_z}, but showing the absolute UV magnitude versus redshift. The dashed and solid curves show the median intrinsic and dust-attenuated UV magnitudes, respectively, computed over the samples of matched galaxies, while the boundaries of the shaded (hatched) regions indicate the minimum and maximum intrinsic (attenuated) magnitudes in these samples. The \colibre{} simulations are in agreement with the observations when dust attenuation is neglected; however, once dust attenuation is included, the simulations significantly underpredict the observed UV luminosities.}
    \label{fig:matching_uv_z}
\end{figure}

Fig.~\ref{fig:matching_uv_z} shows the evolution of the absolute UV magnitudes computed using the \colibre--\textsc{skirt} pipeline (see $\S$~\ref{subsection: radiative transfer} for details). The dashed and solid lines indicate the median intrinsic and dust-attenuated UV magnitudes, respectively, while the boundaries of the shaded (hatched) regions show the minimum and maximum intrinsic (attenuated) UV magnitudes. Unlike the observational constraints shown in the previous figures, the observational data here (white symbols) are direct measurements.

Without dust attenuation, \colibre{} roughly reproduces the observed UV magnitudes for five of the six galaxies, underpredicting the UV luminosity only for the highest-redshift galaxy, MoM-z14 ($z=14.44$), by approximately $0.5$~mag. This is in line with the lower SFR and higher stellar age predicted by the simulation than inferred for MoM-z14 (see Figs.~\ref{fig:matching_SFR_z} and \ref{fig:matching_tage_z}, respectively). In contrast, when dust attenuation is accounted for, \colibre{} significantly underpredicts the UV luminosities of all six observed galaxies, with discrepancies of about $1.5$--$3$~mag. In fact, even the most luminous matched galaxies, corresponding to the upper boundary of the hatched regions, are substantially fainter than the observed galaxies.

\begin{figure}
    \centering
    \includegraphics[width=0.99\linewidth]{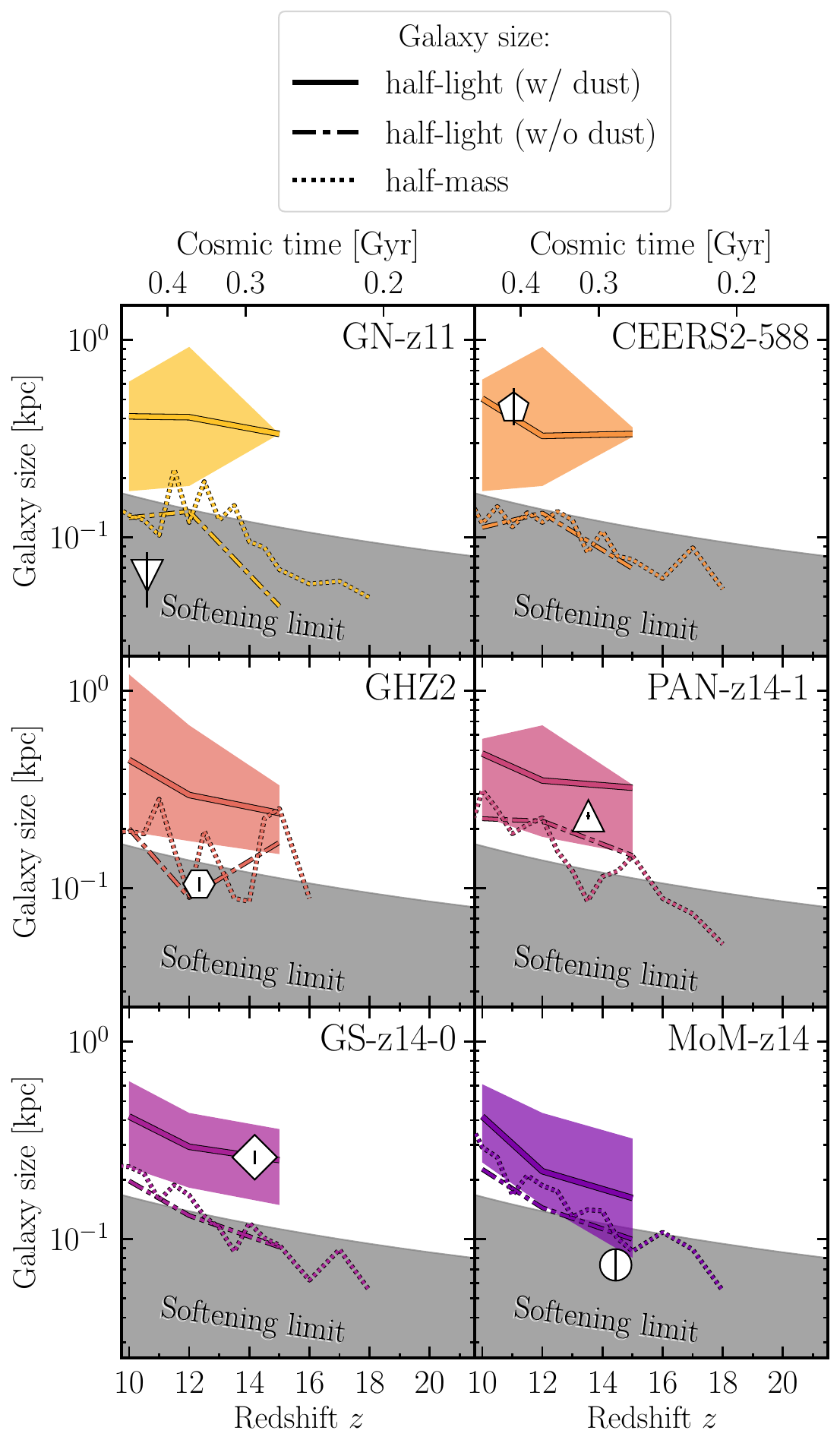}
    \caption{As Fig.~\ref{fig:matching_SFR_z}, but showing galaxy size versus redshift. The three curves with different line styles show the median sizes of the matched simulated galaxies computed using three different definitions: projected half-mass radius (dotted), projected unattenuated half-light radius (dash-dotted), and projected attenuated half-light radius (solid). The boundaries of the shaded regions indicate the minimum and maximum projected attenuated half-light radii. The grey region in the lower part of each panel indicates galaxy sizes below the \colibre{} gravitational softening length (i.e. formally below the resolution limit). \colibre{} is consistent with the observed sizes of all galaxies that are larger than the gravitational softening length.}
    \label{fig:matching_size_z}
\end{figure}

Our combined results from Figs.~\ref{fig:matching_dust_z} and \ref{fig:matching_uv_z} suggest that the discrepancy between the observed UV magnitudes of UV-bright galaxies and those predicted by \colibre{} can be driven by an overabundance of dust in the matched simulated galaxies. \colibre{} exceeds the current observational constraints on dust masses and predicts that dust attenuation can reduce UV luminosities by $1.5$--$3$~mag. At the same time, the spectra of most observed luminous galaxies at $z>10$ indicate little or no dust attenuation \citep[e.g.][]{2026MNRAS.548ag701R}. Together, these findings suggest that high-redshift grain growth in \colibre{} may be overly efficient, leading to excessively strong attenuation. To test this hypothesis, in Section~\ref{section: discussion} we present results from a \colibre{} simulation without grain growth in the ISM, finding that reducing the dust content of simulated galaxies indeed substantially improves the agreement with the observations. Finally, we note that our radiative transfer modelling (\S\ref{subsection: radiative transfer}) does not include nebular continuum emission from star-forming regions. Although insufficient to resolve the discrepancy with the observed UV magnitudes on its own, including this component would brighten the UV magnitudes by up to $\approx0.3$~mag at high redshift \citep{durrant2026cosmologicalsimulationshighredshiftgalaxy}.

\subsubsection{The evolution of galaxy sizes}

Fig.~\ref{fig:matching_size_z} compares the predicted evolution of galaxy size with redshift to the observed half-light radii. The solid and dash-dotted lines indicate the \colibre{} predictions for the projected half-light radii with and without dust attenuation, respectively, while the dotted lines show the projected half-mass radii. As detailed in  $\S$\ref{subsection: radiative transfer}, half-light radii are computed using projected radial luminosity profiles, with the luminosity calculated by redshifting the SED of each galaxy produced by the 3D radiative transfer code \textsc{skirt} into the observer frame and convolving it with the throughput curve of the \textit{JWST} F444W filter. As before, all curves show the medians taken over the sample of \colibre{} galaxies matched to each observed galaxy. We exclude galaxies with fewer than 10 stellar particles from the calculation of the galaxy sizes because these galaxies are poorly resolved. The grey region in the lower part of each panel indicates the regime where simulated galaxy sizes are smaller than the gravitational softening length in the \colibre{} m6 model, implying that the predictions in this regime are unreliable due to insufficient spatial resolution.

\begin{figure*}
    \centering
    \includegraphics[width=0.99\linewidth]{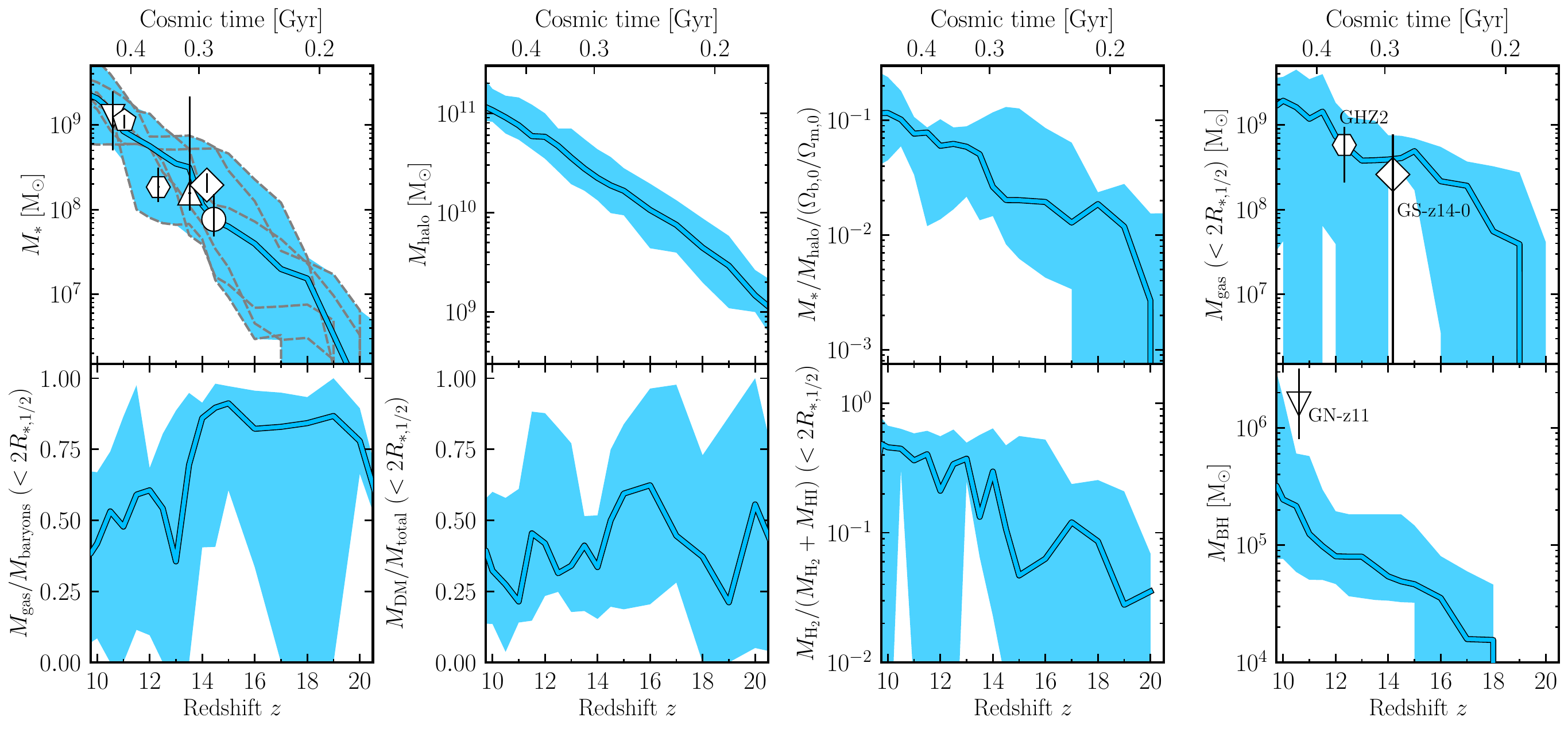}
    \caption{The combined predictions for the evolution of various galaxy and halo properties for the observed galaxies from Table~\ref{table:observations}. The top row, from left to right, shows the galaxy stellar mass, halo mass, galaxy formation efficiency, and total gas mass within $2R_{\rm *,1/2}$. The bottom row, from left to right, shows the total gas mass fraction, DM-to-total mass fraction, molecular gas fraction, and the mass of the central BH. All mass fractions are computed within 3D apertures of size $2R_{\rm *,1/2}$. The thick solid curves indicate the median values, while the boundaries of the shaded regions indicate the minimum and maximum values, all computed from galaxies in the \colibre{} L200m6 simulation selected to have $M_*>10^{7.55}~\mathrm{M_\odot}$ at $z=14$ and $M_*>10^{8.75}~\mathrm{M_\odot}$ at $z=10$. Six galaxies satisfy these criteria, with the grey dashed curves in the top-left panel showing their stellar mass evolution along the main progenitor branch. The white symbols denote observational constraints from Table~\ref{table:observations}, including the gas masses from \citet{2026ApJ..1000..159M} for GHZ2 and GS-z14-0, and the BH mass from \citet{2024Natur.627...59M} for GN-z11.}
    \label{fig:matching_other_quantities_z}
\end{figure*}

We find that the half-mass and unattenuated half-light radii trace each other reasonably well. On average, these sizes increase moderately with decreasing redshift, from about $100$~pc at $z=15$ to $\approx 200$~pc at $z=10$, with their evolution resembling that of the gravitational softening length (i.e. the upper boundary of the grey region in Fig.~\ref{fig:matching_size_z}). At fixed redshift, the attenuated half-light radii are, on average, a factor of $1.5$--$3$ larger than the unattenuated ones, reaching values in the range $200$--$500$~pc. This effect is caused by the high dust column densities in the inner regions of galaxies, which strongly attenuate the rest-frame near-UV emission, making galaxies appear less centrally concentrated.

The sizes of all observed galaxies that are above the gravitational softening length of the \colibre{} simulation (CEERS2-588, PAN-z14-1, and GS-z14-0) are consistent with the \colibre{} attenuated half-light radii. The sizes of the other three galaxies (GN-z11, GHZ2, and MoM-z14), which are below the softening, are smaller than those predicted by \colibre{}. This discrepancy is expected, as gravitational forces are underestimated at scales below the softening length in the simulation, making it increasingly difficult for simulated galaxies to reach sizes significantly smaller than this scale. In Appendix~\ref{appendix: convergence}, we demonstrate that galaxy sizes are indeed smaller in \colibre{} simulations at higher resolution, which adopt a smaller softening length.

Interestingly, for the compact galaxies GHZ2 and MoM-z14, the discrepancy is only a factor of $\approx 2$ and $1.5$, respectively, with the observed sizes still being consistent with the \colibre{} predictions for the dust-free case. In contrast, the discrepancy for GN-z11, which has an observed size of only $\approx 64$~pc \citep{2023ApJ...952...74T}, is more substantial (1~dex). This larger discrepancy may be due not only to the limited resolution of the simulation but also to AGN emission inside the observed source. Specifically, about two-thirds of the observed emission of GN-z11 originates from an unresolved nuclear region \citep{2023ApJ...952...74T}, suggesting significant AGN contamination. Although GN-z11 has not been detected in X-rays, \citet{2024Natur.627...59M} detected [Ne~\textsc{iv}]$\lambda\lambda2422,2424$, N~\textsc{iv}]$\lambda1486$, and C~\textsc{ii}$^{*}\lambda1335$, among other (semi-)forbidden lines, in the \textit{JWST}/NIRSpec spectrum of GN-z11. They estimated a BH mass of $\sim 10^{6}~\mathrm{M_\odot}$ and an AGN luminosity of $\sim 10^{45}~\mathrm{erg}~\mathrm{s}^{-1}$, implying an Eddington ratio of $\approx 5.5$, consistent with a highly accreting central BH. The spectrum of GHZ2 may also be contaminated by an AGN, which could similarly contribute to the more compact observed size of the galaxy. As discussed by \citet{2024ApJ...972..143C}, the large equivalent widths of UV lines (such as C~\textsc{iv} and N~\textsc{iv}]) in the observed spectrum suggest a central AGN as the ionizing source, although both AGN and star-formation models remain compatible with the data.

To sum up, we find that the \colibre{} simulations can reproduce some of the sizes of luminous \textit{JWST} galaxies at $z>10$, provided their sizes are above the gravitational softening length of the simulation, and that dust attenuation can be an important factor driving the observed diversity of galaxy sizes. 

\subsection{The predicted evolution of other galaxy and halo properties}
\label{subsection: extra_predictions}

In this section, we exploit the \colibre{} L200m6 simulation to predict the evolution of additional galaxy and halo properties that have been constrained for only a subset of the six observed galaxies in Table~\ref{table:observations}, if at all.

Having shown that the simulated galaxies matched to the six observed galaxies share broadly similar properties, in this section we present combined predictions rather than predictions for the individual observed galaxies shown in the previous figures. To this end, we select all galaxies from the L200m6 simulation whose stellar mass is $M_*>10^{7.55}~\mathrm{M_\odot}$ at $z=14$ and $M_*>10^{8.75}~\mathrm{M_\odot}$ at $z=10$. Six galaxies in total satisfy these criteria. In the top-left panel of Fig.~\ref{fig:matching_other_quantities_z}, we show the median stellar mass (solid line) and the minimum and maximum stellar masses (shaded region) of the selected sample. This panel validates our selection based on stellar mass at $z=10$ and $z=14$, confirming that the resulting stellar-mass evolution tracks are roughly consistent with the observational constraints on the stellar masses of all six observed galaxies, shown as white symbols in Fig.~\ref{fig:matching_other_quantities_z}. Interestingly, a single SFH track (grey dashed curves in Fig.~\ref{fig:matching_other_quantities_z}) can satisfy multiple observational constraints at different redshifts, implying that galaxies with the properties of our observed sample at different redshifts may, in principle, be progenitors of one another.

For the six selected \colibre{} galaxies, starting from the second panel in the top row and moving clockwise, we further show the evolution of the halo mass, galaxy formation efficiency (i.e. $M_*/M_{\rm halo}$ divided by $\Omega_{\rm b,0}/\Omega_{\rm m,0}$), total gas mass within a 3D aperture of size $2R_{\rm *,1/2}$, central BH mass, molecular gas fraction within $2R_{\rm *,1/2}$, DM-to-total mass fraction within $2R_{\rm *,1/2}$, and gas mass fraction within $2R_{\rm *,1/2}$. We identify the central BH as the most massive BH particle gravitationally bound to the subhalo. For the molecular gas mass, we do not include a contribution from helium.

The white symbols in the top-right panel show observational constraints on the gas mass from \citet{2026ApJ..1000..159M} for GHZ2 and GS-z14-0, while the white symbol in the bottom-right panel shows the observational constraint on the BH mass from \citet{2024Natur.627...59M} for GN-z11. \citet{2026ApJ..1000..159M} estimated the gas mass by subtracting the stellar mass from the dynamical mass, assuming a negligible dark matter contribution within the galaxy. \citet{2024Natur.627...59M} estimated the BH mass using the width of the N~\textsc{iv} line and the relation from \citet{2009ApJ...699..800V}, calibrated using reverberation mapping.

The median halo mass of the simulated galaxies increases with decreasing redshift from $\sim 10^{9}~\mathrm{M_\odot}$ at $z=20$ to $\sim 10^{11}~\mathrm{M_\odot}$ by $z=10$, consistent with Fig.~\ref{fig:matching_Mhalo_z}. The median galaxy formation efficiency increases from about $1$--$2$~per cent at $z=14$ to $10$~per cent by $z=10$. However, the highest galaxy formation efficiencies already reach $\sim 10$~per cent at $z=14$. As we show in Appendix~\ref{appendix: smfs and sfe}, the evolution of the galaxy formation efficiency is driven by two factors. First, at fixed halo mass, the galaxy formation efficiency decreases moderately with time, by up to $\approx 0.5$~dex from $z=17$ to $10$. Second, as galaxies grow over time, they move along the stellar-to-halo mass relation towards higher halo masses, where the galaxy formation efficiency is naturally higher. This effect raises the galaxy formation efficiency by up to $\approx 1.5$~dex between $z=20$ and $10$.

The median gas mass within $2R_{\rm *,1/2}$ grows from $\sim 10^{8}~\mathrm{M_\odot}$ at $z \approx 17$ to $\sim 10^{9}~\mathrm{M_\odot}$ by $z=10$, and is consistent with the observational constraints from \citet{2026ApJ..1000..159M}. The central gas fraction remains at about $80-90$~per~cent for $z\gtrsim 14$ but decreases to $\approx 50$~per cent at lower redshifts. The median DM-to-total mass fraction within $2R_{\rm *,1/2}$ remains in the range of $\approx 20-60$~per~cent at all times. The molecular gas fraction increases gradually with decreasing redshift, rising from a few per cent at $z \approx 20$ to $\approx 40$~per~cent by $z=10$. The dominant fraction of the gas is instead in atomic form (not shown).

Finally, the bottom-right panel reveals that the massive \colibre{} galaxies are seeded with BHs at $15<z<18$. The initial BH mass is equal to the seed mass of $3\times10^4~\mathrm{M_\odot}$, consistent with a relatively massive BH-seed scenario \citep[e.g.][]{2020ARA&A..58...27I}. At high redshift, the BHs in massive haloes grow predominantly through gas accretion, with only a small fraction of their mass growth contributed by mergers with other BHs. A significant fraction of the accreted mass is accreted in the super-Eddington regime \citep{2026MNRAS.550g1148C}, accelerating high-redshift BH growth. By $z=10$, the median BH mass in the matched galaxies reaches $\sim10^{5.5}~\mathrm{M_\odot}$, while the most massive BHs reach $\sim10^6~\mathrm{M_\odot}$, broadly consistent with the observational estimate of \citet{2024Natur.627...59M} for GN-z11.

\section{Discussion}
\label{section: discussion}

Overall, in $\S$\ref{subsection:comparison with JWST}, we demonstrated that, without any rigorous pre-selection, simply matching the observationally inferred stellar masses of the most luminous \textit{JWST} galaxies at $z>10$ to simulated galaxies in \colibre{} at the same redshift is sufficient to reproduce a broad range of properties of the observed galaxies, including their SFRs, stellar ages, gas metallicities, sizes, and gas masses.

That said, there are two notable exceptions to the list of reproduced properties: (i) the dust mass, which \colibre{} appears to overpredict relative to current observational constraints, and (ii) the observed UV magnitudes, which \colibre{} systematically underpredicts once dust attenuation is included. In $\S$\ref{subsection: shortage_uv_bright_gals}, we argue that the agreement with the observed UV magnitudes can be improved by adopting a top-heavy stellar IMF at high redshift, and in $\S$\ref{subsection: overabundance_grains}, we demonstrate that better agreement with both the observed UV magnitudes and the observational constraints on the dust mass can be achieved by suppressing dust grain growth in \colibre{}. Crucially, since we have shown that \colibre{} is consistent with the inferred SFRs, stellar ages, and stellar masses of the observed luminous galaxies at high redshift (Figs.~\ref{fig:matching_Mstar_z}, \ref{fig:matching_SFR_z}, and \ref{fig:matching_tage_z}), neither modifications to the SFHs of simulated galaxies nor departures from the standard $\Lambda$CDM cosmological model appear to be required to reproduce the full range of observed properties of luminous galaxies at $z>10$ investigated in this work.

\subsection{The shortage of UV-bright galaxies}
\label{subsection: shortage_uv_bright_gals}

The insufficient UV brightness of the simulated galaxies observed in Fig.~\ref{fig:matching_uv_z} is consistent with the results of \citet{2026arXiv260506782L}, who showed that \colibre{} systematically underpredicts the observed UVLF at the bright end for $z>7$ when dust attenuation is accounted for, and at $z=15$ even in the dust-free case. At the same time, \colibre{} is consistent with the observationally inferred GSMF up to $z\approx12$ \citep{2026MNRAS.548ag740C}, which may appear surprising given the insufficient number of UV-bright galaxies in the simulations at $z>7$. As shown by \citet{durrant2026cosmologicalsimulationshighredshiftgalaxy}, the discrepancy at the bright end of the observed UVLF at high redshift can be alleviated by replacing the \citet{2003PASP..115..763C} IMF used in the fiducial \colibre{} model with a variable IMF that becomes top-heavy at high natal gas densities, while otherwise resembling a \citet{2001MNRAS.322..231K} IMF. This modification boosts the predicted \colibre{} UVLF at the bright end for $z>7$ into agreement with the observations, while producing a GSMF very similar to that of the fiducial \colibre{} simulation. 

However, as also demonstrated by \citet{durrant2026cosmologicalsimulationshighredshiftgalaxy}, modifying the IMF to reproduce the bright end of the observed UVLF is far from straightforward, as, in addition to boosting the UV luminosity per unit stellar mass formed, a top-heavy IMF significantly alters stellar feedback, metal and dust yields (and hence dust attenuation), and, most importantly, can introduce new tensions with the data. In particular, \citet{durrant2026cosmologicalsimulationshighredshiftgalaxy} found that, at $9<z<12$, the \colibre{} simulations with a variable IMF that becomes top-heavy in high-density environments predict galaxies around the knee of the UVLF that are systematically too bright. Furthermore, because a top-heavy IMF increases the number of CCSNe, these simulations produce more dust, which may exacerbate the tension with the dust constraints shown in Fig.~\ref{fig:matching_dust_z}.

\subsection{The overabundance of dust grains}
\label{subsection: overabundance_grains}

The potential overabundance of dust grains in \colibre{} could indicate either overly efficient early grain growth at high redshift or insufficient removal of dust from star-forming regions by stellar feedback. To test the former hypothesis, we ran a \colibre{} simulation in a (100 cMpc)$^3$ volume at m7 resolution ($m_{\rm gas}=1.47\times 10^{7}~\mathrm{M_\odot}$) in which grain growth through accretion of the ISM gas was switched off\footnote{For this test, we use the m7 \colibre{} resolution rather than the fiducial m6 resolution adopted in this work, as running the same simulation at m6 resolution would be too computationally expensive. We note, however, that the resolution convergence of the properties studied in this work is generally good (see Appendix~\ref{appendix: convergence}).}. While this simulation represents a limiting case, grain growth in the high-redshift Universe may indeed be inefficient, as suggested by the very blue appearance of the UV-bright galaxies studied in this work, as well as by theoretical considerations \citep[e.g.][]{2016MNRAS.463L.112F,2022MNRAS.512..989D}. The different grain growth at high redshift may be driven by the higher CMB temperature, which keeps the dust warmer, thereby reducing the sticking rates of gas-phase species onto dust grains \citep[e.g.][]{2018MNRAS.476.1371C}. Furthermore, owing to resolution limitations, in the fiducial \colibre{} model, the gas density entering the expression for gas accretion onto dust grains (see equation~(1) and the accompanying text in \citealt{2026MNRAS.548ag375S}) is boosted by a factor of up to 100 in order to produce realistic abundances of dust and molecular hydrogen at low redshifts, but this boost may be excessive at high redshift, where the ISM gas densities can be significantly higher than in the local Universe \citep[e.g.][]{2023ApJ...956..139I,2025NatAs...9..155Z}. 

\begin{figure}
    \centering
    \includegraphics[width=0.99\linewidth]{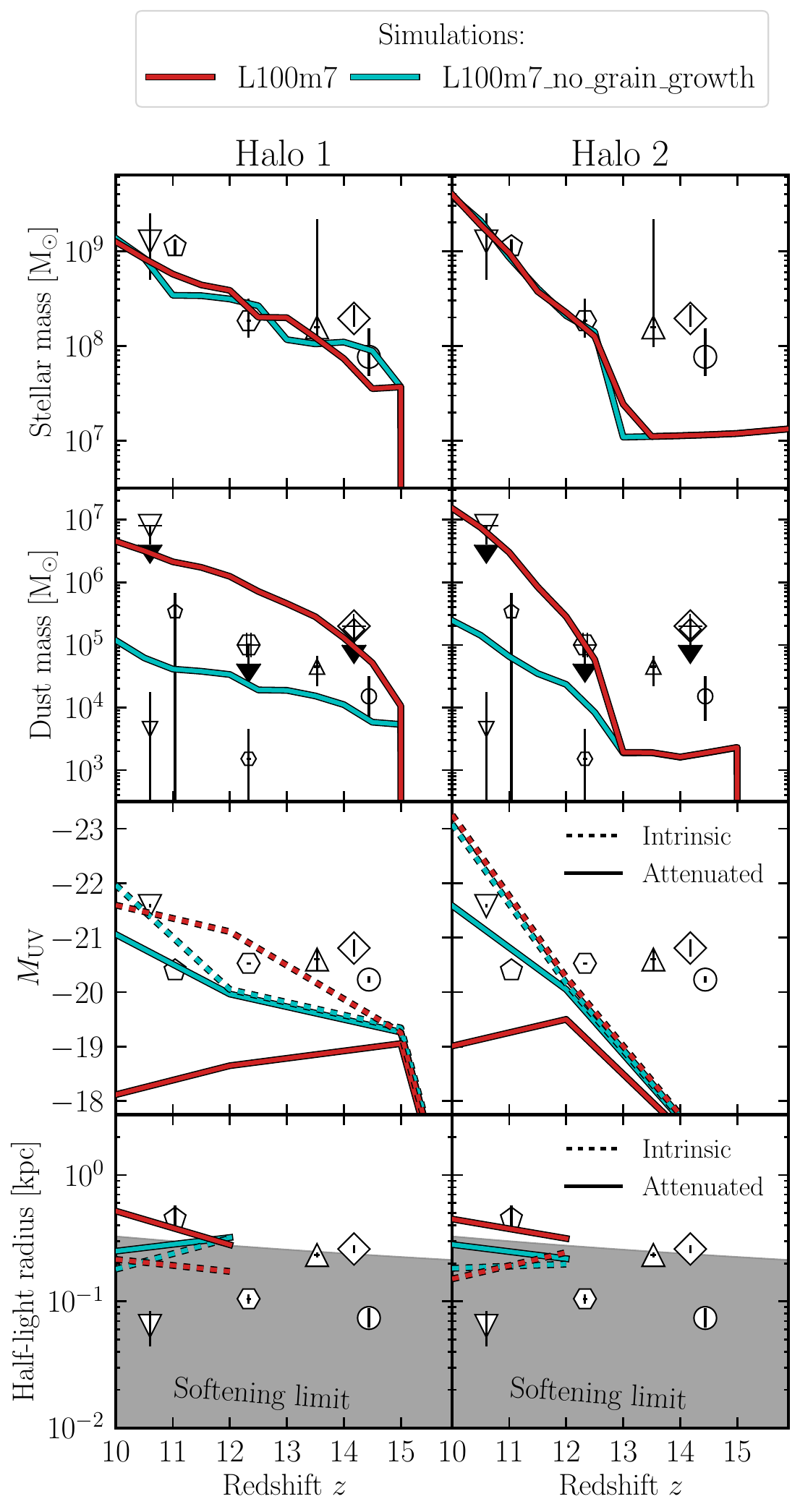}
    \caption{The evolution of stellar mass, dust mass, intrinsic and attenuated absolute UV magnitude, and intrinsic and attenuated projected half-light radius for two massive ($M_{\rm halo} \sim 10^{11}~\mathrm{M_\odot}$ at $z=10$)  haloes in the \colibre{} (100 cMpc)$^3$ volume simulations at m7 resolution. The red curves show the fiducial simulation, in which dust grains are allowed to grow through accretion in the ISM, while the cyan curves show the simulation in which the grain growth is switched off. The intrinsic (attenuated) values are shown by the dashed (solid) curves, while the shaded regions in the bottom panels indicate galaxy sizes below the gravitational softening length. For reference, the white symbols show the observational constraints from Table~\ref{table:observations}. Switching off grain growth reduces the dust masses predicted by \colibre{} at $z \lesssim 13$ by more than 1~dex, resulting in dust-attenuated UV magnitudes that are $\approx 1$–3 mag brighter and attenuated half-light radii that are $\approx 10-50$ per cent smaller.}
    \label{fig:dust_growth}
\end{figure}

In Fig.~\ref{fig:dust_growth}, we compare the simulation with no grain growth with the fiducial \colibre{} simulation (which includes grain growth) in the same volume and at the same resolution. The figure shows the evolution of the stellar mass, dust mass, UV magnitudes, and half-light radii for two massive haloes (with $M_{\rm halo} \sim 10^{11}~\mathrm{M_\odot}$ at $z=10$) that have been cross-matched between the two simulations. These haloes, which are among the most massive in the (100~cMpc)$^{3}$ volume, were selected because they host galaxies with stellar masses similar to the inferred values for the observed galaxies in Table~\ref{table:observations}, making them representative of the galaxy population studied in this work. The stellar and dust masses are measured within 3D apertures of 50~kpc, and we show the half-light radii only at redshifts where the galaxies contain at least five stellar particles.

The properties that are not directly affected by changes in the dust model (stellar mass, intrinsic UV magnitude, and intrinsic half-light radius) are very similar in the two simulations, except for the intrinsic $M_{\rm UV}$ at $z=12$ for the first halo, which is $\approx 1$~mag fainter than in the fiducial simulation, likely due to stochastic run-to-run variations \citep[e.g.][]{2019MNRAS.482.2244K,2023MNRAS.526.2441B}. The dust masses are also initially very similar in the two simulations, as the grains produced by CCSNe have not yet had sufficient time to grow in the ISM. At later times, as the ejected dust grains begin to accrete gas efficiently, the dust mass in the fiducial simulation quickly rises above that in the simulation where the grain growth is not allowed, with the differences monotonically increasing with cosmic time and reaching $\approx 1.5$~dex by $z=10$ for both haloes. This has a dramatic impact on the attenuated UV magnitudes, which become more than 2~mag brighter in the absence of grain growth. The effect on the attenuated half-light radii is less pronounced, with the radii decreasing by $10$--$50$~per~cent when grain growth is switched off.

Overall, this test demonstrates that suppressing the (likely overly efficient) high-redshift grain growth in \colibre{} provides a promising solution for both improving agreement with the observed UV magnitudes and maintaining consistency with the dust-mass constraints of high-redshift galaxies. These results are consistent with those of \citet{2025MNRAS.541.3606T}, who developed a one-dimensional analytical model of galaxy evolution with physically motivated assumptions for dust production and growth, and found that the predicted dust attenuation at high redshift is too strong, leading to a discrepancy with the observed UVLF. Similar conclusions were reached by \citet{2025MNRAS.536.1018L} using the \textsc{Galform} semi-analytical model \citep{2016MNRAS.462.3854L}, for which the predicted attenuated UVLF undershoots the data at $z\gtrsim 12$ unless the dust-to-metal ratios are lowered compared to their original values. Finally, we note that reducing dust attenuation can be achieved not only by suppressing grain growth in the ISM but also by reducing the dust yields from CCSNe, for example if a significant fraction of the freshly formed dust is destroyed by reverse shocks in supernova ejecta \citep[e.g.][]{2007ApJ...666..955N}.

However, one important caveat is that at $z \lesssim 7$, \colibre{} produces fewer dusty galaxies than observed in the real Universe \citep{2026arXiv260726058V}, so suppressing grain growth and/or lowering dust yields to reproduce the $z>10$ observations may further exacerbate this tension. More generally, below $z\approx 8$, the very high inferred dust masses of observed galaxies \citep[e.g.][]{2026MNRAS.545f1897A,2026MNRAS.546f2284B} demand faster grain growth and higher dust yields, the opposite of what $z\gtrsim 10$ data suggest, which is a well-known tension \citep[e.g.][]{2025A&A...694A.286F,2026ApJ..1003..170S}. While resolving this tension is beyond the scope of this work, one possible solution could be a steep acceleration of grain growth below $z\approx 10$, for example if the initially large ($\sim0.1~\mu\mathrm{m}$) grains expected from CCSNe \citep[e.g.][]{2007ApJ...666..955N,2020ApJ...902..135S} are progressively shattered into smaller grains, which can grow more rapidly due to their larger surface-to-volume ratios \citep{2026OJAp....959986N}.

An alternative solution has been proposed by \citet{2024A&A...684A.207F}, who constructed an attenuation-free model that reproduces the observed blue colours of \textit{JWST} galaxies and the cosmic UV luminosity density at high redshift without invoking unusually strong dust destruction rates or unusually low stellar dust yields. Instead, in this model, freshly produced dust is evacuated from star-forming, UV-emitting regions to larger, kpc-scale distances by radiation-driven outflows, making the galaxies appear dust-free and UV-bright. By comparing the solid and dotted curves at fixed redshift in Fig.~\ref{fig:matching_dust_z}, one can see that for the majority of \colibre{} galaxies, the dominant fraction of the dust mass instead remains concentrated within twice the half-mass radius at most times, which, according to Fig.~\ref{fig:matching_size_z}, corresponds to scales of only a few hundred parsecs. Although the presence of the dip around $z=12$--$13$ in all solid curves does indicate the evacuation of dust from $\sim 100$~pc to kpc scales by feedback-driven outflows generated by the first stars, the overall fraction of such events in the matched galaxy samples is small (as can be judged from the shallow depth of the dip).

As shown by \citet{2026MNRAS.546ag268B}, radiation-pressure feedback in \colibre{} is relatively modest, making it unlikely to drive strong outflows capable of evacuating dust from high-redshift galaxies. The reason for this behaviour is likely twofold. First, the momentum of each photon is assumed to be transferred to the gas and dust only once upon absorption, neglecting re-emitted infrared photons, which could otherwise substantially boost the total momentum transfer in dense, optically thick environments. Second, owing to the limited resolution of the simulation, the cumulative momentum imparted by radiation pressure from young stellar populations is distributed over a relatively large gas mass ($\sim 10^{7}~\mathrm{M_\odot}$), reducing its effectiveness in driving (dusty) winds.

\section{Conclusions}
\label{section: conclusions}

We have studied the properties of the most massive galaxies at $z>10$ in the \colibre{} simulations of galaxy formation \citep{2026MNRAS.548ag375S,2026MNRAS.548ag300C}. The \colibre{} model captures the multiphase nature of the ISM \citep{2025MNRAS.543..891P}, includes a prescription for the formation and evolution of interstellar dust \citep{2026MNRAS.545f2040T}, and uses SN and AGN feedback whose strength was calibrated to reproduce the observationally inferred GSMF and size--stellar mass relation in the local Universe. No $z>0$ constraints were applied during the calibration, so that all high-redshift galaxy properties are genuine predictions of the model.

We used the \colibre{} L200m6 simulation, corresponding to a $(200~\mathrm{cMpc})^3$ volume and m6 resolution (gas and DM particle masses of $m_{\rm gas}\approx m_{\rm dm}\sim10^6~\mathrm{M_\odot}$), which provides a good balance between a sufficiently large volume, necessary to build a representative sample of massive galaxies at high redshift, and sufficiently high resolution, allowing an early onset of star formation. We first investigated the properties of massive galaxies at $z>10$, selected at $z=10$ to have stellar masses $M_* \geq 10^8~\mathrm{M_\odot}$. Identifying $515$ such galaxies, we divided them into four 0.5-dex stellar-mass bins and tracked them back to the onset of star formation along their main progenitor branches. We found that these massive galaxies experience rapid early stellar mass build-up, with galaxies selected in higher $z=10$ stellar-mass bins reaching higher stellar masses and residing in more massive haloes at fixed redshift (Fig.~\ref{fig:evolution_mstar}). The first star particles in the galaxies from the most massive bin ($10^{9.5}\leq M_*/\mathrm{M_\odot}<10^{10}$) form as early as $z=20$, with their stellar mass increasing by up to three orders of magnitude within just $\approx 300$~Myr of cosmic time, from $M_* \approx 10^{6.5}~\mathrm{M_\odot}$ at $z=20$ to $M_*\approx 10^{9.5}~\mathrm{M_\odot}$ by $z=10$. The host haloes of these most massive galaxies have masses of $10^{10.2-10.6}~\mathrm{M_\odot}$ at $z=14$, increasing to $\approx10^{11.2}~\mathrm{M_\odot}$ by $z=10$.

Having studied the evolution of the stellar and halo masses of massive \colibre{} galaxies at $z>10$, we then compared a subset of these galaxies with the most luminous and well-studied spectroscopically confirmed galaxies observed by \textit{JWST} at $z>10$. We compiled a sample of six observed galaxies with available spectroscopic data: GN-z11, CEERS2-588, GHZ2, PAN-z14-1, GS-z14-0, and MoM-z14 (see Table~\ref{table:observations} for details). These galaxies span redshifts $10.6 < z < 14.44$ and have measured UV magnitudes $-21.58 \leq M_{\rm UV} \leq -20.23$ and inferred stellar masses in the range $10^{7.9} < M_*/\mathrm{M_\odot} < 10^{9.1}$. For each observed galaxy, we identified six simulated counterparts that best reproduce the observationally inferred stellar mass at the observed redshift. Our main results are as follows:

\begin{itemize}

\item The matched simulated galaxies undergo rapid stellar mass growth, with stellar masses increasing on average by about 3~dex within just $\approx 300$~Myr of cosmic time (Fig.~\ref{fig:matching_Mstar_z}). For a given observed galaxy, the SFHs of the matched counterparts can differ significantly, with variations in stellar mass at fixed redshift (sufficiently far from the redshift of the observation) of 1-2 dex.

\item Using the matched samples of simulated galaxies, we predicted that GN-z11 and CEERS2-588 reside in host haloes of mass $M_{\rm halo}\approx10^{10.9\text{--}11.0}~\mathrm{M_\odot}$ and $M_{\rm halo}\approx10^{10.7\text{--}11.2}~\mathrm{M_\odot}$, respectively, GHZ2 in $M_{\rm halo}\approx10^{10.4\text{--}10.6}~\mathrm{M_\odot}$, PAN-z14-1 in $M_{\rm halo}\approx10^{10.1\text{--}10.7}~\mathrm{M_\odot}$, GS-z14-0 in $M_{\rm halo}\approx10^{10.0\text{--}10.6}~\mathrm{M_\odot}$, and MoM-z14 in $M_{\rm halo}\approx10^{9.9\text{--}10.5}~\mathrm{M_\odot}$ (Fig.~\ref{fig:matching_Mhalo_z}).

\item The matched \colibre{} galaxies are consistent with the SFRs (Fig.~\ref{fig:matching_SFR_z}) and stellar ages (Fig.~\ref{fig:matching_tage_z}) inferred for the observed galaxies, even though these quantities were not used in the matching procedure. The only exception is MoM-z14, whose SFR and stellar age are, respectively, higher and lower than those predicted by \colibre{}, highlighting the strong correlation between these two quantities. At $z\approx12$, \colibre{} predicts a typical mass-weighted stellar age of $\approx50~\mathrm{Myr}$, decreasing to $\approx30~\mathrm{Myr}$ by $z\approx14$.

\item Most of the stellar mass in the matched \colibre{} galaxies is formed in situ (Fig.~\ref{fig:insitu-exsitu_decomposition}). The ex-situ component contributes, on average, less than 15~per~cent of the total stellar mass at all redshifts, with a few exceptions where it reaches 30--50~per~cent at $12<z<16$.

\item The simulated galaxies undergo rapid chemical enrichment, with gas-phase metallicities in the ISM reaching $\sim 10$~per~cent of the solar value by $z \approx 15-17$ (Fig.~\ref{fig:matching_Z_z}). For the simulated galaxies matched to GN-z11 and CEERS2-588, the metallicity further increases to solar by $z = 10$, whereas for the remaining four observed galaxies it remains sub-solar. The observationally inferred gas-phase metallicities of all six observed galaxies are broadly reproduced.

\item \colibre{} predicts very rapid dust grain growth. The dust mass increases from $\sim 10^{3-4}~\mathrm{M_\odot}$ at $z \approx 18$ to $\sim 10^{6-7}~\mathrm{M_\odot}$ by $z = 10$, with a large fraction of the dust remaining concentrated within twice the stellar half-mass radius at most times (Fig.~\ref{fig:matching_dust_z}). Although the dust masses of the matched simulated galaxies tend to be systematically higher than those inferred from the observations, these discrepancies can be alleviated by suppressing the dust growth rate at high redshift (Fig.~\ref{fig:dust_growth}), which is boosted by up to a factor of 100 in \colibre{} to compensate for the limited resolution.

\item The predicted UV magnitudes are in agreement with the observations when dust attenuation is neglected, except for the highest-redshift object (MoM-z14, $z = 14.44$), whose UV magnitude is $\approx 0.5$ mag brighter than that of the brightest matched \colibre{} galaxy at the same redshift. In contrast, when dust attenuation is accounted for, the simulation systematically underpredicts the observed UV luminosities for all six galaxies in the observed sample. These discrepancies can be alleviated by suppressing grain growth in the simulations at high redshift (Fig.~\ref{fig:dust_growth}) and/or by replacing the \citet{2003PASP..115..763C} IMF with a variable IMF that becomes top-heavy at high redshift (as shown by \citealt{durrant2026cosmologicalsimulationshighredshiftgalaxy}).

\item The sizes of the matched simulated galaxies span $\approx 50-500$~pc, depending on redshift and whether dust attenuation is included, with the lower limit reflecting the gravitational softening scale (Fig.~\ref{fig:matching_size_z}). \colibre{} reproduces the sizes of observed galaxies that are larger than the gravitational softening length of the simulation.

\item \colibre{} reproduces the existing constraints on the gas masses of GHZ2 and GS-z14-0 and is broadly consistent with the constraint on the BH mass of GN-z11 (Fig.~\ref{fig:matching_other_quantities_z}). \colibre{} also predicts that less than about 50~per cent of the ISM gas in the matched galaxies is in molecular form, that the central DM mass fraction remains roughly within the $\approx 20-60$~per cent range, and that the central gas fraction lies between 40 and 90~per cent, depending on redshift. The median values of the galaxy formation efficiency are below 3~per cent at $z>14$, but rise to $\sim 10$~per cent by $z=10$, and for some individual galaxies they can reach 10~per cent already at $z=14$.
\end{itemize}

Taken together, these results provide strong evidence that the properties of high-$z$ galaxies observed by \textit{JWST}, including the most luminous objects, can be naturally reproduced within the standard framework of galaxy formation physics and the $\Lambda$CDM cosmological model. We note, however, that our conclusions are based on comparing \colibre{} to only those observed galaxies that have been spectroscopically confirmed, reaching a maximum redshift of $z=14.44$ \citep{2026OJAp....956033N}. Very recently, \citet{2025ApJ...991..179P} identified several galaxy dropout candidates at $z \approx 17$ and $z \approx 25$ in the deepest \textit{JWST}/NIRCam data to date, with measured UV magnitudes of $-19 < M_{\rm UV} < -17$ and inferred stellar masses $\sim 10^{6-8}~\mathrm{M_\odot}$. If confirmed, these galaxies would represent a new milestone in the study of the early Universe and become a critical benchmark for testing models of galaxy formation at high redshift. Obtaining deeper \textit{JWST}/NIRCam and MIRI data, together with future \textit{JWST}/NIRSpec and ALMA follow-up observations, will be crucial for gaining new insights into, and placing tighter constraints on, the physical properties of galaxies during cosmic dawn.

\section*{Acknowledgements}

EC thanks Filip Hu\v{s}ko, Piyush Sharda, Matthieu Schaller, Kai Wang, and Xu Zhao for useful discussions. This work used the DiRAC@Durham facility managed by the Institute for Computational Cosmology on behalf of the STFC DiRAC HPC Facility (www.dirac.ac.uk). The equipment was funded by BEIS capital funding via STFC capital grants ST/K00042X/1, ST/P002293/1, ST/R002371/1 and ST/S002502/1, Durham University and STFC operations grant ST/R000832/1. DiRAC is part of the National e-Infrastructure. AD acknowledges an STFC doctoral studentship. AP was supported by funding from the European Research Council (ERC) under the European Union's Horizon 2020 research and innovation programmes (grant agreement no.\ 818085 GMGalaxies). EC acknowledges support from STFC consolidated grant ST/X001075/1. RAC acknowledges support from STFC grants ST/Y002482/1 and ST/Y001907/1.

\section*{Data Availability}

The data underlying this article will be shared on reasonable request to the corresponding author. The public version of the \textsc{Swift} code can be found on \href{http://www.swiftsim.com}{www.swiftsim.com}. The \textsc{Swift} modules related to the \colibre{} model of galaxy formation will be integrated into the public version after the public release of \colibre. The \textsc{chimes} astrochemistry code is publicly available at \href{https://richings.bitbucket.io/chimes/home.html}{https://richings.bitbucket.io/chimes/home.html}. 

\bibliographystyle{mnras}
\bibliography{main} 

\appendix

\section{Resolution convergence}
\label{appendix: convergence}

\begin{figure}
    \centering
    \includegraphics[width=0.95\linewidth]{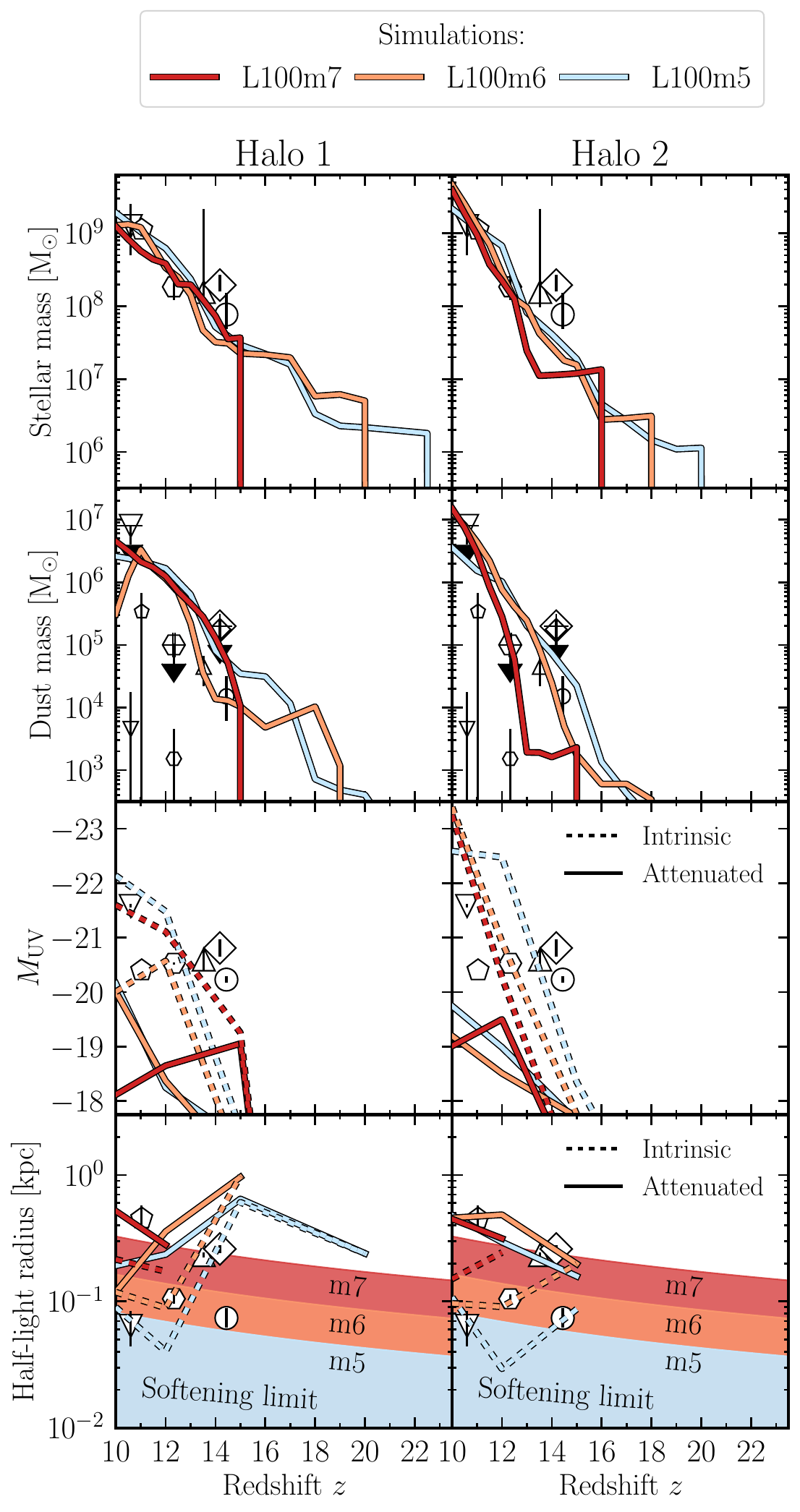}
    \caption{Numerical convergence test for two massive haloes ($M_{\rm halo} \sim 10^{11}~\mathrm{M_\odot}$ at $z=10$) in (100~cMpc)$^{3}$ \colibre{} simulations at three different resolutions (colours). From top to bottom: the evolution of stellar mass, dust mass, intrinsic and dust-attenuated absolute UV magnitude, and intrinsic and dust-attenuated projected half-light radius. The intrinsic (attenuated) values are shown by the dashed (solid) curves, while the shaded regions in the bottom panels indicate the gravitational softening limit for each resolution. For reference, the white symbols show observational constraints on the galaxies from Table~\ref{table:observations}. Stellar mass, dust mass, and attenuated half-light radii are well converged across all three resolutions at $z\lesssim13$, whereas the intrinsic half-light radii systematically decrease with increasing resolution.}
    \label{fig:convergence}
\end{figure}

Throughout this work, we have used the L200m6 \colibre{} simulation for our analysis, which has a volume of (200~cMpc)$^{3}$ and m6 resolution, corresponding to a mean gas particle mass of $m_{\rm gas}=1.84\times10^{6}~\mathrm{M_\odot}$. In this appendix, we test how well our main results converge with numerical resolution. To this end, we compare predictions from the \colibre{} simulations at m7, m6, and m5 resolutions, all using a volume of (100~cMpc)$^{3}$, which is the largest box available at m5 resolution.

Fig.~\ref{fig:convergence} shows the evolution of the galaxy stellar mass, dust mass, intrinsic and attenuated UV magnitude, and intrinsic and attenuated projected half-light radius for two massive haloes ($M_{\rm halo} \sim 10^{11}~\mathrm{M_\odot}$ at $z=10$), selected from and cross-matched between the three simulations. These are the same two haloes whose properties were shown in Fig.~\ref{fig:dust_growth}. The galaxies hosted by these haloes have stellar masses of $M_*\sim 10^{9}~\mathrm{M_\odot}$ at $z=10$, making them representative of the galaxy population studied in this work. As in Fig.~\ref{fig:dust_growth}, the stellar and dust masses are measured within 3D apertures of 50~kpc, and we show the half-light radii only at redshifts where the galaxies contain no fewer than five stellar particles.

The stellar and dust masses are broadly converged at $z \lesssim 13$ across all three resolutions, with differences of roughly a factor of a few and no significant systematic deviations. At $z \gtrsim 13$, the m7 simulation begins to deviate most strongly with increasing redshift, as expected because the stellar mass formed in the selected haloes approaches and eventually falls below the mass of a single stellar particle at m7 resolution ($\sim 10^7~\mathrm{M_\odot}$), resulting in sharp drop-offs in the predicted stellar and dust masses due to insufficient resolution.

The convergence of $M_{\rm UV}$, including the relative difference between attenuated and intrinsic values, is reasonable, although the deviations between different resolution levels are larger than for the stellar and dust masses. This likely reflects the greater stochasticity in the number of young stellar particles, to which $M_{\rm UV}$ is particularly sensitive. We note, however, that \citet{2026arXiv260506782L} found the $M_{\rm UV}$--$M_*$ relation to be well converged in a statistical sense over $7 < z < 15$. Finally, the bottom row shows that intrinsic half-light radii systematically decrease with increasing resolution at all redshifts, roughly in proportion to the gravitational softening length (shaded regions in Fig.~\ref{fig:convergence}), whereas the attenuated half-light radii, all of which lie above the corresponding softening scales owing to the strong attenuation of galaxies' central regions, are converged relatively well.

Overall, these results provide evidence that the main findings of this work based on integrated galaxy properties are not driven by numerical effects, while confirming that the only spatially resolved properties investigated here, galaxy sizes, should be interpreted with more caution and should generally not be expected to match observational data in the regime below the gravitational softening scale of the simulation.

\section{The impact of SED-fitting assumptions on stellar mass and ages}
\label{appendix: SED assumptions}

\begin{figure*}
    \centering
    \includegraphics[width=0.99\linewidth]{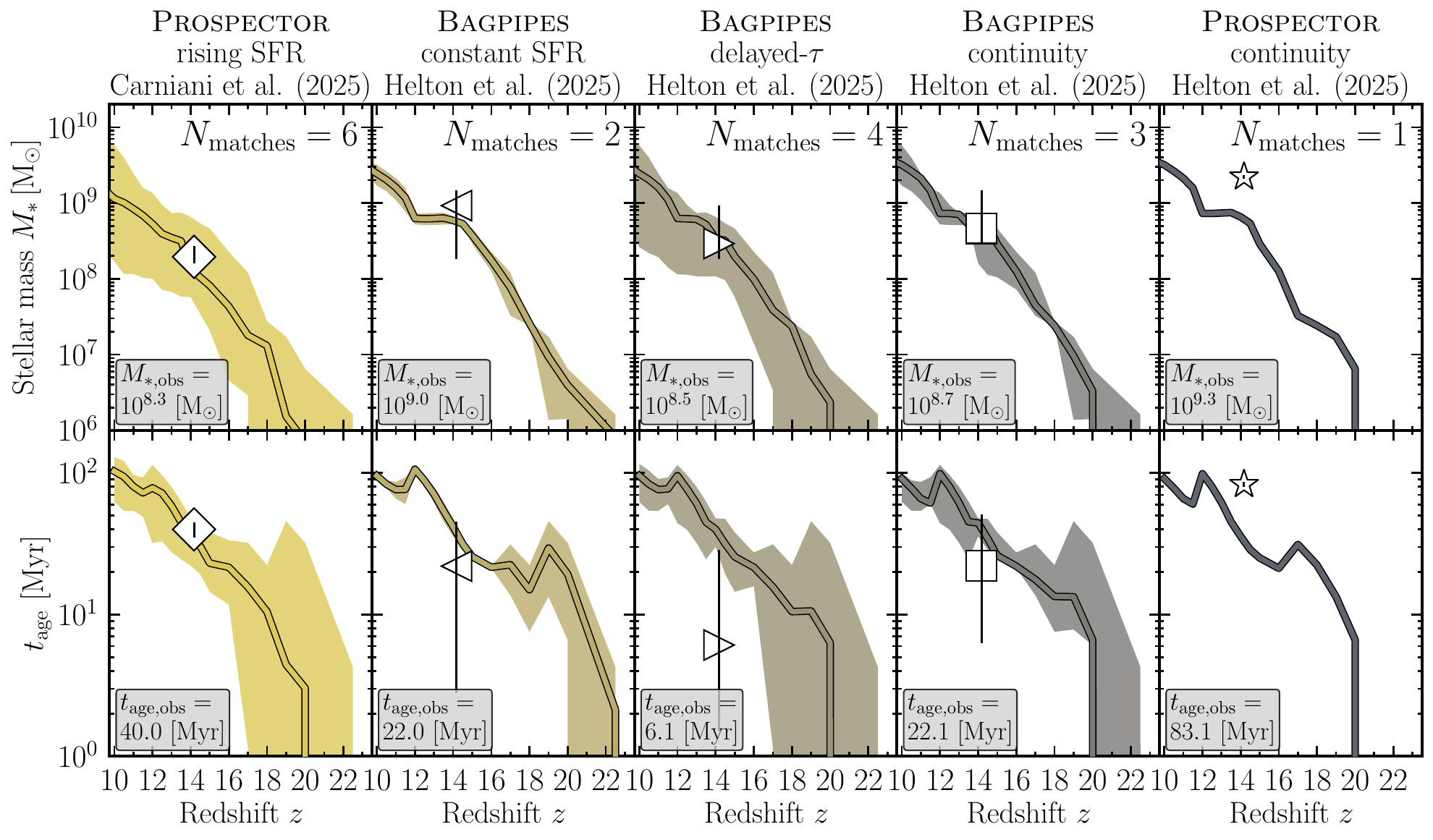}
    \caption{The observationally inferred stellar masses (top row) and mass-weighted stellar ages (bottom row) for GS-z14-0 using different SED-fitting pipelines. The left column shows the constraints from \citet{2025A&A...696A..87C} based on \textsc{Prospector} and a rising SFH prior, used in the main part of this work. Columns two to five show the constraints from \citet{2025NatAs...9..729H} for, respectively, \textsc{Bagpipes} with a constant SFR prior, \textsc{Bagpipes} with a delayed-$\tau$ SFH, \textsc{Bagpipes} with a continuity prior, and \textsc{Prospector} with a continuity prior. For each case, we select simulated galaxies from the \colibre{} L200m6 simulation using the matching algorithm described in Section~\ref{subsection: matching}. The solid curves show the median stellar masses and mass-weighted ages computed over the samples of matched \colibre{} galaxies, and the shaded regions indicate the minimum and maximum values in these samples. \colibre{} is consistent with all observational estimates for GS-z14-0, except for the case based on \textsc{Prospector} with a continuity prior, in which \colibre{} strongly underpredicts both the inferred stellar mass and age.}
    \label{fig:GS14-variations}
\end{figure*}

In this appendix, we quantify the sensitivity of the results presented in this work to the assumptions adopted in SED-fitting algorithms used to infer the stellar masses and ages of observed galaxies. We perform this test for the second-highest-redshift galaxy, GS-z14-0, whose stellar mass and age have been reported using multiple SED-fitting algorithms and SFH priors, with differences between the inferred values exceeding 1~dex.

Specifically, in the main part of this work, we adopted the stellar mass and age inferred by \citet{2025A&A...696A..87C}, who used the Bayesian inference framework \textsc{Prospector} to derive these quantities. Their analysis employed the Flexible Stellar Population Synthesis (\textsc{fsps}) code \citep{2009ApJ...699..486C} with MIST stellar evolutionary tracks and isochrones \citep{2016ApJ...823..102C}, the dust attenuation law of \citet{2013ApJ...775L..16K}, nebular emission from \citet{2017ApJ...840...44B}, and the rising SFH prior \citep{2025MNRAS.537.1826T}, yielding $M_*\approx10^{8.29}~\mathrm{M_\odot}$ and $t_{\rm age}\approx40$~Myr.

In Fig.~\ref{fig:GS14-variations}, we compare these estimates with those of \citet{2025NatAs...9..729H}, who used the Bayesian SED-fitting code \textsc{Bagpipes}, adopting the MILES stellar library \citep{2011A&A...532A..95F}, PARSEC stellar evolutionary tracks and isochrones \citep{2012MNRAS.427..127B}, the nebular emission model of \citet{2017ApJ...840...44B}, and the dust attenuation law of \citet{2000ApJ...533..682C}. They assumed three SFH priors -- constant SFH, delayed-$\tau$, and continuity -- which yielded stellar mass estimates of $M_* \approx10^{9.0}$, $10^{8.5}$, and $10^{8.7}~\mathrm{M_\odot}$, and mass-weighted ages of $t_{\rm age}=22.0$, $6.1$, and $22.1$~Myr, respectively. In addition, \citet{2025NatAs...9..729H} fitted the same data using the \textsc{Prospector} pipeline, based on the \textsc{fsps} library with MIST stellar evolutionary tracks and isochrones \citep{2016ApJ...823..102C} and a continuity SFH prior, finding $M_* \approx10^{9.38}~\mathrm{M_\odot}$ and $t_{\rm age}=83.1$~Myr. The differences between the lowest and highest stellar masses and ages reported by \citet{2025NatAs...9..729H} illustrate the well-known mass--age degeneracy: similar observed SEDs can often be reproduced by either a relatively young, low-mass stellar population or an older, more massive stellar population. Finally, we note that \citet{2025NatAs...9..729H} used photometric data from \textit{JWST}/NIRCam and \textit{JWST}/MIRI together with spectroscopic data from \textit{JWST}/NIRSpec, whereas \citet{2025A&A...696A..87C} additionally incorporated ALMA observations of GN-z14-0.

Fig.~\ref{fig:GS14-variations} shows the combined set of five constraints on the stellar mass and age of GS-z14-0 from \citet{2025A&A...696A..87C} and \citet{2025NatAs...9..729H} (white symbols in different columns). For each version of the constraints, we looked for simulated galaxies from the \colibre{} L200m6 simulation, and performed the matching as described in $\S$\ref{subsection: matching}. Because the inferred stellar masses reported by \citet{2025NatAs...9..729H} are very high, we find fewer than six simulated counterparts matching their values with our fiducial stellar mass tolerance of $0.6$~dex. Note also that only stellar mass, not age, is used to select counterparts, leaving age as an independent cross-check. The solid lines indicate the evolution of the median stellar mass (top row) and mass-weighted stellar age (bottom row), computed over the samples of matched simulated galaxies. As in previous figures, the boundaries of the shaded regions indicate the minimum and maximum values of these two quantities in the matched samples. The number of matched counterparts, which satisfy our $0.6$~dex tolerance on the deviation from the observed stellar mass, is shown in the top panel of each column.

We find that \colibre{} has counterparts not only for our fiducial set of constraints on GS-z14's stellar mass and age from \citet{2025A&A...696A..87C}, but also for those from \citet{2025NatAs...9..729H} using the \textsc{Bagpipes} code with constant, delayed-$\tau$, and continuity SFH priors, although for the delayed-$\tau$ case, the predicted stellar age is only marginally in agreement with the data. However, \colibre{} struggles to match the inferred values of stellar mass and age from \citet{2025NatAs...9..729H} based on \textsc{Prospector} and the continuity SFH prior, shown in the right-most column. In this case, only one matched galaxy is identified, whose stellar mass at the observed redshift of GS-z14-0 is $\approx 0.5$~dex lower than the observationally inferred value. The stellar age of this simulated galaxy is also inconsistent: $30$~Myr versus the observed value of $\approx 80$~Myr.

These results agree with \citet{2025NatAs...9..729H}, who discuss that no current cosmological simulations can form galaxies at the redshift of GS-z14-0 with such high stellar masses and ages, and argue that if the physical properties of GS-z14-0 inferred by \textsc{Prospector} with the continuity prior are correct, this would have important implications for our understanding of galaxy formation in the early Universe.

\section{Comparison with other UV-bright galaxies at high redshift}
\label{appendix:comparison_to_other_bright_galaxies}

\begin{figure*}
    \centering
    \includegraphics[width=0.8\linewidth]{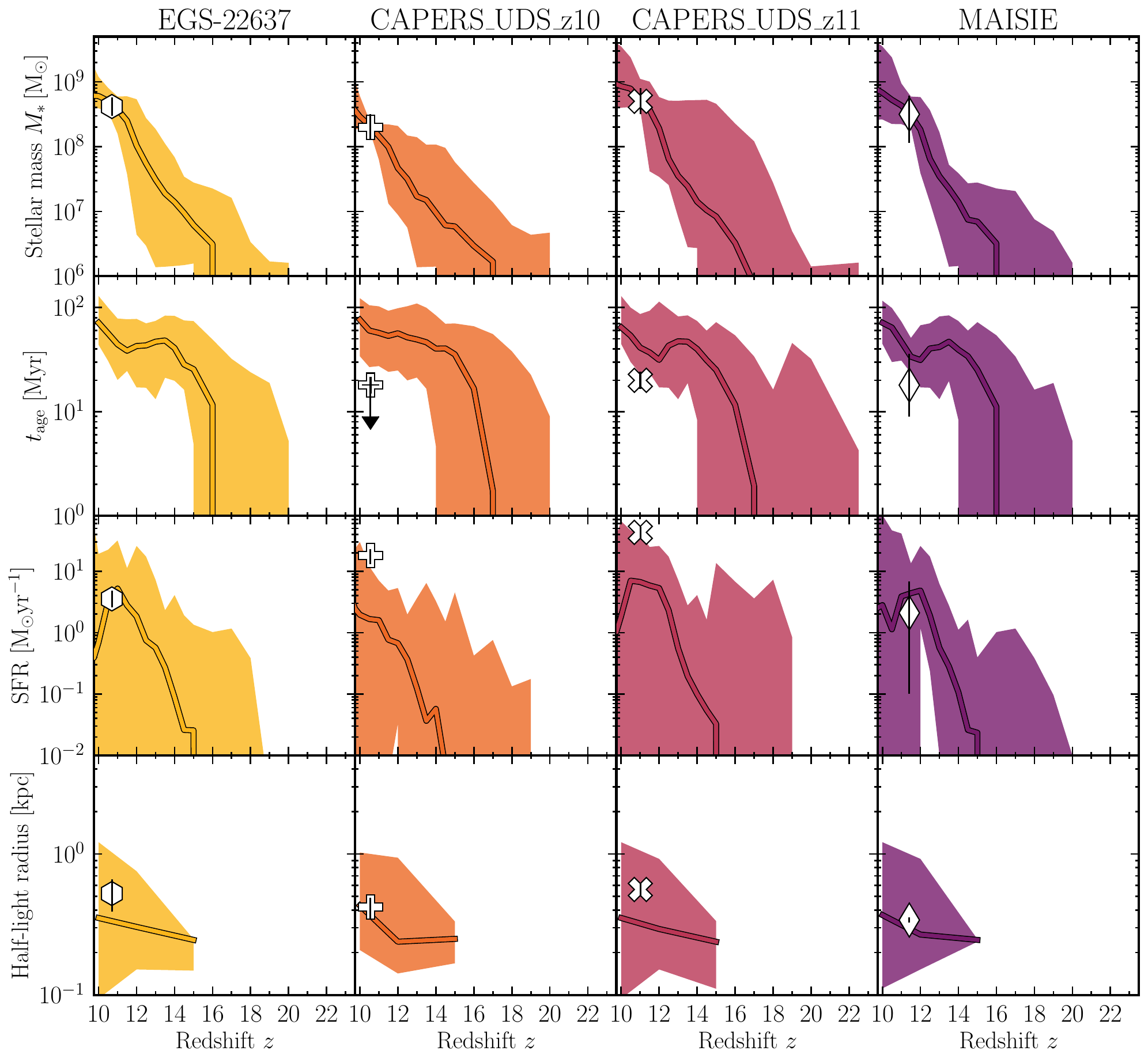}
    \caption{Comparison with four bright galaxies at $z>10$ that were not included in our fiducial observed sample but satisfy $M_{\rm UV}<-20.2$: EGS-22637, CAPERS\_UDS\_z10, CAPERS\_UDS\_z11, and MAISIE (orange squares in Fig.~\ref{fig:selection}). Each observed galaxy is matched with 24 \colibre{} galaxies from the L200m6 simulation, requiring the stellar mass difference at the observed redshift to be no more than 0.6~dex. From top to bottom, we show \colibre{} predictions for the stellar mass, mass-weighted stellar age, SFR, and projected attenuated half-light radius. The thick solid lines indicate the evolution of the median values computed over the matched samples, while the boundaries of the shaded regions show the minimum and maximum values in these samples. The white symbols indicate the observationally inferred values (where available). The names of the observed galaxies are indicated at the top of each column. All observed galaxies are well matched by \colibre{} in terms of their stellar mass at the observed redshift, and their other properties are also broadly reproduced, except for the inferred age of CAPERS\_UDS\_z10, which is at least a factor of $1.5$ lower than the age of the youngest matched \colibre{} galaxy at the same redshift.}
    \label{fig:comparison_with_other_gals}
\end{figure*}

In the main part of this work, to build a sample of luminous high-redshift galaxies for comparison with the \colibre{} simulations, we selected observed galaxies that (i) have absolute UV magnitudes $M_{\rm UV}< -20.2$, (ii) lie at redshifts $z>10$, and (iii) are spectroscopically confirmed. We then further reduced the resulting sample to the six most luminous galaxies at their respective redshifts (see $\S$\ref{subsection: observational data}). For completeness, in this appendix we present \colibre{} predictions for four other spectroscopically confirmed galaxies that also satisfy our initial selection criteria ($M_{\rm UV}< -20.2$ and $z>10$).

These four galaxies are EGS-22637 \citep{2026MNRAS.548ag701R}, CAPERS\_UDS\_z10 \citep{2025ApJ...988L..10K}, CAPERS\_UDS\_z11 \citep{2025ApJ...988L..10K}, and MAISIE \citep{2022ApJ...940L..55F,2023Natur.622..707A}, all denoted by the orange squares in Fig.~\ref{fig:selection}. They have redshifts in the range $10.5<z<11.4$ and inferred stellar masses of $10^{8.3}<M_*/\mathrm{M_\odot}<10^{8.7}$. We match simulated galaxies from the \colibre{} L200m6 simulation to these four observed galaxies using the same stellar mass-based matching algorithm as for our main observed sample (\S\ref{subsection: matching}). Because the observationally inferred stellar masses of these galaxies are substantially lower than those of the galaxies in our main sample at comparable redshifts (see Table~\ref{table:observations}), for each of these four galaxies, we find at least 12 simulated counterparts with stellar masses exceeding the observationally inferred value within our fiducial tolerance of $0.6$~dex. We therefore construct the matched sample for each of the four galaxies by selecting the 24 counterparts within our fiducial tolerance that have the smallest stellar-mass deviations, with 12 counterparts having stellar masses above and 12 below the observationally inferred value.

Fig.~\ref{fig:comparison_with_other_gals} compares \colibre{} predictions with these four additional observed galaxies, with the columns, from left to right, corresponding to EGS-22637, CAPERS\_UDS\_z10, CAPERS\_UDS\_z11, and MAISIE. From top to bottom, the figure shows the evolution of the stellar mass, mass-weighted stellar age, SFR, and projected attenuated half-light radius. The thick solid lines indicate the median values computed over the matched samples of 24 galaxies, while the shaded regions indicate the minimum and maximum values in these samples. The white symbols indicate the observationally inferred values, where available.

As expected from the matching procedure, all four observed galaxies are well matched by the \colibre{} galaxies in terms of their stellar masses at the observed redshifts. Their SFRs and half-light radii are also broadly reproduced. The stellar ages of CAPERS\_UDS\_z11 and MAISIE are marginally reproduced, while the inferred age of CAPERS\_UDS\_z10, which is an upper limit, is at least a factor of $1.5$ lower than that of the youngest matched \colibre{} galaxy at the same redshift.

We note that the SED fitting for CAPERS\_UDS\_z11 and CAPERS\_UDS\_z10 was performed using \textsc{Bagpipes} with a bursty-continuity prior \citep[see][for details]{2025ApJ...988L..10K}. As shown by \citet{2022ApJ...927..170T}, the bursty continuity prior favours younger stellar ages and SFHs consistent with the rapidly rising cosmic SFR density, whereas adopting the continuity prior can increase the inferred stellar ages by about a factor of three (see also Fig.~\ref{fig:GS14-variations}). Since both sets of inferred values are compatible with the observational data, the large differences in the inferred stellar ages merely reflect the substantial uncertainties in the derived quantity. We therefore conclude that the tension between the \colibre{} predictions and the inferred age of CAPERS\_UDS\_z10 can likely be resolved by adopting a different SFH prior such as the continuity prior, which is as reasonable a choice as the bursty continuity prior. As concluded by \citet{2022ApJ...927..170T}, better constraints on the stellar ages (and other physical properties) of high-redshift galaxies require additional multi-wavelength and deeper observations.

\section{The evolution of star formation rates and galaxy formation efficiency of massive galaxies}
\label{appendix: smfs and sfe}

\begin{figure*}
    \centering
    \includegraphics[width=0.99\linewidth]{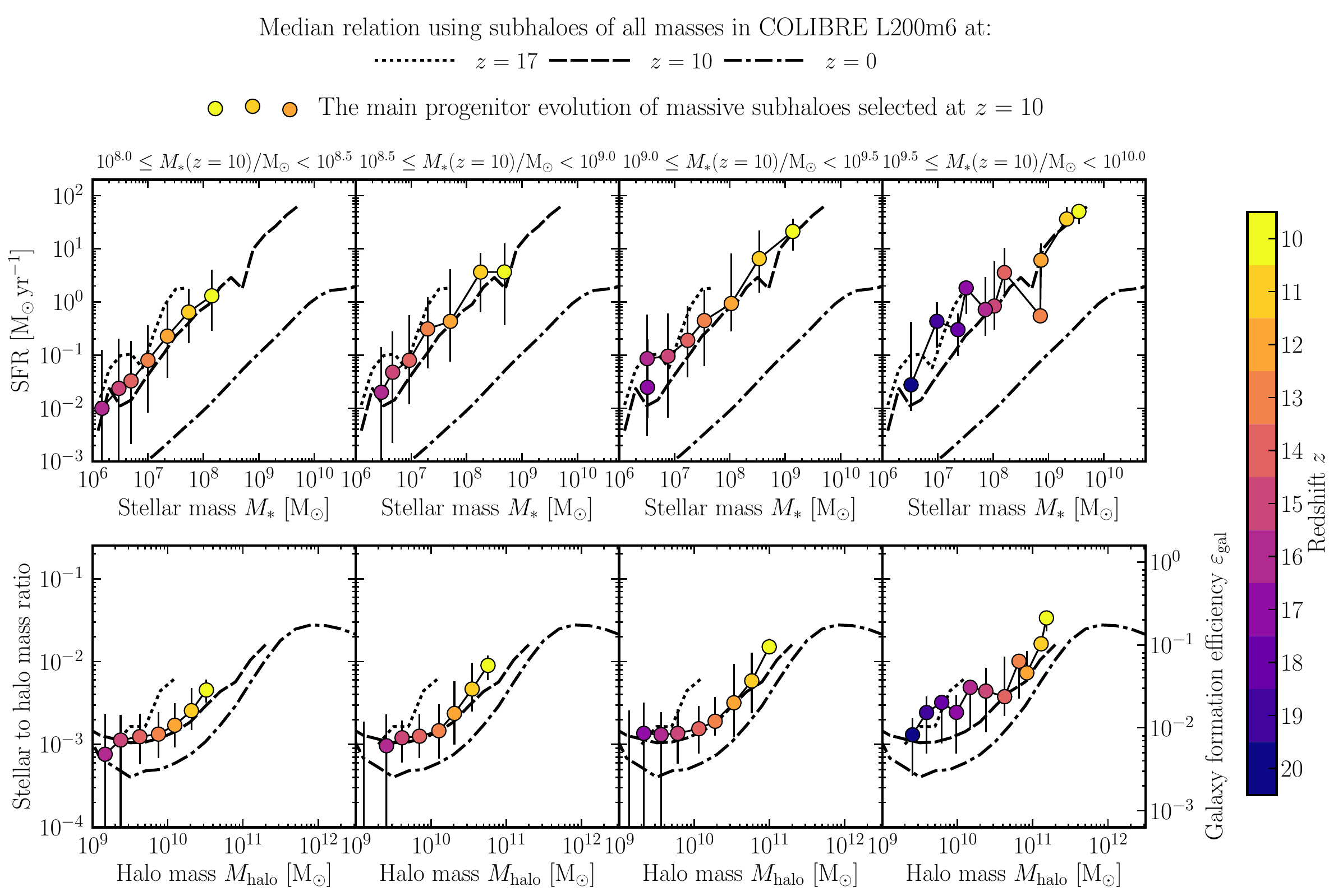}
    \caption{The redshift evolution of the relation between SFR and stellar mass (top row) and between stellar-to-halo mass ratio and halo mass (bottom row) in the \colibre{} L200m6 simulation. The $x$ and $y$ positions of the coloured circles indicate the median values over the sample of massive galaxies selected in four $z=10$ stellar mass bins, as labelled at the top of each column. The colour of the circles denotes the redshift between $z=10$ and $20$, corresponding to the main progenitors of the galaxies selected at $z=10$, and their error bars show the 16$^{\rm th}$ and 84$^{\rm th}$ percentiles. For reference, we also show the median SFR--$M_*$ relation for all star-forming galaxies ($\mathrm{sSFR}>0.2/t_{\rm H}(z)$; top panels) and the median SHMR for all central subhaloes (bottom panels) in the L200m6 simulation at redshifts $z=17$, $10$, and $0$ (dotted, dashed, and dash-dotted black lines, respectively).}
    \label{fig:sfr_and_shmr}
\end{figure*}

Fig.~\ref{fig:sfr_and_shmr} shows the evolution of the SFR--$M_*$ relation and the stellar-to-halo-mass relation for the same massive galaxies selected at $z=10$ as in Fig.~\ref{fig:evolution_mstar} (i.e. galaxies with stellar masses of at least $10^8~\mathrm{M_\odot}$ at $z=10$). Different columns correspond to different $z=10$ stellar mass bins. In each panel, the circular symbols are positioned at the median $x$ and $y$ values computed over the corresponding sample of massive galaxies tracked along their main progenitor branches. Their colour indicates redshift (from $z=20$ to $10$), while the error bars show the $16^{\rm th}$ to $84^{\rm th}$ percentiles of the corresponding $y$-value distribution.

For reference, the black curves indicate the median SFR--$M_*$ relation for all star-forming galaxies (defined as $\mathrm{sSFR}\equiv\mathrm{SFR}/M_* > 0.2 / t_{\rm H}(z)$ where $t_{\rm H}(z)$ is the Hubble time at redshift $z$) in the top row and the median stellar-to-halo mass relation (SHMR) for all central subhaloes in the bottom row, both taken from the \colibre{} L200m6 simulation at redshifts $z=17$ (dotted), $z=10$ (dashed) and $z=0$ (dashed-dotted).

The star-forming main sequence (SFMS), given by the median SFR--$M_*$ relation for star-forming galaxies, has a slope that depends only marginally on redshift, with the SFR being roughly proportional to $M_*$. In contrast, the normalization of the SFMS is a strong function of redshift, changing by 2~dex between $z=0$ and $10$ and by a further $\approx 0.5$~dex between $z=10$ and $17$. The shallower change in the latter case is expected, as the evolution of the SFMS normalization correlates with the Hubble time: the ratio of the Hubble times at $z=17$ and $10$ is about a factor of 2, whereas between $z=10$ and $0$ it is $\approx 20$. The SFRs of massive galaxies selected at $z=10$ evolve along the tracks bracketed by the $z=10$ and $z=17$ SFMSs, increasing roughly linearly with growing stellar mass. Specifically, for the most massive $z=10$ bin ($10^{9.5}\leq M_*/\mathrm{M_\odot}<10^{10}$), the SFR and $M_*$ increase from $\sim 10^{-1.5}~\mathrm{M_\odot}~\mathrm{yr}^{-1}$ and $\sim 10^{6.5}~\mathrm{M_\odot}$ at $z = 20$ to $\sim 10^{1.5}~\mathrm{M_\odot}~\mathrm{yr}^{-1}$ and $\sim 10^{9.5}~\mathrm{M_\odot}$ at $z=10$, respectively.

Unlike the normalization of the SFMS, the normalization of the SHMR for the entire population of central subhaloes decreases only moderately with redshift: by up to $\approx 0.5$~dex both from $z=17$ to $10$ and from $z=10$ to $0$, with the strongest evolution near $M_{\rm halo}\sim 10^{10}~\mathrm{M_\odot}$. This trend implies more efficient star formation at very high redshift, which emerges naturally in \colibre, even though the \colibre{} star formation prescription assumes a fixed star formation efficiency per free-fall timescale of 1 per cent \citep{2024MNRAS.532.3299N}. The stellar-to-halo mass ratio of massive galaxies selected at $z=10$ increases monotonically with cosmic time as they move along the SHMR from lower to higher masses. The growth between $z=20$ and $10$ is larger for higher $z=10$ stellar-mass bins, increasing from $\approx 0.8$~dex for the lowest-mass bin ($10^{8} \leq M_*/\mathrm{M_\odot}<10^{8.5}$) to $\approx 1.5$~dex for the highest-mass bin ($10^{9.5} \leq M_*/\mathrm{M_\odot}<10^{10.0}$). The massive galaxies in all four bins roughly follow the \colibre{} SHMR for central subhaloes at $z=10$, except for the highest-mass bin at $16<z<20$, which instead follows the $z=17$ relation. This trend is expected, as the $z=17$ relation is composed primarily of subhaloes that host galaxies from the highest $z=10$ mass bin. As the redshift approaches $z=10$, the stellar-to-halo mass ratios in all bins begin to slightly exceed the median relation for the full population of central subhaloes at the same halo mass and redshift. This behaviour is also expected, as our massive galaxies are selected based on stellar mass at $z=10$, biasing their stellar-to-halo mass ratios high relative to the median relation at that redshift.

Finally, the stellar-to-halo mass ratio can be converted into a galaxy formation efficiency by dividing it by the cosmic baryon fraction, $\Omega_{\rm b,0}/\Omega_{\rm m,0}$. The corresponding values are shown on the right-hand $y$-axis of the rightmost panel. We find that the galaxy formation efficiency increases with decreasing redshift and, at fixed redshift, with the stellar mass used for the $z=10$ selection. Specifically, at $z=14$, galaxies selected in the $10^{9}\leq M_*/\mathrm{M_\odot}<10^{9.5}$ and $10^{9.5}\leq M_*/\mathrm{M_\odot}<10^{10}$ bins have median formation efficiencies of about 1 and 3 per cent, respectively, and reside in haloes of mass $M_{\rm halo}\sim 10^{10}~\mathrm{M_\odot}$ and $10^{10.5}~\mathrm{M_\odot}$. By $z=12$, the corresponding efficiencies increase to 2 and 5 per cent, while the host halo masses increase to $M_{\rm halo}\sim 10^{10.5}~\mathrm{M_\odot}$ and $10^{11}~\mathrm{M_\odot}$. By $z=10$, the efficiencies further rise to roughly 10 and 20 per cent, and the host halo masses become $M_{\rm halo}\sim 10^{11}~\mathrm{M_\odot}$ and $2\times 10^{11}~\mathrm{M_\odot}$. The latter efficiency is comparable to the peak efficiency at $z=0$, achieved in haloes with $M_{\rm halo}\sim 10^{12}~\mathrm{M_\odot}$, as indicated by the peak of the black dash-dotted line (see also \citealt{2019MNRAS.488.3143B}).

\bsp	
\label{lastpage}
\end{document}